\documentclass[12pt]{article}

\pdfoutput=1

\usepackage{latexsym}
\usepackage{epsfig,amssymb,euscript}
\usepackage{amsmath}
\usepackage{mathrsfs}
\usepackage{ccaption}             
\usepackage[usenames]{color}

\usepackage[
      colorlinks=true,
      linkcolor=blue,
      urlcolor=blue,
      filecolor=black,
      citecolor=red,
      pdfstartview=FitV,
      pdftitle={},
        pdfauthor={Simon A. Gentle, Mukund Rangamani, Benjamin Withers},
        pdfsubject={},
        pdfkeywords={},
        pdfpagemode=None,
        bookmarksopen=true
      ]{hyperref}

  \captionnamefont{\bfseries}
  \captiontitlefont{\small\sffamily}
  \captiondelim{: }
  \hangcaption

 \newenvironment{list1}{
  \begin{list}{$\bullet$}{%
      \setlength{\itemsep}{0.02in}
      \setlength{\parsep}{0.05in} \setlength{\parskip}{0in}
      \setlength{\topsep}{0.1in} \setlength{\partopsep}{0in} 
      \setlength{\leftmargin}{0.17in}}}{\end{list}}
\newenvironment{list2}{
  \begin{list}{$\bullet$}{%
      \setlength{\itemsep}{2mm}
      \setlength{\parsep}{0.03in} \setlength{\parskip}{0in}
      \setlength{\topsep}{0.in} \setlength{\partopsep}{0in} 
      \setlength{\leftmargin}{0.3in}}}{\end{list}} 

\def\sec#1{\S\ref{#1}}
\def\fig#1{Fig.\,\ref{#1}}
\def\req#1{(\ref{#1})}
\def\App#1{Appendix \ref{#1}}

\makeatletter
\@addtoreset{equation}{section}

\def\be{\begin{equation}}
\def\ee{\end{equation}}
\def\bea{\begin{eqnarray}}
\def\eea{\end{eqnarray}}

\definecolor{purple}{rgb}{0.8,0.3,0.5}
\definecolor{green}{rgb}{0.1,0.8,0.2}
\definecolor{orange}{rgb}{0.8,0.5,0.1}

\def\SAdS#1{Schwarzschild-AdS$_{#1}$}
\def\AdS#1{AdS$_{#1}$}
\def\RNAdS#1{RN-AdS$_{#1}$}

\title{\bf A soliton menagerie in AdS}

\author{\normalsize Simon A.~Gentle, Mukund Rangamani and Benjamin Withers\\ \\
        \small \it Centre for Particle Theory \& Department of Mathematical Sciences, \\
        \small \it Science Laboratories, South Road, Durham DH1 3LE, United Kingdom. \\ \\
        \normalsize\href{mailto:s.a.gentle@durham.ac.uk}{\texttt{s.a.gentle}}\texttt{, }\href{mailto:mukund.rangamani@durham.ac.uk}{\texttt{mukund.rangamani}}\texttt{, }\href{mailto:b.s.withers@durham.ac.uk}{\texttt{b.s.withers@durham.ac.uk}}}
\date{December 2011}

\begin{document}

\setlength{\baselineskip}{16pt}

\maketitle
\begin{picture}(0,0)(0,0)
\put(350,260){DCPT-11/55}
\end{picture}
\vspace{-36pt}
\thispagestyle{empty}                        

\begin{abstract}
We explore the behaviour of  charged scalar solitons in asymptotically global AdS$_4$ spacetimes. This is motivated in part by attempting to identify under what circumstances such objects can become large relative to the AdS length scale. We demonstrate that such solitons generically do get large and in fact in the planar limit  smoothly connect up with the zero temperature limit of planar scalar hair black holes. In particular, for given Lagrangian parameters we encounter multiple branches of solitons: some which are perturbatively connected to the AdS vacuum and surprisingly, some which are not. We explore the phase space of solutions by tuning the charge of the scalar field and changing scalar boundary conditions at AdS asymptopia, finding intriguing critical behaviour as a function of these parameters. We demonstrate these features not only for phenomenologically motivated gravitational Abelian-Higgs models, but also for models that can be consistently embedded into eleven dimensional supergravity.
\end{abstract}

\pagebreak
\setcounter{page}{1}

\tableofcontents

\section{Introduction}
\label{sec:Introduction}

Gravity in AdS spacetimes is full of surprises. The presence of a negative cosmological constant manifests itself as an attractive harmonic gravitational potential well on test particles, leading one to visualise  (global) AdS spacetimes as a covariant gravitational box. One well-known consequence of this feature is the fact that the canonical ensemble for gravitational degrees of freedom is well-defined with AdS boundary conditions \cite{Hawking:1982dh}.\footnote{Recall that asymptotically AdS spacetimes have timelike boundaries, rendering them non-globally hyperbolic, thereby necessitating specification of boundary conditions at AdS ${\mathscr I}$.} Another recent discovery is the fact that AdS supports  non-trivial geon solutions \cite{Dias:2011ss}, which correspond to steady state non-dissipative condensates of bulk gravitons.

Some of these surprising features, such as black hole thermodynamics, are well understood using the AdS/CFT correspondence as a necessary consequence of the dual field theory dynamics.  Nevertheless, as exemplified by the geon solutions, in recent years various new surprises have been uncovered, mostly in the process of trying to model interesting exotic configurations in the boundary CFT using the gravitational description. Of primary interest to our considerations is the fact that asymptotically AdS spacetimes admit solitonic configurations with non-trivial matter profiles.

To motivate the discussion of solitons in AdS, let us hark back to the view of AdS as a gravitational box and consider linearised matter fields in this geometry. It is a simple matter to show that the eigenenergies of the single particle states are quantised $\omega  = \Delta + 2\,n+l$, where $n \in {\mathbb Z}_+$ denotes the harmonic level and $l \in {\mathbb Z}_+$ the rotational quantum number. $\Delta$ is a zero-point energy from the bulk perspective and translates into the conformal dimension of the corresponding CFT operator on the boundary. One naively imagines that it would be impossible to macroscopically populate the single particle energy levels to form a coherent condensate of such fields (assuming they are bosons); the harmonic gravitational potential will cause the particles to collapse to the center of AdS and we should end up with a black hole.

Many examples prove this intuition to be incorrect. For one it is possible to balance the gravitational interaction by matter repulsion (e.g., by considering charged matter as we discuss) or more simply by tuning the boundary conditions for fields at AdS ${\mathscr I}$. Since both of these ingredients will play a role in our considerations below, we now briefly review previous discussions of solitons in AdS.

The earliest construction of solitons in AdS spacetimes was achieved in a very simple model of scalar fields coupled to gravity \cite{Hertog:2004rz} by allowing for non-trivial boundary conditions for the scalar field. These boundary conditions correspond to deformations of the boundary CFT by multi-trace operators \cite{Berkooz:2002ug,Witten:2001ua}. Once one is willing to relax boundary conditions, solitons can be made to order; one demands that there be a soliton and works out what choice of boundary condition would allow it to exist. Such constructions go by the name of designer gravity \cite{Hertog:2004ns} and have been extensively investigated in the context of positive energy theorems for AdS spacetimes with multi-trace boundary conditions, \cite{Amsel:2006uf,Faulkner:2010fh}. 

A different construction of solitons is to consider charged scalar fields in AdS and allow the charge repulsion to compensate against the gravitational attraction. In fact, the equations for  BPS configurations preserving various fractions of supersymmetries in diverse dimensions have been known for a while \cite{Chong:2004ce} and admit charged supersymmetric solitonic solutions. More recently, charged solitons which asymptote to global AdS$_5$ were constructed in a perturbative expansion in the scalar amplitude in \cite{Basu:2010uz} within a simple phenomenological Einstein-Maxwell-scalar theory with negative cosmological constant. Subsequently, \cite{Bhattacharyya:2010yg}  and \cite{Bobev:2010de} constructed a family of charged scalar solitons within a particular five dimensional truncation of Type IIB supergravity to five dimensions (which is contained within the ansatz of  \cite{Chong:2004ce}).\footnote{The truncation in question can equivalently be described  as a truncation of $\mathcal{N}=8$ $SO(6)$ gauged supergravity in 4+1 dimensions, where equal charges were turned on in $U(1)^3 \subset SO(6)$.}

Our interest in this paper is with the global charged solitons in asymptotically \AdS{4} spacetimes motivated by a 
multitude of observations. First of all, an interesting question is whether global charged AdS solitons can get `large' i.e., is it possible for the soliton size to be parametrically larger than the length scale $\ell$ set by the cosmological constant? This question is not purely academic; one of the surprises about gravity in AdS is that when the characteristic size of gravitational configurations $R$ (be they solitons or even black holes) becomes large compared to the AdS scale $\ell$ we can consider a scaling regime which takes us to solutions in planar (Poincar\'e) AdS. This is of course  well known in the context of asymptotically AdS black holes and plays an important role in modelling thermal physics of the dual CFT on Minkowski space. In many of the known examples of  solitons in AdS this does not happen. For example, smooth charged solitons constructed in \cite{Bhattacharyya:2010yg} are always at most of AdS scale and do not admit passage into the planar limit.\footnote{We are simplifying a little here; the statement is strictly true for solitons which carry equal charges under all three $U(1)^3 \subset SO(6)$. Soltions which are charged under a single $U(1)$ do get arbitrarily large -- we are interested in understanding under what circumstances this can happen.} This behaviour seems at face value consistent with the known results for self-gravitating fluids in AdS \cite{Page:1985em,Hubeny:2006yu,deBoer:2009wk,Arsiwalla:2010bt}. However, in the case of self-gravitating fluids it is known that there is a critical dimension beyond which the solitons can become parametrically large in AdS units \cite{Hammersley:2007rp,Vaganov:2007at}, indicating that weakening the gravitational attraction can lead to large solitons (amusingly this critical dimension is around $11$). A natural question then is whether one  can have a different control parameter which allows solitons to get arbitrarily large. Indeed one such parameter presents itself; the charge of the bulk scalar field which controls the strength of the charge repulsion.

A second motivation for our consideration is the fact that the proclivity of charged scalar fields in AdS to condense,  as originally shown in \cite{Gubser:2008px,Hartnoll:2008vx} (in the context of planar AdS), plays a key role in the holographic modelling of superfluids. Unburdened by the no-hair theorems which beset asymptotically flat black holes,  planar AdS black holes can be unstable to the formation of scalar hair as these authors demonstrated. In \cite{Hartnoll:2008kx} it was shown that black branes with charged scalar hair can be used to model the condensed phase of a superfluid.

In these studies and applications of AdS/CFT to finite density systems (see \cite{Hartnoll:2009sz,Herzog:2009xv,Horowitz:2010gk,Hartnoll:2011fn} for reviews of these developments) one is interested in the behaviour of the CFT in Minkowski spacetime in the grand canonical ensemble, i.e., at non-zero temperature and charge chemical potential (both of which are held fixed). On the other hand, investigating such theories with global AdS boundary conditions allows us to identify the CFT states corresponding to the condensation of bulk charged fields. Moreover, we can look at fixed charge sectors of the dual field theory and ask what the spectrum of CFT states looks like from the bulk viewpoint i.e, chart out the microcanonical phase diagram.  These considerations partly motivated the earlier work of global charged solitons \cite{Basu:2010uz, Bhattacharyya:2010yg}, wherein it was argued that the global solitons are the true ground states of the theory with fixed charge. Naively one would have expected that the ground state with fixed charge corresponds to a charged black hole in AdS, but the fact that the scalars can condense implies that one exits the black hole phase and enters the phase of the charged solitons below a critical mass.

However, the previous discussion of scalar solitons remaining bounded in size turns out to imply that the spectrum of smooth solitons terminates on a critical solution with maximal mass (this happens even for the supersymmetric solitons of \cite{Bhattacharyya:2010yg}). Equivalently in the microcanonical phase diagram one has a critical charge for the CFT states. While one can certainly crank up the charge of the state beyond this value, it was conjectured in  \cite{Bhattacharyya:2010yg} that for larger values of CFT charge, one ends up with a singular global bulk soliton. Given our earlier observation regarding the strength of the bulk scalar interactions (equivalently the charge quantum number of the dual CFT operator), it is interesting to ask what happens to the microcanonical phase diagram as we dial this parameter. 

Returning to the context of the holographic superfluids, it is widely believed that solitons provide the zero-temperature limit of these superfluid black branes.  For example, charged scalar configurations which asymptote to planar AdS$_4$ were constructed numerically for a restricted range of parameters in \cite{Gubser:2009cg,Horowitz:2009ij} and were shown to back up this hunch.  Once again we have a poser: if for choice of bulk Lagrangian one does not have a large global soliton, then how is it possible that we obtain a planar solution with a condensed scalar? One possibility is that the planar solution descends from a singular global soliton as for instance argued in \cite{Bhattacharyya:2010yg}, but it is not clear if this is generic. More to the point, we need to address what class of singular global solutions are admissible . After all, if one was only interested in constructing solutions with desired AdS asymptotics and was willing to be agnostic about the interior core region, then the problem is trivial. We would have solutions for any choice of boundary data by simply integrating in the asymptotic data to a timelike singularity.  In this paper we provide numerical and (some) analytical evidence to connect global and planar solitons with hairy black holes and branes more generally.

A final motivation for our considerations comes from  studies of superfluid black branes within consistent truncations of supergravity; the first examples of such in 3+1 dimensions were given in \cite{Gauntlett:2009dn}.  Many more truncations of $\mathcal{N}=8$ $SO(8)$ gauged supergravity in 3+1 dimensions were studied recently in \cite{Donos:2011ut}, where it was found that the behaviour and existence of solutions was highly dependent on the details of the truncation.  For example, it was shown that within the truncation where equal charges were turned on in $U(1)^4 \subset SO(8)$ (which is the \AdS{4} analog of the truncation studied in \cite{Bhattacharyya:2010yg}), the hairless extremal Reissner-Nordstr\"om black brane dominates the grand canonical ensemble at low temperatures. This is despite the fact that one has new branch of hairy black hole solutions -- these however turn out to be subdominant as they have larger free energy. The situation is in strong contrast to other truncations such as the one originally investigated in \cite{Gauntlett:2009dn}.\footnote{Such behaviour has been noticed earlier in \cite{Gubser:2009fk} (we thank Sean Hartnoll for alerting us to this) and has also been encountered in other \AdS{5} gauged supergravity truncations  \cite{Aprile:2011uq}. These examples have an important lesson to impart: the linearised/probe analysis, as discussed for example in \cite{Denef:2009tp}, is at best only useful to indicate where new solution branches exist. It cannot tell us whether the new branch of solutions is actually the true saddle point configuration.} During our study of solitons in these truncations we will also find clues to help explain this behaviour. 
  
One issue with consistent truncations is the lack of parameters that one can dial. While in phenomenological bottom-up models one has the advantage of using the bulk Lagrangian parameters or equivalently positing the spectrum of the dual CFT by hand, in supergravity consistent truncations these are a-priori fixed.\footnote{Note that in contrast to popular misconception, existence of a consistent truncation does not imply that the dual field theory is known. A supergravity consistent truncation only gives restricted information about the dual CFT spectrum: the quantum numbers of a few chiral operators at best.} However, the fact that we are interested in asymptotically AdS spacetimes comes to our rescue; we can relax boundary conditions on some fields (and we shall mostly do so only for scalar fields) to take advantage of multi-trace relevant operators (if the CFT admits such). For the examples we study this dial is present and as in the recent investigations of \cite{Faulkner:2010gj,Iqbal:2011aj} we can use this dial to explore the microcanonical phase structure (albeit of the deformed CFT).  To a certain extent we will see that relaxing the boundary condition on scalar fields is tantamount to reducing the effect of the gravitational attraction; its effect is qualitatively similar to increasing the charge repulsion or going up to higher dimensions.
 
 With these motivations in mind, we undertake a study of charged scalar solitons in \AdS{4} in three specific examples of bulk theories: (i) a phenomenological Abelian-Higgs model with a massive charged scalar, (ii) the consistent truncation of \cite{Gauntlett:2009dn} and (iii) a consistent truncation of 11-dimensional supergravity originally studied in \cite{Chong:2004ce} (and more recently in \cite{Donos:2011ut}). We find a bevy of surprising results, most notably the fact that the spectrum of global charged solitons is a lot richer than what previous investigations would suggest. 
 
 In particular, we show that even in theories where one encounters a family of solitonic solutions with a maximal mass, there appears to be a second branch of solitons, completely disconnected from the former branch.\footnote{Similar features have been observed for charged solitons  with V-shaped potentials in asymptotically flat spacetimes \cite{Kleihaus:2009kr}.} A distinguishing feature of the new branch of solutions is that they cannot be ascertained by perturbation analysis about global AdS as in \cite{Basu:2010uz}; in these solutions a macroscopic amount of scalar is turned on. Moreover, in this new branch the mass of the soliton is unbounded from above. As we tune parameters of the bulk Lagrangian or indeed modify boundary conditions, we see new critical behaviour; the branches of solitons merge and subsequently split, leading to a rather rich structure.  We not only tackle the behaviour of the global solitons, but also go on to show how these smoothly join up with the planar solutions constructed as zero temperature ground states for superfluids. Furthermore, we also show where these solutions lie in microcanonical phase diagram, which itself exhibits interesting features.
 Whilst it is tempting to conjecture that the solitons we construct are the ground states for the CFT on $   {\mathbb R} \times {\bf S}^2$ at a fixed value of charge, this statement appears to depend sensitively on the theory in question. In particular, it can transpire in certain cases that one encounters global black holes with or without scalar hair which are lighter than a soliton at a given value of the charge \cite{Basu:2010uz,Dias:2011tj}. Our focus will be primarily on solitons and we only briefly touch upon the nature of global black holes. 
  
The outline of this paper is as follows. We begin in \sec{sec:Theories} with the basic set-up of the problem and describe the various  theories we consider along with a discussion of the general features common to all examples. Subsequently in  \sec{sec:Results} we present a summary of our main results as a compendium for quick reference. We then move on to illustrate the details of our constructions in the following three sections: \sec{sec:BU} deals with the phenomenological Abelian-Higgs model, while \sec{sec:GSW} and \sec{sec:DG} contain details of the $SU(3)$ and $U(1)^4$ consistent truncations of 11-dimensional (11D) supergravity.  We conclude in \sec{sec:Discussion} with a discussion and our thoughts for further work.

{\em Note added:} Reference \cite{Dias:2011tj} which appears simultaneously with our work on the archive
explores in a comprehensive manner the microcanonical phase diagram of a phenomenological Abelian-Higgs model in \AdS{5}. This work includes a detailed discussion of charged global scalar hair black holes, which we do not undertake in this paper.


\section{Generalities}
\label{sec:Theories}


In this section we outline the general features common to all the theories we consider so as to set the stage for our discussion of solitons. We describe the bulk theories and the ansatz  we work with,  along with details of the choice of boundary conditions  at our disposal. After a brief preview of known analytic solutions, we quickly review the planar limit of global solitons which forms a crucial component of our considerations in later sections.

\subsection{Theories, Ans\"atze and boundary conditions}
\label{sec:tab}

We will begin our exploration of gravitating global solitons in the familiar territory of phenomenological models for superfluid instabilities \cite{Hartnoll:2008kx}. This family of theories contains all the ingredients we require to build global soliton solutions, whilst providing some useful dials with which we can navigate the space of solutions. In detail, the family of theories are parameterised by the mass $m_\psi$ and charge $q$ of the scalar field
\begin{equation}\label{SHHH}
S = \frac{1}{16\pi G_4} \int d^4x \sqrt{-g} \left(R + \frac{6}{\ell^2} - \frac{1}{4}\hat{F}^2 - \left|d\psi - i q \hat{A} \ell^{-1} \psi\right|^2 - m_\psi^2 |\psi|^2\right),
\end{equation}
supplemented by appropriate boundary terms. $\hat{F}=d\hat{A}$ is the Maxwell field strength and $\ell$ is the AdS length.  We will find it convenient to work with manifestly real and gauge invariant quantities, which can be achieved at the level of the action by absorbing the phase $\alpha$ of the scalar $\psi  = \phi\, e^{i\alpha}$ into the gauge field, $A \equiv \hat{A} - q^{-1} \ell  d\alpha$,
\begin{equation}\label{Sbottomup}
S = \frac{1}{16\pi G_4} \int d^4x \sqrt{-g} \left(R  - \frac{1}{4}F^2 - \left(\partial\phi\right)^2 -\frac{1}{\ell^2}q^2  \phi^2 A^2 - \frac{1}{\ell^2}\left(-6 + m_\phi^2\ell^2\,\phi^2\right)\right),
\end{equation}
where $\phi$ is real and $m_\phi \equiv m_\psi$. The equations of motion of this new action are the same as those arising from \eqref{SHHH}.\footnote{The equation of motion for the phase $\nabla_\mu \left(A^\mu \phi^2\right) = 0$ follows from the Maxwell equations, and is a consequence of conservation of the global $U(1)$ current, $J^\mu \propto \phi^2A^\mu  $. Recall that $\alpha$ is contained in $A$ in a gauge invariant manner.} As we will see momentarily, the advantage of this form is that the action may be more readily compared with consistent truncations. 

Working with phenomenological models is instructive, but has a number of drawbacks. There is the concern that any interesting features might be artefacts of the chosen action. Indeed, one could consider infinitely many alterations of the action \eqref{Sbottomup}. In a phenomenological context it is also more difficult to discuss consequences for any CFT dual. For these reasons we also adopt two consistent truncations of 11D supergravity with the same field content. In previous work \cite{Bhattacharyya:2010yg}, global soliton solutions were constructed within a truncation of $\mathcal{N}=8$ $SO(6)$ gauged supergravity in 4+1 dimensions, where equal charges were turned on in  $U(1)^3 \subset SO(6)$. Here we will be interested in a lower dimensional analogue, employing truncations of 3+1 dimensional $\mathcal{N}=8$ $SO(8)$ gauged supergravity. Superfluid black branes for various truncations of this $SO(8)$ theory were considered in \cite{Donos:2011ut}.\footnote{These truncations were originally investigated in \cite{Chong:2004ce}.} We will consider the equally charged truncation $U(1)^4 \subset SO(8)$ \cite{Donos:2011ut} and we will consider also a single field truncation first studied in \cite{Gauntlett:2009dn} (the M-theory superconductor construction). Whilst this latter single field truncation admits a $SU(4)$ invariance, it is part of a larger $SU(3) \subset SO(8)$ truncation \cite{Bobev:2010ib} explored in \cite{Donos:2011ut}, as such we shall refer to this as the $SU(3)$ truncation.

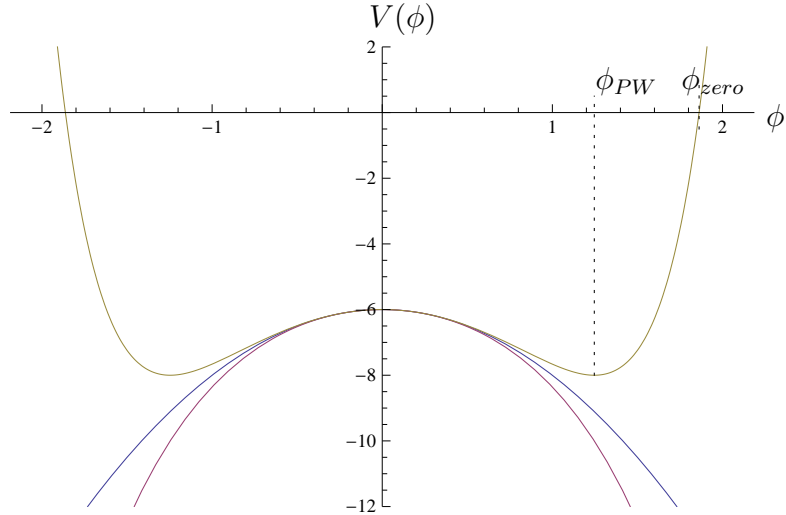
\begin{figure}[h!]
\begin{center}
\includegraphics[width=0.6\columnwidth]{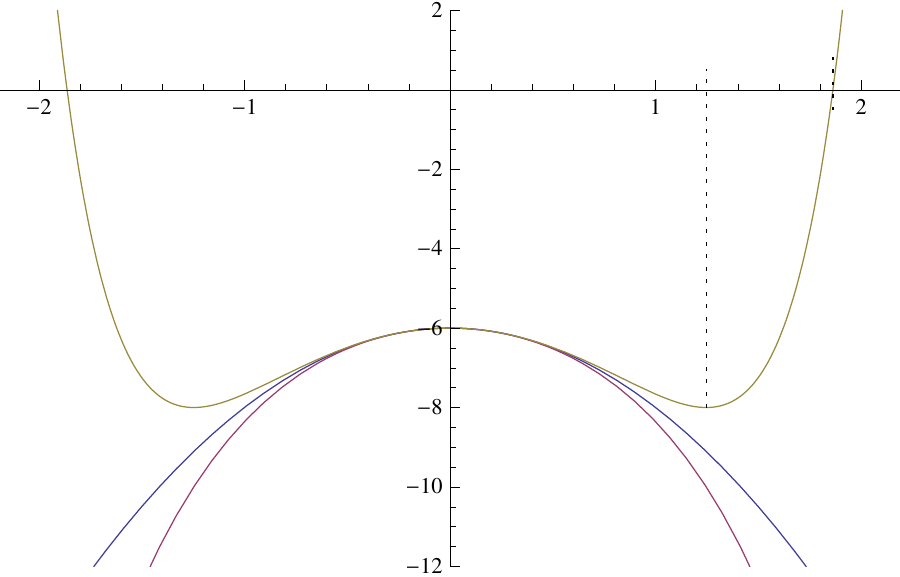}
\setlength{\unitlength}{0.1\columnwidth}
\begin{picture}(0.3,0.4)(0,0)
\put(-3,4){\makebox(0,0){$V(\phi)$}}
\put(0,3.2){\makebox(0,0){$\phi$}}
\put(-1.2,3.5){\makebox(0,0){$\phi_{PW}$}}
\put(-0.5,3.5){\makebox(0,0){$\phi_{zero}$}}
\end{picture}
\end{center}
\caption{The scalar potentials for the three models we consider. The $SU(3)$ truncation has the potential with a global minimum at $\phi_{PW}$ and a zero at $\phi_0$. The Ableian-Higgs potential and the $U(1)^4$ truncation have potentials that are unbounded from below, with the latter being exponential (thus steeper) while the former is quadratic.}\label{Vphiplot}
\end{figure}

Moving forward we employ a bulk action which is sufficiently general to encompass each of these theories:
\begin{equation}\label{Sgeneral}
S_{bulk} = \frac{1}{16\pi G_4} \int d^4x \sqrt{-g} \left(R  - \frac{1}{4}F^2 - \left(\partial\phi\right)^2 - \frac{1}{\ell^2}Q(\phi) A^2 -\frac{1}{\ell^2}V(\phi)\right).
\end{equation}
Equations of motion for this action as well as the required boundary terms are presented in Appendix \ref{sec:generalEoms}. The specific cases of the charge-coupling $Q(\phi)$ and potential $V(\phi)$ for the theories of interest are summarised in Table \ref{theorytable}.
\begin{table}[h!]
\begin{center}
\begin{tabular}{  c | c | c | c | c}
   Theory & $Q(\phi)$ & $V(\phi)$ & mass$^2$$\,\ell^2$ & charge  \\
   \hline 
  Phenomenological \cite{Hartnoll:2008kx} & $q^2  \phi^2$ & $-6 + m_\phi^2\ell^2\,\phi^2$ & $m_\phi^2\ell^2$ & $q$ \\
   $SU(3) $ \cite{Gauntlett:2009dn}& $\frac{1}{2} \,\sinh^2{\sqrt{2}\phi}$ & $\cosh^2{\frac{\phi}{\sqrt{2}}}\,\left(-7+\cosh{\sqrt{2}\phi}\right)$ & $-2$ & $1$ \\
   $U(1)^4$ \cite{Donos:2011ut}& $\frac{1}{2} \,\sinh^2{\frac{\phi}{\sqrt{2}}}$ & $-2\left(2+\cosh{\sqrt{2}\phi}\right)$ & $-2$ & $\frac{1}{2}$
\end{tabular}
\end{center}
\caption{Components of the action \eqref{Sgeneral} for the theories under consideration, together with the mass and charge for the scalar field fluctuations about the $\phi=0$ extremum  of $V(\phi)$.\label{theorytable}}
\end{table}
For the remainder of this work, when discussing the phenomenological theory we will specialise to the case $m^2_\phi\, \ell^2=-2$.\footnote{Many of the features we encounter generalise to other values of $m^2_\phi\, \ell^2$. Heuristically we expect the properties of the solutions we consider to depend on the charge to mass ratio of the scalar field, which after all controls the balance of Maxwell repulsion versus gravitational attraction.} For orientation we show the scalar potentials for the three cases we consider in \fig{Vphiplot}; some of these features will play a role in our analysis later. We will set the AdS length scale $\ell = 1$ in what follows.
%

\subsubsection{Ansatz for global solitons}
\label{s:gsanz}
Throughout this paper we will employ the following metric ansatz (preserving $ \mathbb{R}_t\times SO(3)$) for spherically symmetric, static global solutions: 
\begin{equation}
ds^2 = - g(r)e^{-\beta(r)} dt^2+ \frac{dr^2}{g(r)}  + r^2 d\Omega_2^2
\end{equation}
where $d\Omega_2^2$ is the line element for a unit ${\bf S}^2$. For the vector and scalar field we take $A = A(r) dt$ and $\phi = \phi(r)$. Thus for the solutions of interest $\xi = \partial_t$ is a timelike Killing vector field and since ${\cal L}_\xi \phi = {\cal L}_\xi A =0$ our solutions are globally static solitons. 

With this ansatz, the boundary of global AdS is located at $r\to \infty$, where the fields admit the asymptotic expansion 
\begin{subequations}
\begin{align}
A(r) &= \mu - \frac{\rho}{r}+\ldots    \label{Aasym}\\
\phi(r) &= \frac{\phi_1}{r} + \frac{\phi_2}{r^2}+\ldots \label{phiasym}\\
\beta(r) &= \beta_\infty + \ldots \label{betaasym}\\
g(r) &= r^2 + 1+ \frac{\phi_1^2}{2} -\frac{g_1}{r}+\ldots  \label{gasym}
\end{align}
\label{asymfalloffs}
\end{subequations}
for each of the theories listed in Table~\ref{theorytable}. We will use coordinate freedom in $t$ to set $\beta_\infty = 0$. We will require in most cases that the solutions are regular at the origin in the IR, which determines the following small $r$ expansion near $r=0$:
\begin{subequations}
\begin{align}
A(r) &= A_c +\ldots    \label{Acore}\\
\phi(r) &= \phi_c+\ldots \label{phicore}\\
\beta(r) &= \beta_c + \ldots \label{betacore}\\
g(r) & = 1 + \ldots \label{gcore}
\end{align}
\label{corefalloffs}
\end{subequations}
In this expansion $g(r)$ is completely determined and the expansion proceeds in powers of $r^2$ as required.

\subsubsection{Boundary conditions for the scalar}
\label{sec:bcscalar}

Since the scalar field $\phi$ has mass $m_\phi^2=-2$ which is the conformal mass term in \AdS{4} we can not only impose the standard ($\phi_1\ {\rm fixed}$) and alternate ($\phi_2\  {\rm fixed}$) boundary conditions for the scalar, but we can also consider more general multi-trace boundary conditions. For the standard and alternate boundary conditions we denote the CFT operator dual to $\phi$ as ${\mathcal O}_{\phi_1}$ and ${\mathcal O}_{\phi_2}$, with dimensions $\Delta =2$ and $\Delta =1$ respectively. Note that as a result we have $\langle\,{\mathcal O}_{\phi_1}\,\rangle = \phi_2$ etc.. In the $U(1)^4$ truncation, turning on the dimension 1 operator is a supersymmetric deformation of the CFT \cite{Chong:2004ce}. As described in \cite{Berkooz:2002ug,Witten:2001ua} the multi-trace boundary conditions are in general given by picking a local functional $\phi_2(\phi_1)$. Of particular interest to us will be the relevant double-trace operator $({\mathcal O}_{\phi_2})^2$ which we will turn on to deform the CFT,
\begin{equation}
\delta S_{CFT} \propto \int d^3x\,\varkappa\,  {\mathcal O}_{\phi_2}^2
\end{equation}
which is achieved by choosing 
\begin{equation}
\phi_2(\phi_1) = \varkappa \, \phi_1
\label{dtbc}
\end{equation}	
Such a deformation will be useful for exploring the structure of solutions arising in consistent truncations. Previous work involving the use double-trace deformations in the context of holographic superfluids is \cite{Faulkner:2010gj}. 

\subsubsection{Conserved charges from asymptotics}
\label{sec:consecharges}

Once we have fixed our boundary conditions, we can immediately read off the conserved charges. From the fall-off of the scalar field we learn that for standard and alternate boundary conditions, the vacuum expectation value (vev) of the dual operator is
\begin{equation}
\langle\,{\mathcal O}_{\phi_1}\,\rangle = \phi_2 \;\; (\Delta =2 , \varkappa \to \infty)    \ , \qquad \langle\,{\mathcal O}_{\phi_2}\,\rangle = \phi_1 \;\; (\Delta = 1 , \varkappa =0)   
\label{vevst}
\end{equation}	
while for double-trace deformations,
\begin{equation}
\qquad \langle\,{\mathcal O}_{\phi_2}\,\rangle = \phi_1 \;\; ( \varkappa \neq 0)   
\label{vevdt}
\end{equation}	
The fall-off of the gauge field tells us that that boundary chemical potential is set by $\mu$, while $\rho$ gives the charge density. The mass density of our solution is obtained from the fall-off of the metric and the scalar field (see \App{sec:generalEoms}); for double-trace boundary conditions \req{dtbc} we have:
\begin{equation}
m = g_1 +\frac{3}{2} \, \varkappa \, \phi_1^2 \,,
\label{mg1phi}
\end{equation}	
while for solutions with vevs for single-trace operators (i.e., either $\phi_1 \neq 0$ or $\phi_2 \neq 0$) 
\begin{equation}
m = g_1\,.
\label{msingtr}
\end{equation}	
%

\subsection{Exact solutions}\label{sec:exact}

For each of the theories in Table~\ref{theorytable} we have simple solutions at $\phi=0$, where  $V'(0)=Q'(0)=Q(0)=0$. At this point in moduli space we have a family of exact solutions, the Reissner-Nordstr\"om-\AdS{4} (henceforth \RNAdS{4}) black holes,
\begin{equation}
g(r) = r^2+1-\frac{m}{r}+\frac{\rho^2}{4r^2},\quad \beta(r) = 0,\quad A(r)=\mu-\frac{\rho}{r} \qquad \text{at}\quad\phi(r) = 0 \label{globalschw}
\end{equation}
which for $m=\rho=0$ is global AdS$_4$ with radius of curvature $\ell$ (here set to 1). Fluctuations of the scalar field about this point have masses and charges given by the last two columns in Table \ref{theorytable}.

\subsubsection{Other vacua in $SU(3)$}

For the phenomenological theory and the $U(1)^4$ equal charged truncation, $\phi =0$ is the only extremum of $V(\phi)$. However, for the single scalar truncation of $SU(3)$ there are two more located at $\phi = \phi_{PW} =  \pm \sqrt{2}\text{arccosh}{\sqrt{2}}$.\footnote{This vacuum was shown to be perturbatively unstable \cite{Bobev:2010ib}.  We anticipate that the properties of the solutions constructed and conclusions drawn in this paper will not depend on this fact. One could instead study a perturbatively stable alternative, for example the single field truncation of \cite{Bobev:2011rv}. We thank Nikolay Bobev for discussions on this issue.} At this point we have another one-parameter family of exact solutions,
\begin{equation}
g(r) = \frac{4}{3}r^2+1-\frac{m}{r},\quad \beta(r) = 0,\quad A(r)=0 \qquad \text{at}\quad\phi(r) = \pm \sqrt{2}\,\text{arccosh}{\sqrt{2}}\label{globalpw}
\end{equation}
which for $m=0$ is global AdS$_4$ with radius of curvature $\frac{\sqrt{3}}{2}$.

\subsubsection{Singular neutral solitons in $U(1)^4$}

We also find the following one parameter family of singular solutions in the $U(1)^4$ truncation, parameterised by the $\Delta = 1$ scalar fall off, $\phi_1$. The planar limit of this singular global solution will be discussed later and will play an important role in the supersymmetric solutions of this theory.
\begin{equation}
g(r)  = r^2 + 1+\frac{\phi_1^2}{2}+\frac{\phi_1^2}{2 r^2},\quad e^{-\beta(r)} g(r)=  r^2+1,\quad A(r)=0 \quad \phi(r) = \sqrt{2}\,\text{arcsinh}{\frac{\phi_1}{\sqrt{2}r}}\label{globalsingulardg}
\end{equation}
This solution corresponds to a neutral soliton with no conserved charges, but has a non-trivial vev $\langle{\cal O}_{\phi_2}\rangle$ -- from the asymptotic scalar fall-off it is clear that there is no deformation in the dual CFT. The solution above corresponds to a designer gravity soliton where the  $\Delta =1$ operator ${\cal O}_{\phi_2}$ of the CFT spontaneously acquires a vev breaking the $U(1)$ global symmetry on the boundary.
 
\subsection{The planar limit of global AdS solutions}
\label{sec:planarlimit}

In this paper we will be faced with families of global solutions; indeed, we have already encountered some examples in \sec{sec:exact}. We will be able to construct planar solutions by taking a limit of these global families of solutions, if the appropriate limit exists. This will allow us to make comparisons with solutions in planar AdS, many of which are already constructed. 

Heuristically, the planar limit can be visualised as arising when the object of interest (say the soliton) becomes parametrically larger in size when compared with the AdS scale $\ell$ (here set to $1$). In a sense, the object is reaching out towards the AdS boundary and hence one can zoom into a local patch.  As we do so, the effects of curvature become less significant and the boundary begins to look planar. Of course, in the process of zooming we must ensure that masses, charges or vevs can be made sufficiently large, sufficiently fast in order that we do not end up in vacuum Poincar\'e AdS. The goal of this section is to quantify this scaling limit.

Consider the simple example of the global \SAdS{4} solution, the $\rho=0$ case of \eqref{globalschw}:
\begin{equation}
ds^2 = -r^2 \left(1+\frac{1}{r^2} - \frac{m}{r^3}\right) dt^2 + \frac{dr^2}{r^2 \left(1+\frac{1}{r^2} - \frac{m}{r^3}\right)} +  r^2 d\Omega_2^2
\end{equation}
This is a one-parameter family of solutions; a physical measure of the mass can be constructed by taking an appropriate ratio with the curvature scale of the boundary, for example. For a fixed solution on this branch, there is no analogue of the Poincar\'e AdS scaling symmetry, which applied as a coordinate scaling
\begin{equation}
r\to \lambda \,r,\qquad t \to \lambda^{-1}t
\end{equation}
alters the metric in the following way
\begin{equation}
ds^2 = -r^2 \left(1+\frac{1}{\lambda^2r^2}- \frac{m}{\lambda^3r^3}\right) dt^2 +\frac{dr^2}{r^2\left(1+\frac{1}{\lambda^2r^2}- \frac{m}{\lambda^3r^3}\right)} + r^2 \lambda^2 d\Omega_2^2.
\end{equation}
Taking $\lambda$ finite does not change the boundary curvature; after all, we are only making a change of coordinates. However, if we take the singular limit $\lambda \to \infty$ of this coordinate transformation, then the boundary curvature becomes zero. In this limit, curvature and mass terms disappear and we are left with Poincar\'e AdS. If however, we simultaneously move along the branch of solutions,
\begin{equation}
r\to \lambda \,r,\qquad t \to \lambda^{-1}t,\qquad m\to \lambda^3\, m, \qquad  \lambda^2\,d\Omega_2^2 \to d{\bf x}_2^2
\end{equation}
then in the $\lambda\to\infty$ limit a mass term survives
\begin{equation}\label{planarschw}
ds^2 = -r^2 \left(1- \frac{m}{r^3}\right) dt^2 + \frac{dr^2}{r^2 \left(1 - \frac{m}{r^3}\right)} +  r^2 \,d{\bf x}_2^2
\end{equation}
namely, we have recovered planar \SAdS{4}. Note that each choice of $m$ in \eqref{planarschw} is equivalent up to coordinate transformations, so it is a zero-parameter family. Another way of saying this is that $m$ has scaling dimension $3$, and without another parameter we cannot construct a quantity invariant under the scaling symmetry (where we now must include $x^i$ rescaling). It is important to emphasise that in order to obtain a non-trivial solution in the planar limit it was necessary to have a branch of solutions with an unbounded asymptotic coefficient. In doing so we have moved from a one-parameter family of global solutions to a single planar solution. Similarly, we may obtain planar solutions by applying this procedure to the global solutions \eqref{globalpw} and \eqref{globalsingulardg}.

More generally we can consider a family of global solutions characterised by their asymptotic expansion \eqref{asymfalloffs}. For concreteness we focus on a one-parameter family of solutions parameterised by $v$. The asymptotic coefficients on these solutions will be functions of $v$, and in general we will have five of them,\footnote{$\beta_\infty$ may be set to zero by coordinate freedom in $t$.}
\begin{equation}
m(v),\quad \mu(v),\quad \rho(v),\quad \phi_1(v),\quad \phi_2(v).
\end{equation}
with scaling dimensions $3,1,2,1,2$ respectively. If any one of these is unbounded from above, it is easy to show that a planar solution exists with a non-trivial asymptotic expansion. The global \SAdS{4} solution considered above is one example. 

The number of asymptotic charges that survive in the planar limit depends not only on whether that parameter is unbounded, but also on how fast it grows along the branch compared with other coefficients. Let us work with scaling dimension $1$ quantities, 
\begin{equation}
\tilde{c}_{I}(v) \equiv c_{I}(v)^\frac{1}{\Delta(c_I)},
\end{equation} 
where $\Delta(c_I)$ denotes the scaling dimension of $c_I$. We require at least one coefficient to be unbounded along the branch, and in this regime we may study their growth assuming a power-law $v$ dependence,
\begin{equation}
v \frac{d}{dv}\tilde{c}_{I}(v) = p_I\,\tilde{c}_{I}(v).
\end{equation}
In particular we can choose our parameter along the branch, $v$, to coincide with $\tilde{c}_I$ for the largest $p_I$. Then, in this new parameterisation, the coefficient $\tilde{c}_I$ will survive the planar limit if it scales as 
\begin{equation}
v \frac{d}{dv}\tilde{c}_{I}(v) = \,\tilde{c}_{I}(v).
\end{equation}
as we send $v\to \infty$.

Let us take the example of an unbounded mass, $m^{1/3}$, growing fastest out of all five coefficients. We know at least that $m$ survives the limiting process. In order for the other coefficients to survive in the planar limit, they must grow as
\begin{equation}
c_I(m) \propto m^{\frac{\Delta(c_I)}{3}}.
\end{equation}
For example, the $\Delta =1$ vev must be unbounded and grow $\phi_1 \propto m^\frac{1}{3}$ at large $m$, if we are to have a dimension $1$ vev in the planar theory. If at least two coefficients survive in the planar limit, they can be used to construct the physical parameters of the planar solution. For example, $m\,\phi_1^{-3}$ is a quantity with scaling dimension $0$ and is therefore a physical parameter on the space of planar solutions.

The above results trivially generalise to higher dimensions: note that $m$, the mass density, has scaling dimension $d$ in \AdS{d+1}. From this we can infer an important point:  for a conserved $U(1)$ charge $\rho$ it is easy to show using the above that $\rho \propto m^{\frac{d-1}{d}}$ in \AdS{d+1}. On the other hand the BPS equations for supersymmetric solutions (relevant in particular for $d\leq 6$) imply that $m\propto \rho$. Thus if we had a global \AdS{d+1} supersymmetric solution and consider scaling it to the planar limit, we will necessary end up with an $m=0$ planar solution. Note that this general argument, based on scalings alone, does not exclude the possibility that a supersymmetric global solution can connect with a planar solution with $m=0$ and $\rho\neq 0$ and in fact we will see an example of this in \sec{s:dgbps}. Of course, with such scalings there is a danger that the resulting solution with planar symmetry can be horribly unphysical: a case in point is the scaling limit of $m = \rho$  \RNAdS{4} solution which leads to a zero mass, but non-zero charge planar singular solution. 

In the following sections we will see examples of branches of solutions where all asymptotic coefficients are bounded, and therefore do not connect with the planar limit. We will also see cases where all coefficients grow in exactly the right way to survive taking the planar limit and an example where only the vev of the $\Delta =1$ operator ${\cal O}_{\phi_2}$,  i.e., $\phi_1$, survives. The latter example will actually correspond to a supersymmetric solution scaled up to the planar limit consistent with the observation above.


\section{Summary of results}\label{sec:Results}


To guide the reader through the rest of the paper, we summarise the key results of our findings in this section, postponing the details to later sections. For brevity we denote the CFT quantities characterizing the bulk gravitational solution (parameterised by the core value of the scalar $\phi_c$) collectively via ${\cal X} = \{ m, \rho, \langle {\mathcal O} \rangle \}$, where $m(\phi_c)$ is the ADM (boundary) energy density, $\rho(\phi_c)$ the Maxwell charge density and $\langle {\mathcal O} \rangle (\phi_c)$ the vev for the dual CFT operator.  We will refer to this data as the asymptotic charges of the solution since they correspond to the one-point function of conserved operators ($T_{\mu\nu}$ and $J_\mu$) and of $\mathcal{O}$ and serve to characterise the CFT state. 

\subsection{Solitons in the Abelian-Higgs theory}
\label{sec:sumBU}

The simplest bulk Lagrangian is the Abelian-Higgs model coupled to gravity \eqref{Sbottomup}. We fix the mass of the scalar field $m^2_\phi= -2$ and study the space of solutions as a function of $q$. We will focus on $q > q_{ERN}$ so that the solitons we find are lighter than the corresponding \RNAdS{4} solution at fixed total charge.

We find the following spectrum of smooth global solitons in the theory (the results we quote are independent of the choice of boundary condition for the scalar field):
\begin{list1}
\item For any value of $q \neq 0$, there is a branch of smooth solitons that is continuously connected to the global \AdS{4} vacuum. These solutions can be parameterised in terms of the core value of the scalar $\phi_c$. Solutions with $\phi_c \ll 1$ can be determined in a perturbation expansion around global \AdS{4} as originally described in \cite{Basu:2010uz} (cf., \App{sec:pertsolitons}). Going beyond small core scalar values we find:
\begin{list2}
\item[(i)] For $q < q_c$ we find that the branch of solutions connected to the \AdS{4} vacuum has bounded asymptotic charges which  are non-monotone functions of $\phi_c$. They attain maximum values at some finite value of $\phi_c$ and undergo damped oscillations as a function of $\phi_c$ about an asymptotic value of the charge, ${\cal X}^{(1)}_\infty$. This is similar to the behaviour seen for stars in AdS \cite{Page:1985em,Hubeny:2006yu,Hammersley:2007rp,Vaganov:2007at} and also for charged solutions in \AdS{5} 
\cite{Bhattacharyya:2010yg}. In the context of radiation stars, the solutions are damped to a singular solution with infinite core density, which exhibits an invariance under a scaling symmetry \cite{Sorkin:1981wd, Vaganov:2007at, Heinzle:2003ud}. In this sense we say that the large $\phi_c$ solutions are controlled by an attractive fixed-point under the action of a scaling, or for brevity, \emph{attractor solutions}.\footnote{We emphasise that we have not investigated the scaling characteristics at large $\phi_c$ of the solutions presented, though we will study the other scaling limit, viz., the planar limit, introduced in \sec{sec:planarlimit} extensively.}

\item[(ii)] For $q > q_c$ the behaviour of the smooth solitons is drastically different: we now find that the asymptotic charges are  monotonically increasing functions of $\phi_c$. 
\end{list2}
\item The presence of a critical charge $q_c$ which demarcates two different large $\phi_c$ behaviours for solitons smoothly connected to the \AdS{4} vacuum indicates that there could be a new branch of solutions which is non-perturbative in $\phi_c$. A-priori it is not clear that such solutions are smooth, but we have constructed a smooth branch of solutions numerically. In fact, one finds:
\begin{list2}
\item[(i)] For $q < q_c$ there is a new branch of smooth solitons wherein the asymptotic charges diverge,
i.e., ${\cal X}(\phi_c) \to \infty$. These solutions exist in some domain $\phi_c \in (\phi_c^\star, \infty)$.\footnote{While we have not done so, it would be intriguing  to try predict the value of $\phi_c^\star$ by careful examination of the perturbative construction of solutions to ascertain when such non-perturbative effects set in.}  
The solutions curve is multi-branched, with the charges diverging at large $\phi_c$. 
Curiously, for some $\phi_c > \phi_c^\star$ this branch has features in common with the bounded branch, here exhibiting a global \emph{minimum} charge at some finite $\phi_c > \phi_c^\star$ and  oscillate about an asymptotic attractor solution ${\cal X}^{(2)}_\infty$. 
\item[(ii)] The case $q = q_c$ is special: we find that the second branch of solutions described above appears to be tangent at its minimal asymptotic charge with the branch of solutions emanating from the \AdS{4} vacuum (at its maximum). For larger values of $\phi_c$ we see two branches of solutions asymptoting to different conserved charges, ${\cal X}^{(1)}_\infty$ and ${\cal X}^{(2)}_\infty$ respectively. This is the first instance when the perturbative solutions become connected to those which attain ${\cal X} \to \infty$.
\item[(ii)] For $q_c< q < q_\infty$ we find multiple solution branches depending sensitively on the precise value of $q$, in addition to the one connecting the \AdS{4} vacuum to large charge solitons. For $q \gtrsim q_c$ we find that the second branch  is not well parameterised by $\phi_c$ -- for some values of $\phi_c$ there are at least two solutions with different asymptotic charges. In this regime, the curve ${\cal X}(\phi_c)$ appears to connect the two large core scalar values ${\cal X}^{(1)}_\infty$ and ${\cal X}^{(2)}_\infty$ respectively. Increasing $q$ we find that there is a succession of mergers and splits in the family of solutions  -- local minima of the higher branch   ${\cal X}^{(2)}_\infty$ merge with the local maxima of the lower branch ${\cal X}^{(1)}_\infty$ and split off into bubbles. The process of `bubble nucleation' in solution space appears to occur for each min-max pair,  at a sequence of charges $q_i$ ($q_1 = q_c$ above), and terminates at some value of  $q> q_c$, which we denote as $q_\infty$. 
\end{list2}
\item For $q_c< q_\infty< q$ there is a single branch of solutions; this is continuously connected to the \AdS{4} vacuum and has unbounded conserved charges.
\end{list1}

There are a few salient features of the solutions we should note here: the solution branches wherein ${\cal X}\to \infty$ have scaling regimes which connect them to the planar \AdS{4} solutions. In all cases we are able to verify that the resulting planar solutions, obtained by implementing the scalings outlined in \sec{sec:planarlimit}, smoothly merge onto the planar zero-temperature hairy black holes constructed earlier in \cite{Horowitz:2009ij}. We will henceforth refer to these solutions as {\em planar solitons}.

We also investigate the behaviour of global black holes in these theories, including their planar limits. Armed with this data we are able to conjecture a microcanonical phase diagram for solitons of the dual field theory. Of particular significance is the fact that the phase diagram is determined for generic $q < q_\infty$ by (at most) four distinguished solutions:
\begin{list1}
\item The global AdS vacuum, which controls the small $ {\cal X}$ behaviour through the small solitons
\item Two large $\phi_c$ attractor solutions whose conserved charges are $ {\cal X}^{(1)}_\infty$ and $ {\cal X}^{(2)}_\infty$. In particular, the extremal values of charges ${\cal X}^{(1)}_{max}$ and ${\cal X}^{(2)}_{min}$ along the branch set the end-points of the phase curves of the microcanonical ensemble.
\item The planar soliton attained for $\lim_{\phi_c\to\infty} {\cal X}(\phi_c)$ which characterises the large charge limit and smoothly matches onto the zero temperature superfluid solutions in planar \AdS{4}.
\end{list1}

\subsection{Solitons in the $SU(3)$ consistent truncation}
\label{s:sumGSW}
The first model which is embedded as a consistent truncation is the $SU(3)$ truncation described in \eqref{Sgeneral} and Table \ref{theorytable}. Here  the scalar charge and mass are fixed to be $m^2_\phi = -2$ and $q = 1$. This implies that the only freedom left at our disposal is the choice of quantization of the scalar field $\phi$. We study the behaviour of the global solitons in the theory as a function of scalar boundary conditions, which we take to be of the double-trace  form \eqref{dtbc}, parameterised by $\varkappa$.

Let us first note that the scalar potential $V(\phi)$ in this case has three distinguished points: $\phi =0$ which is the \AdS{4} vacuum of interest, $\phi = \phi_{PW} $ where it attains a gobal minimum and $\phi = \phi_{zero}$ where it vanishes; see \fig{Vphiplot}. These points play an important role in the behaviour of the solitons.
The global solutions of this $SU(3)$ model exhibit different behaviour depending on the choice of the scalar boundary condition:
\begin{list1}
\item For the standard boundary condition $\phi_1 =0$ ($\Delta =2$), we find behaviour similar to the critical case $q= q_c$. In particular, 
\begin{list2}
\item[(i)] There are two branches of solitons which intersect near $\phi_c = \phi_{zero}$. We take this to indicate that the perturbative branch of solitonic states is connected smoothly to the large soliton branch. 
\item[(ii)] There are solutions for larger values of $\phi_c$, $\phi_c \gtrsim \phi_{zero}$ with multi-valued behaviour for ${\cal X}(\phi_c)$. While we think they are controlled by some scaling-attractor solution, the exponential nature of the potential makes this hard to establish with any degree of certainty.
\item[(iii)] The planar soliton is attained in the limit $\phi_c \to \phi_{PW}$. 
\end{list2}
\item For the alternate boundary condition $\phi_2 = 0$ ($\Delta =1$) there is a single branch of solitons which connects the \AdS{4} vacuum to the planar limit. These are all obtained for $\phi_c \in [0,\phi_{PW}]$.
\item Turning on double-trace deformations we find that $\varkappa \neq 0$ allows us to smoothly interpolate between the two cases above (which have $\varkappa \to \infty$ and $\varkappa =0$ respectively). It appears that there are two soliton branches for any $\varkappa \in (0,\infty)$:
\begin{list2}
\item[(i)] One branch connects the \AdS{4} vacuum to the planar solitons. The latter are again attained as $\phi_c \to \phi_{PW}$. For $\varkappa $ larger than some critical value $\varkappa_\star$ this branch of solitons becomes multivalued.  This occurs for $\phi_c \in (\phi_{PW},\phi_{zero})$.  For all finite values of $\varkappa$ along this branch of solitons $\phi_c$ stays bounded from above by $\phi_{zero}$. 
\item[(ii)] There are solutions for values of $\phi_c \gtrsim \phi_{zero}$ as in the case of the $\Delta =2$ boundary condition with at least two solutions with different conserved charges for given $\phi_c$.
 \end{list2}
\end{list1}

The fact that the planar limit is attained as $\phi_c \to \phi_{PW}$ in all cases can be understood as the solitonic solutions becoming domain wall spacetimes, interpolating between the \AdS{4} vacuum at the origin and the Pope-Warner vacuum at $\phi_{PW}$. In fact, we find that the global domain walls are almost thin-shell like in this limit. Also, the presence of the zero of the potential at $\phi_{zero}$ appears to control the large core scalar behaviour. 

The microcanonical phase diagram has a phase boundary given by a $m(\rho)$ bounding the admissible soliton solutions for fixed charge. This is the curve of smooth global solitons for a given choice of boundary conditions. This phase boundary is clearly smooth for $\varkappa \in [0,\varkappa_\star]$. Despite there being some multi-valued behaviour of ${\cal X}(\phi_c)$ for $\varkappa > \varkappa_\star$, the phase boundary appears to remains smooth. Solutions which arise from the region $\phi_c > \phi_{PW}$ all seem to lie above the phase boundary and are subdominant in the microcanonical ensemble. 

\subsection{Solitons in the $U(1)^4$ consistent trucation}
\label{sec:sumDG}

The second model which is embedded as a consistent truncation is the $U(1)^4$ truncation described in \eqref{Sgeneral} and Table \ref{theorytable}. Here again the scalar charge and mass are fixed: $m^2_\phi = -2$ and $q = \frac{1}{2}$. We study the behaviour of solitons as a function of the double-trace coupling $\varkappa$ introduced through the boundary condition \eqref{dtbc}. When $\varkappa = 0$, i.e., for the $\Delta =1$ boundary condition, the corresponding solutions can be supersymmetric; the ground state of the system saturates the BPS bound. The main results of our analysis are described below:
\begin{list1}
\item First we turn to the case where  $\varkappa \neq 0$ where we find a single solution branch which is of course connected to the \AdS{4} vacuum. As in the case of the Abelian-Higgs model, the conserved charges appear to be bounded from above. ${\cal X}(\phi_c)$ appears to be non-monotone and controlled by an attractor solution for large $\phi_c$. An interesting feature of this branch however, is that for some range of $\phi_c$ we find three solutions with distinct values of the conserved charges; the solution curve ${\cal X}(\phi_c)$ folds back onto itself. This feature is present for finite values of $\varkappa$ and gets less pronounced as $\varkappa \to \infty$. It is plausible that the large core scalar behaviour is controlled by an attractor. 
\item  $\varkappa = 0$ is special: this is the situation where we can have supersymmetric BPS states (preserving $\frac{1}{8}$ of the supercharges). Now we have more than one branch of solutions!
\begin{list2}
\item[(i)]
 The branch of solutions which is connected to global \AdS{4} is BPS with $m = \rho$ and has monotone behaviour of the conserved charges as a function of $\phi_c$. We find numerically for $\phi_c \gg 1$ that 
\begin{equation}
{\cal X}(\phi_c) \propto  \sqrt{\phi_c }\,,
\label{}
\end{equation}	
whose slow growth, as we shall demonstrate, is a symptom of a critical branch of solutions demarcating the boundary between non-monotone and unbounded growth of  ${\cal X}(\phi_c)$. These solutions can be determined directly from a BPS equation, which is a second order non-linear ODE for the system originally derived in \cite{Chong:2004ce}.
\item[(ii)] In addition there is a second branch of solutions which are not connected to the \AdS{4} vacuum. These solutions respect the BPS bound, but don't saturate it -- in the microcanonical phase diagram they reside above the BPS line and co-exist with global hairy black holes. These solutions are a relic of the folding-back feature noted for $\varkappa \neq 0$: in the limit $\varkappa \to 0$ the fold pinches off and leaves behind this second branch of solutions. 
\end{list2}
\end{list1}

It is rather curious that the only planar limit of this model is attained for $\varkappa =0$ in the limit $\phi_c \to \infty$.  This solution in itself turns out to be peculiar: it is analytical, neutral and is a singular planar soliton! This is actually consistent with earlier analysis of \cite{Donos:2011ut} who showed that the planar hairy black holes in the $U(1)^4$ theory behave peculiarly: they exist only for temperatures {\em larger} than a critical $T_c$ (which is the value predicted by a linear analysis).

Once again we are able to describe the microcanonical phase diagram for this theory. For the $\Delta =1$ boundary condition the supersymmetric solitons are as expected the true ground states of the system. Similar statements apply when $\varkappa \neq 0$.


\section{Phenomenological Abelian-Higgs models}\label{sec:BU}

In this section we present results for the phenomenological case of the action \eqref{Sgeneral}, whose potential $V(\phi)$ and coupling $Q(\phi)$ are given in the first row of Table \ref{theorytable}. As discussed we will work in the ansatz outlined in \sec{s:gsanz}, whose large $r$ behaviour is given by \eqref{asymfalloffs}. The behaviour of the fields in the IR ($r \to 0)$ will depend on the specific solution considered. 

\subsection{Global solitons}
\label{bu:global}

We will now construct horizon-free regular global soliton solutions whose IR boundary conditions are given by the  core expansion \eqref{corefalloffs}. We pick a theory by fixing $q$.  The near core expansion contains $3$ undetermined coefficients, whilst the asymptotic expansion contains $5$. The total differential order of the equations of motion is $6$ and so we can a-priori expect to find families of solutions with $2$ parameters. However, we need to specify the scalar boundary condition as well, and as explained earlier this involves giving a local functional relation $\phi_2(\phi_1)$. Once chosen this reduces the asymptotic data count to $4$ and so, for a given choice of $\phi_2(\phi_1)$) we will find one parameter branches of solutions within a given theory.

We first consider situations where we only have vevs for the single trace operators of the CFT. Since we fix $m_\phi^2 = -2$ we have to make a choice of the CFT spectrum. We either have an operator ${\cal O}_{\phi_2}$ with  $\Delta = 1$ or an operator ${\cal O}_{\phi_1}$ with $\Delta = 2$; the two cases correspond to the choices $\phi_2 =0$ and $\phi_1 =0$ respectively. 

\subsubsection{Condensates of $\Delta = 2$ operator ${\cal O}_{\phi_1}$}
\label{sec:phi2sol}

Let us begin by considering solutions which are accessible within perturbation theory about the vacuum global AdS$_4$ solutions (the $m=\rho=0$ case of \eqref{globalschw}). We demonstrate how these can be constructed analytically in \App{sec:pertsolitons}; here we present the results of direct numerical integration of the equations of motion. From the results of the perturbative construction, we learn that for small solitons $m = \frac{\rho}{q}$, while small extremal \RNAdS{4} black holes have $m  = \rho$. This implies that for $q < q_{ERN} \equiv 1$, the solitons we construct will be heavier than \RNAdS{} black holes in a fixed charge sector \cite{Basu:2010uz}. We focus on the regime $q> q_{ERN}$ to illustrate some of the new features in the space of soliton solutions.

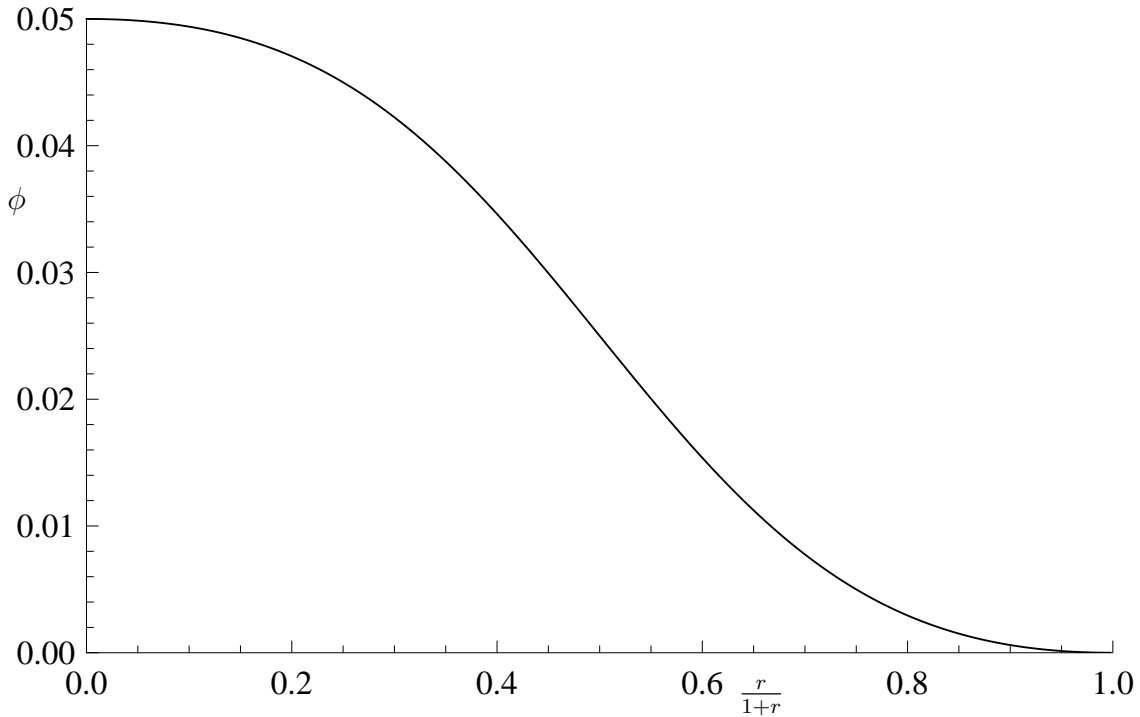
\begin{figure}[h!]
\begin{center}
\includegraphics[width=0.9\columnwidth]{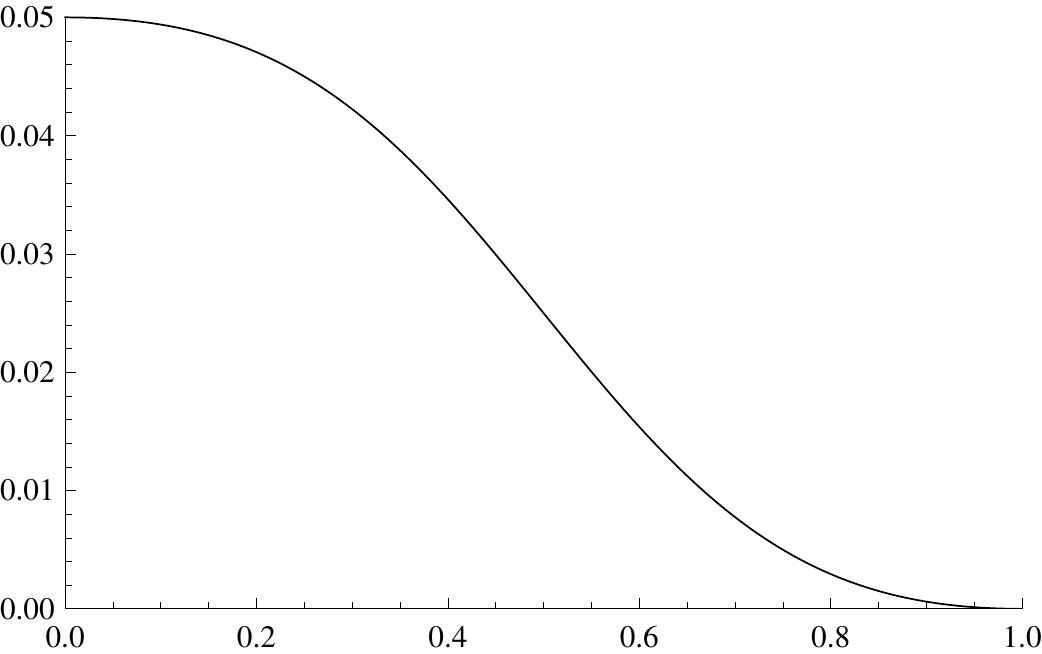}
\setlength{\unitlength}{0.1\columnwidth}
\begin{picture}(1.0,0.45)(0,0)
\put(-4.0,4.5){\makebox(0,0){$\phi$}}
\put(2.0,0.5){\makebox(0,0){$\frac{r}{1+r}$}}
\end{picture}
\caption{$\phi(r)$ for a representative low mass global soliton with $\Delta =2$ boundary condition. For this solution we have chosen $q^2=1.2$ and fixed the point along the branch by fixing $\phi_c=0.05$. At this point we find asymptotic data: $q\mu = 2.001(4), \rho = 0.002(1), m=0.001(9), q\phi_2 = 0.053(6)$ and other near core data: $qA_c = 1.998(4), \beta_c = 0.002(5)$.}\label{bu:orientation}
\end{center}
\end{figure}

For orientation in  \fig{bu:orientation} we present the radial profile of $\phi$ for a representative global soliton in this regime. 
To obtain this solution we have picked a theory by fixing $q$, and picked a value for $\phi_c$ which specifies a position along the one-parameter family of solutions. 

The  solution constructed in \fig{bu:orientation} has no nodes in the scalar profile, since we have tried to macroscopically populate the single particle ground state of the scalar field. As explained in \sec{sec:Introduction}, the linear scalar wave equation in global AdS has a discrete spectrum of eigenstates with energies $\omega = \Delta + 2 \, n$ (setting $l = 0$ for spherically symmetric configurations). We are going to focus on solitons which have $n =0$; there are other solutions at the same value of $q$ and $\phi_c$ with higher numbers of nodes in the radial profile, corresponding to populating the excited oscillator states. Such solutions have lower masses and can also be constructed perturbatively (cf., \cite{Basu:2010uz}). We will not study such solutions here.\footnote{At fixed $\phi_c$ higher node solutions have lower masses; however holding  $\rho$ fixed leads to these being heavier than the zero node solutions.}

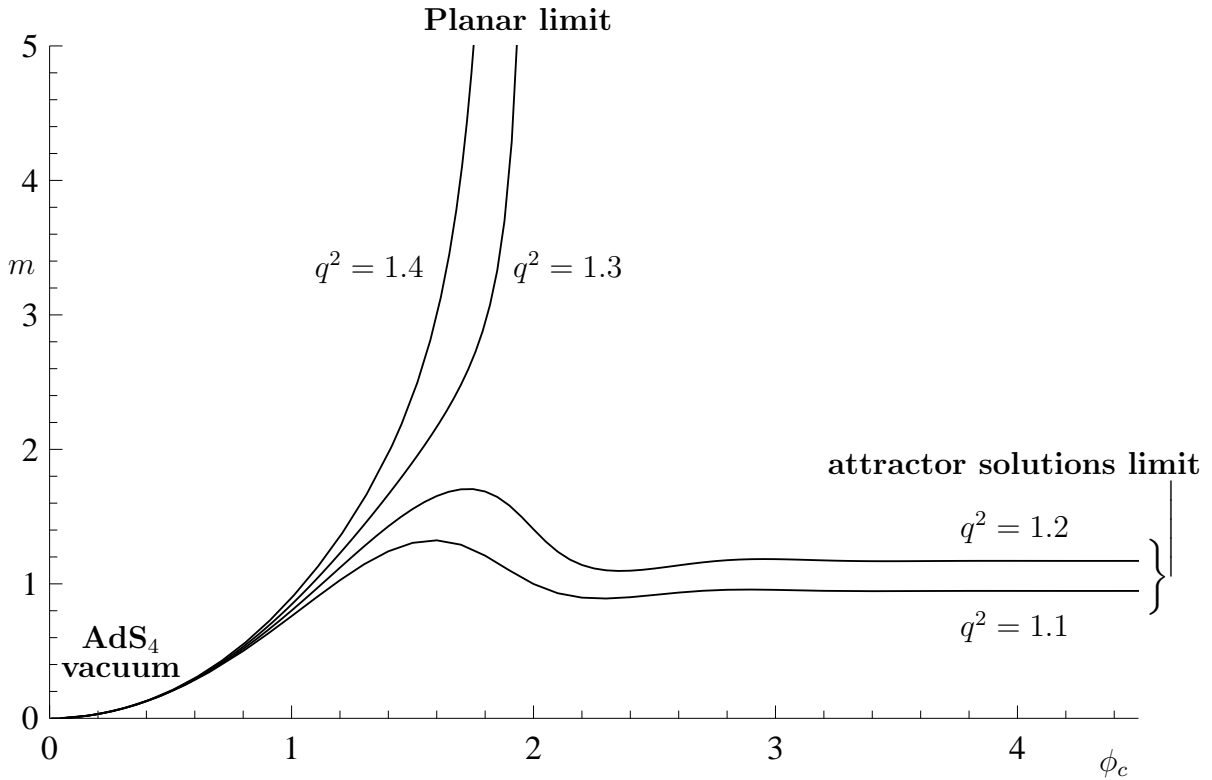
\begin{figure}[h!]
\begin{center}
\includegraphics[width=0.9\columnwidth]{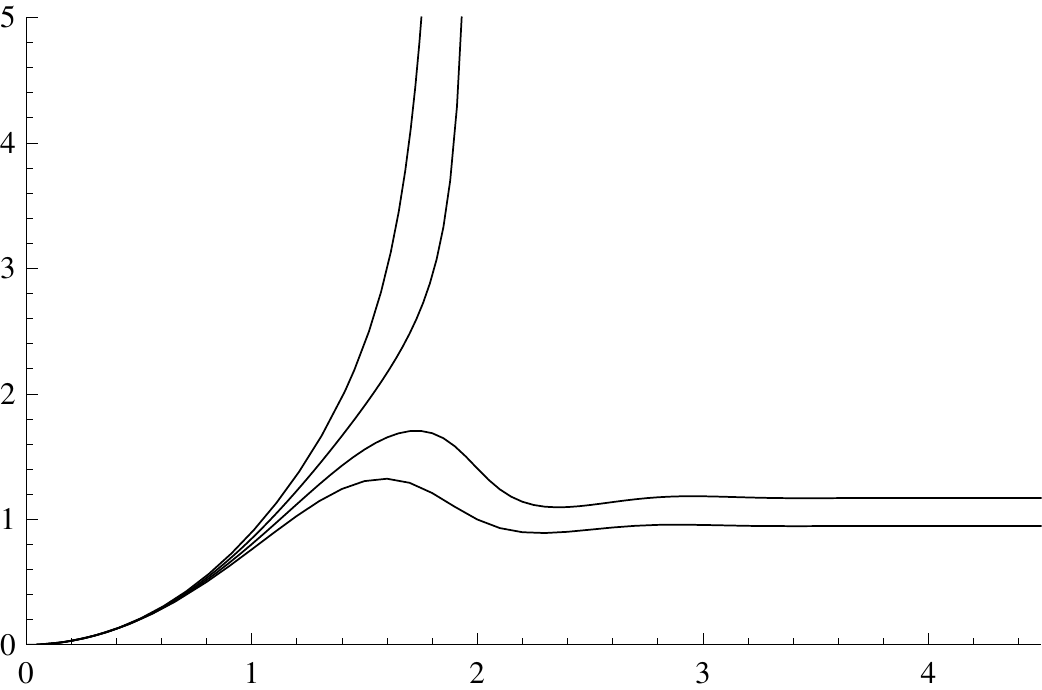}
\setlength{\unitlength}{0.1\columnwidth}
\begin{picture}(1.0,0.45)(0,0)
\put(4.8,0.5){\makebox(0,0){$\phi_c$}}
\put(-4.0,4.5){\makebox(0,0){$m$}}
\put(4.0,1.6){\makebox(0,0){$q^2=1.1$}}
\put(4.0,2.4){\makebox(0,0){$q^2=1.2$}}
\put(0.4,4.5){\makebox(0,0){$q^2=1.3$}}
\put(-1.2,4.5){\makebox(0,0){$q^2=1.4$}}
\put(5.15,2){\makebox(0,0){$\bigg\}$}}
\put(5.26,2.39){\makebox(0,0){$\Bigg|$}}
\put(4.0,2.92){\makebox(0,0){\textbf{attractor solutions limit}}}
\put(-3.2,1.5){\makebox(0,0){\textbf{AdS$_4$}}}
\put(-3.2,1.25){\makebox(0,0){\textbf{vacuum}}}
\put(0.0,6.5){\makebox(0,0){\textbf{Planar limit}}}
\end{picture}

\caption{$m(\phi_c)$ for branches which are accessible in perturbation theory about the vacuum solutions, for fixed values of $q$, as labelled. The mass is either bounded or unbounded depending on whether $q^2$ is greater or less than the critical value, $q_c^2 \simeq 1.259$. }\label{buglobalconnected}
\end{center}
\end{figure}

Starting with the zero node solutions we can consider moving along the branch to larger $\phi_c$ values within a fixed theory (fixed $q$). As we do we leave the realm of solutions which may be approximated by perturbation theory. Outside the perturbative regime we find that the solutions strongly depend on the charge parameter $q$. This is illustrated by mass, $m$, along the branch and is shown in  \fig{buglobalconnected}.
We find that $m$ is either bounded or unbounded depending on the value of $q$, and that there is a critical value $q=q_c$ at which there is a transition between the two behaviours. Numerically we find
\begin{equation}
q_c^2 \simeq 1.259\ , \qquad \phi_1 = 0  \ \; (\text{equivalently}\; \Delta =2)
\label{}
\end{equation}	
The nature of the transition is not clear from considering only these branches of solutions and we will address this momentarily.

For $q<q_c$ all asymptotic charges ${\cal X} = \{m,\rho,\phi_2\} $ and $\mu$ are bounded, non-monotone functions of $\phi_c$. We know from the discussion of  \sec{sec:planarlimit} that this branch of solutions therefore does not connect with a non-vacuum solution in the planar limit (i.e., implementing the scaling lands us exactly at Poincar\'e \AdS{4}). One can also show numerically that the solutions attain a maximal conserved charge at some finite value of $\phi_c$, which we label as ${\cal X}^{(1)}_{max}$. Moreover, as discussed in \sec{sec:sumBU} at large $\phi_c$ they exhibit decaying oscillations for each of the conserved charges, towards the attractor ${\cal X}^{(1)}_\infty$ as $\phi_c\to\infty$.

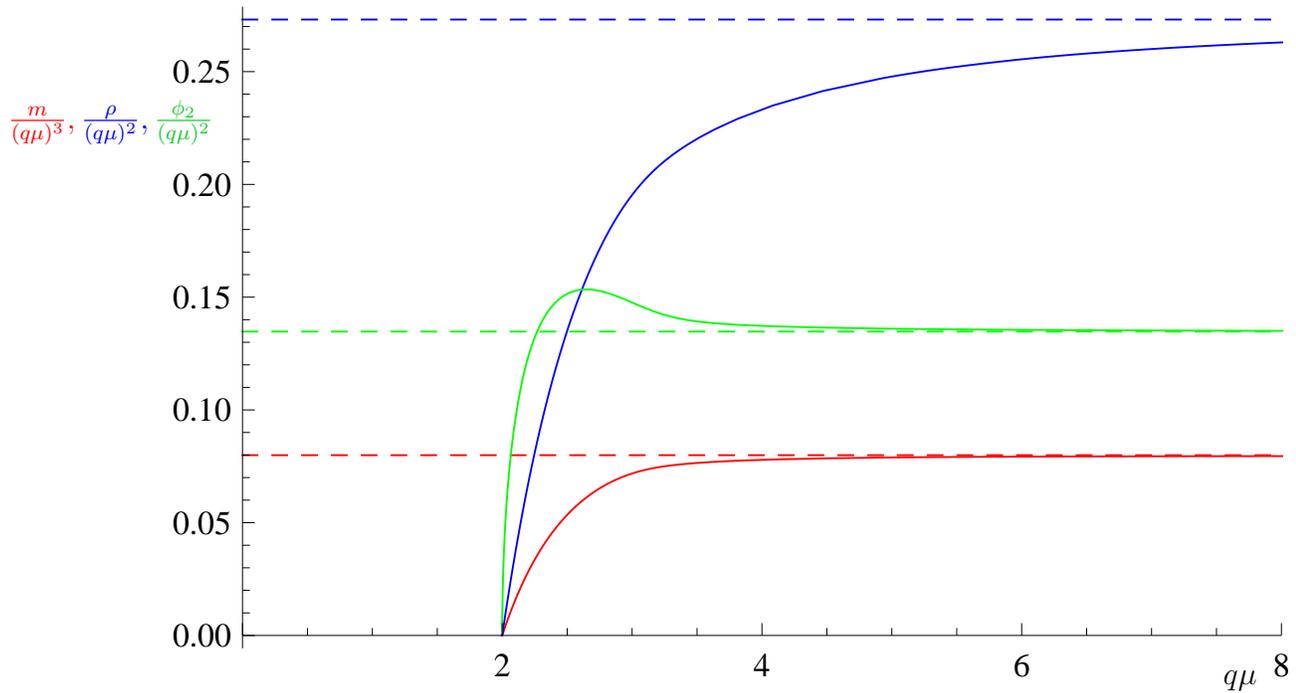
\begin{figure}[h!]
\begin{center}
\includegraphics[width=0.9\columnwidth]{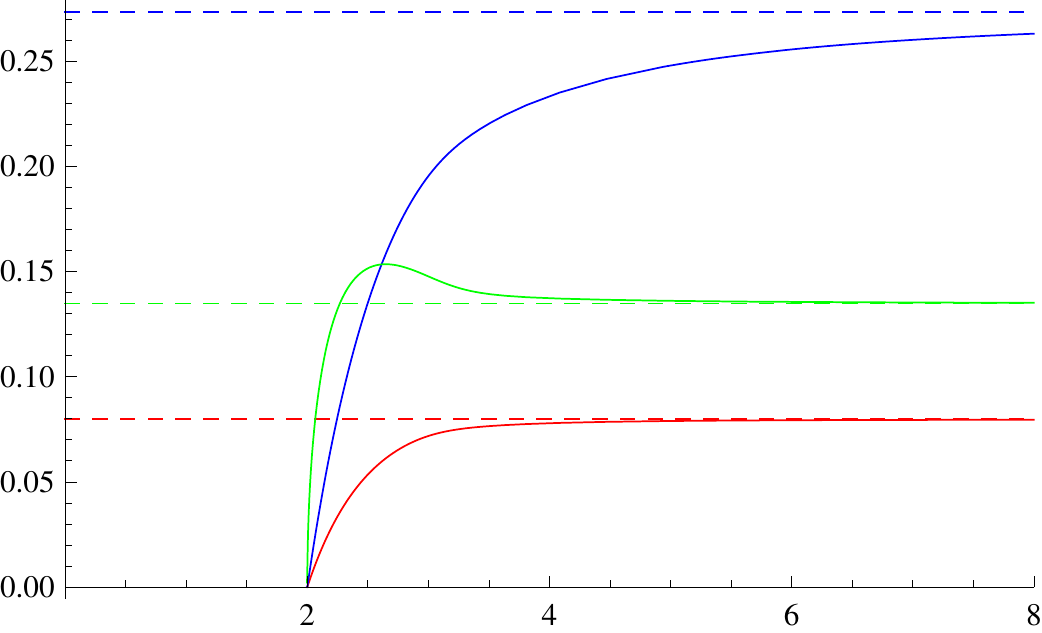}
\setlength{\unitlength}{0.1\columnwidth}
\begin{picture}(1.0,0.45)(0,0)
\put(4.6,0.5){\makebox(0,0){$q \mu$}}
\put(-4.5,5.0){\makebox(0,0){$\color{red}{\frac{m}{(q\mu)^3}},\color{blue}{\frac{\rho}{(q\mu)^2}},\color{green}{\frac{\phi_2}{(q\mu)^2}}$}}
\end{picture}
\caption{(a) The scaling invariants \eqref{dim2invariants} for the branch connected to global AdS$_4$ in theories where $q>q_c$. Here as a representative example we have chosen $q^2=1.3$. Red is the $m$ invariant, green is the $\phi_2$ invariant and blue is the $\rho$ invariant. The dashed lines indicate these quantities for planar hairy black hole solutions at low temperature. (b) $\phi$ profiles for various points along the branch showing agreement with $\phi$ profiles for planar hairy black holes at low temperature.}\label{bupertunbounded}
\end{center}
\end{figure}

For theories with $q>q_c$ all asymptotic coefficients are unbounded; the branch of solutions then connects with a non-vacuum solution in the planar limit. One natural candidate planar solution is the zero-temperature limit of a planar black hole with $\langle{\cal O}_{\phi_1}\rangle = \phi_2 \neq 0$. Parameterising the branch by $\mu$, if all coefficients are to survive the planar limit we must have that each of the following dimensionless quantities
\begin{equation}
m\mu^{-3}, \qquad \rho\mu^{-2}, \qquad \phi_2\mu^{-2}.\label{dim2invariants}
\end{equation}
become constant at large $\mu$. These quantities are plotted along the unbounded branch in \fig{bupertunbounded}; they  indeed asymptote to constant values at large $\mu$. Furthermore, comparison with the corresponding quantities for planar hairy black holes at low temperature in the same theory reveals good agreement between the large $\mu$ global solutions and the low $T/\mu$ planar hairy black holes. We thus have planar solitons along this branch.

One might expect that something happens to the family of hairy planar black hole solutions as one lowers $q$ below $q_c$ given the qualitative changes observed in the global solitons constructed so far. However, the planar hairy black hole solutions do continue to exist for $q<q_c$ with no obvious qualitative change. This suggests that there are other global soliton branches which we have so far not constructed; branches which are not perturbatively connected to the global \AdS{4} vacuum. A-priori these could be singular, which would be a bit disappointing.\footnote{One can always construct singular solitons by simply integrating in the boundary data. The crucial physical question is which of these are good solutions.} However, we find that there are smooth global solitons which become planar even for $q<q_c$.
We now present a more complete picture of the regular global soliton solutions at fixed $q$, including branches of solutions which cannot be constructured in perturbation theory around vacuum global \AdS{4}.

\begin{figure}[h!]
\begin{center}
\includegraphics[width=1.01\columnwidth]{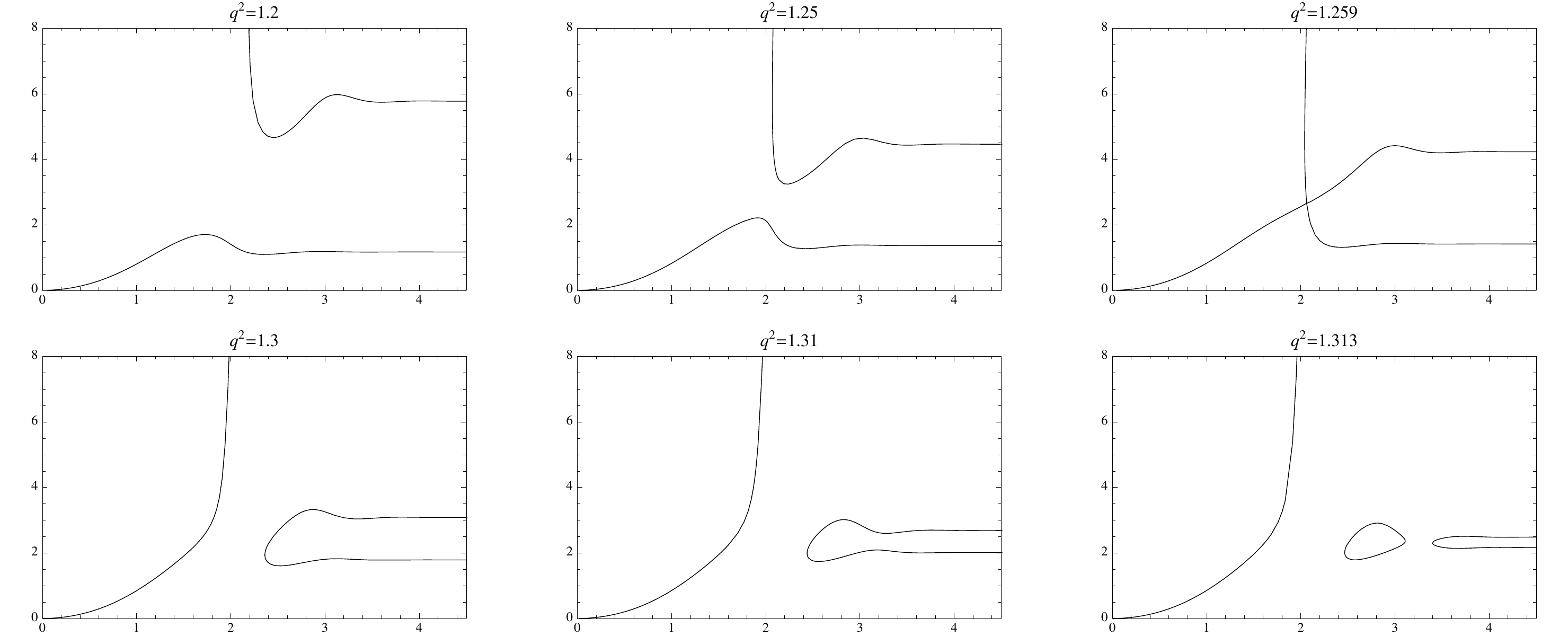}
\setlength{\unitlength}{0.1\columnwidth}
\begin{picture}(1.0,0.45)(0,0)
\end{picture}
\caption{$m(\phi_c)$, for a range of fixed $q$ illustrating the existence of a planar limit even when there is a branch with a mass bounded from above. Each maximum of the bottom branch connects with the corresponding minimum of the top branch as $q$ is increased.}\label{bu:tiles}
\end{center}
\end{figure}

In \fig{bu:tiles} we present all branches of zero node solutions found for a range of fixed $q$ values.
At low $q<q_c$, we find a new branch of solutions which are not connected to the vacuum solution, exhibiting a minimum $m$ for the branch. There is now a new attractor solution which we label by conserved charges ${\cal X}_\infty^{(2)}$ which controls the behaviour of this branch at large $\phi_c$. The minimal charge solution along this branch is denoted as ${\cal X}^{(2)}_{min}$. We also note that the scalar field at the core is bounded from below along this new branch of solutions (see \fig{bu:tiles}).

The presence of this second attractor solution has interesting implications on the phase space of solutions. As $q$ is increased we note that the attractor values of the charges start to move closer to each other. Moreover, the first maximum of the lower branch and the first minimum of the upper branch, i.e., the extrema occurring at the smallest $\phi_c$ on the corresponding branches, move together, meeting at our critical value, i.e., ${\cal X}^{(1)}_{max}(q_c) = {\cal X}^{(2)}_{min}(q_c)$, which defines this critical  parameter $q_c$.  At this point an unbounded branch of solutions connected to the AdS$_4$ vacuum is born. 

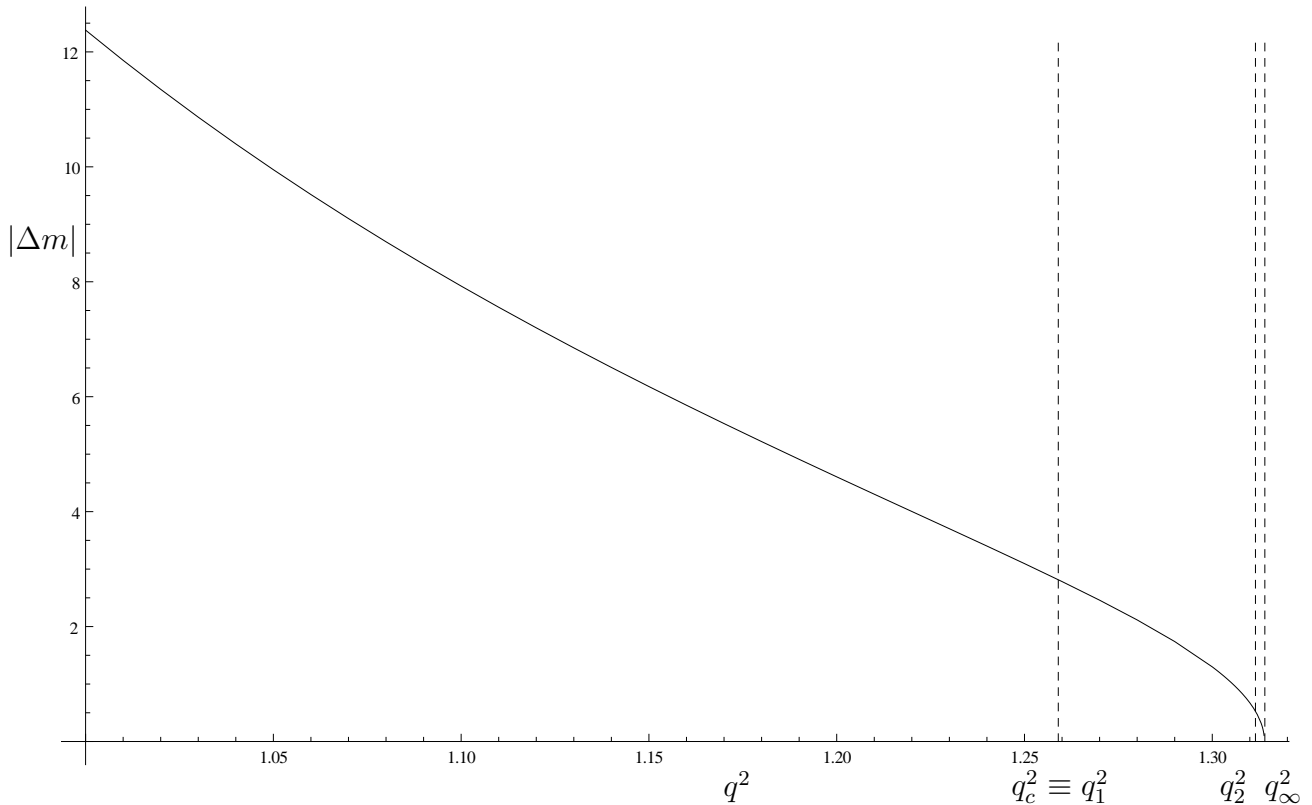
\begin{figure}[h!]
\begin{center}
\includegraphics[width=0.99\columnwidth]{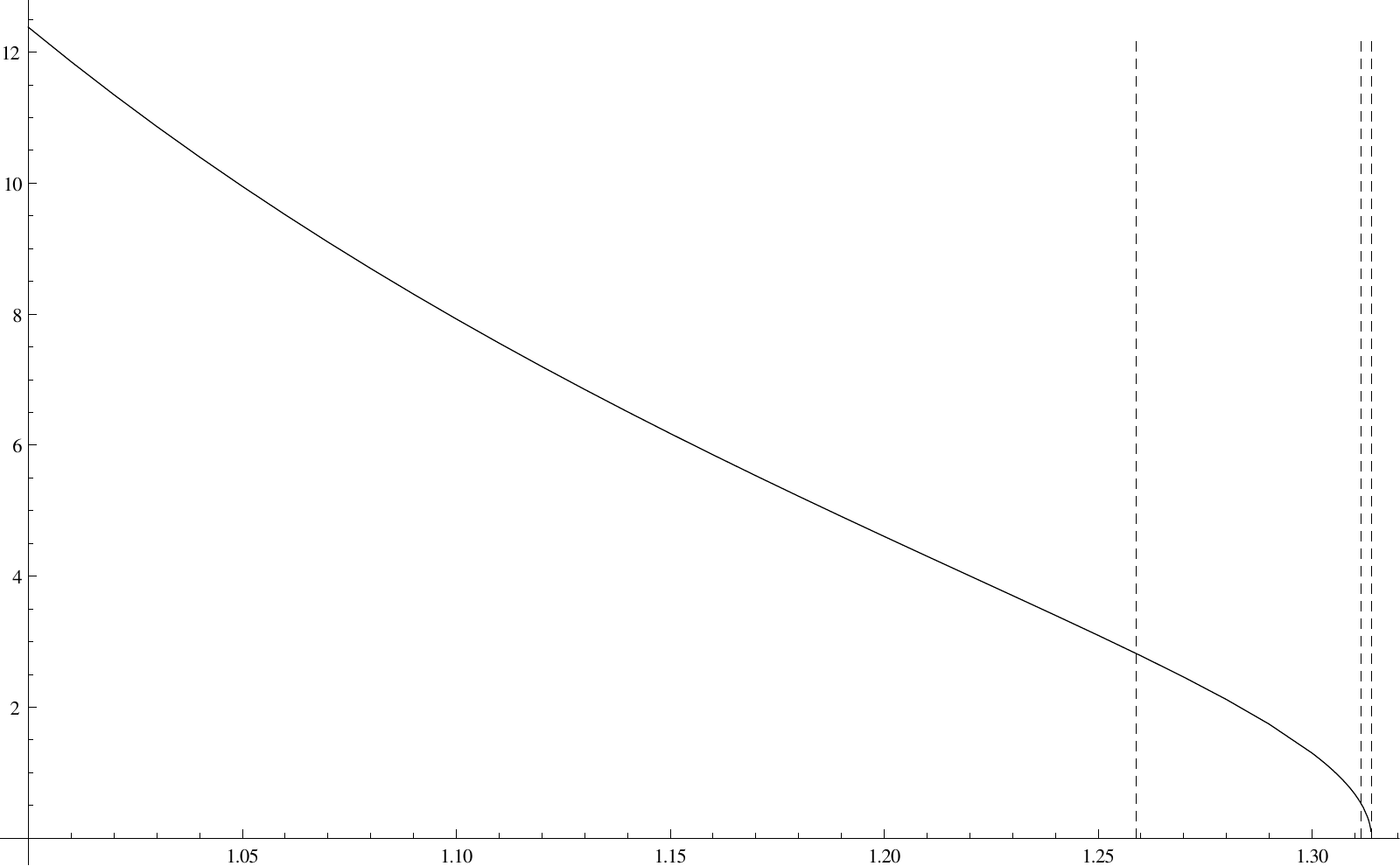}
\setlength{\unitlength}{0.1\columnwidth}
\begin{picture}(1.0,0.45)(0,0)
\put(-4.6,4.7){\makebox(0,0){$|\Delta m|$}}
\put(1,0.3){\makebox(0,0){$q^2$}}
\put(3.6,0.3){\makebox(0,0){$q_c^2 \equiv q_1^2$}}
\put(5.0,0.3){\makebox(0,0){$q_2^2$}}
\put(5.4,0.3){\makebox(0,0){$q_\infty^2$}}
\end{picture}
\caption{Difference in mass for the two large $\phi_c$ solutions as a function of $q^2$. At $q_1^2$ the two branches of solution meet, as demonstrated in the third pane of \fig{bu:tiles}. Above $q_\infty^2$ the large $\phi_c$ solutions cease to exist. We expect to see an infinite sequence of $q_i$ values between $q_2^2$ and $q_\infty^2$, corresponding to the nucleation of closed solution branches.}
\label{bu:annihilation}
\end{center}
\end{figure}

Subsequently, we find that there are further critical values of $q$ which we label by $q_i$ (where $q_{i+1}>q_{i}$ with $q_1 \equiv q_c$), where the $i^{\rm th}$ maximum of the bottom branch connects with the $i^{\rm th}$ minimum of the top branch, leaving closed branches (bubbles) in parameter space as illustrated in  \fig{bu:tiles}. We have checked this behaviour up to $q_2$. Since the solutions for large $\phi_c$ are governed by the damped oscillations about the attractors, we expect an asymptotic sequence of bubble creations in solution space. However, the two attractor solutions which control the asymptotic behaviour, merge together and move off (into the complex domain) above $q_\infty^2\simeq 1.313(8)$. Effectively, by this point the oscillatory branches have ceased to exist; smooth global solitons only exist for some bounded domains in the $\phi_c$ plane. This behaviour is illustrated in \fig{bu:annihilation}.

\begin{figure}[h!]
\begin{center}
\includegraphics[width=0.99\columnwidth]{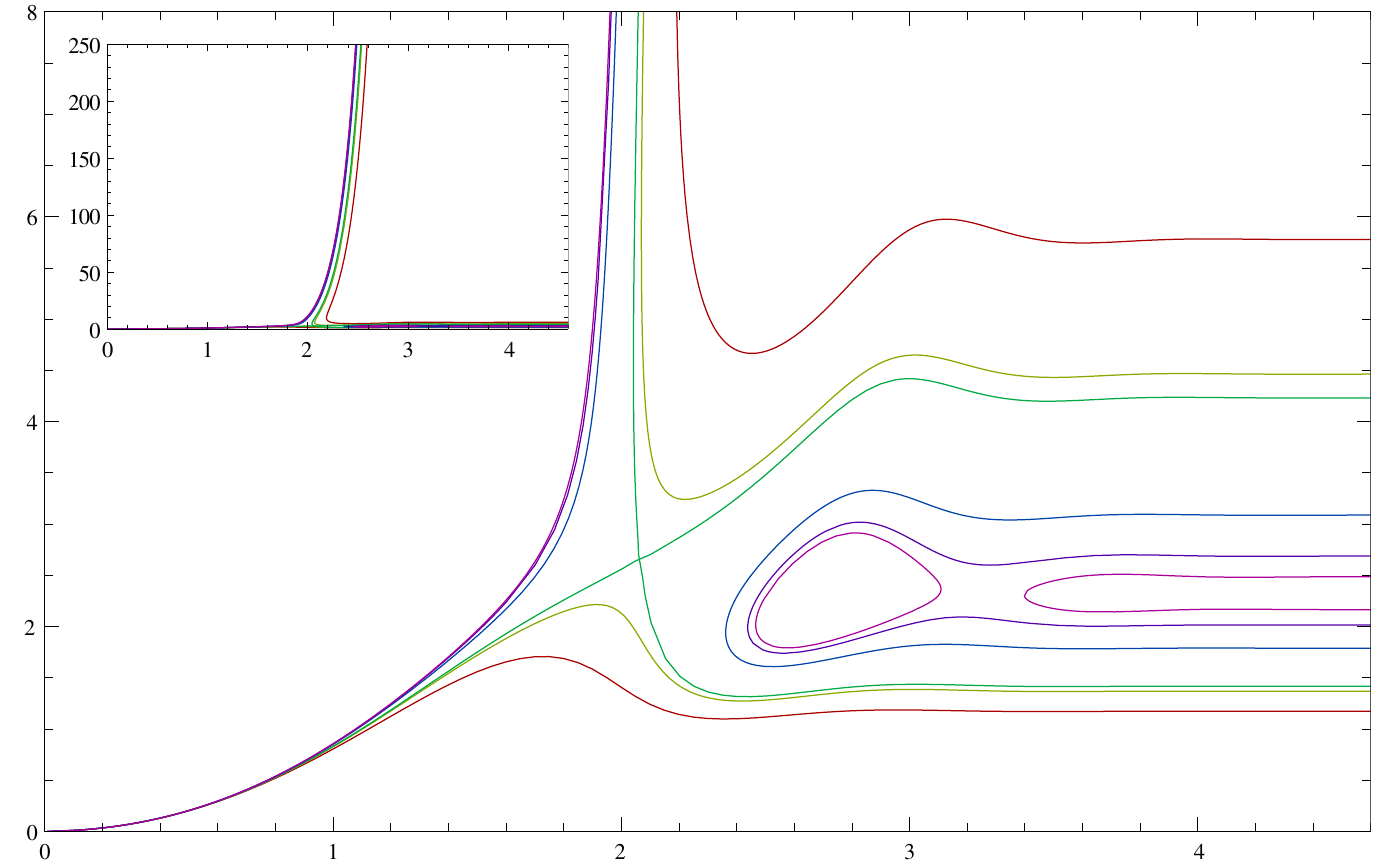}
\setlength{\unitlength}{0.1\columnwidth}
\begin{picture}(1.0,0.45)(0,0)
\put(-4.6,4.5){\makebox(0,0){$m$}}
\put(4,0.3){\makebox(0,0){$\phi_c$}}
\end{picture}
\caption{Contours of the function $q(m,\phi_c)$. Colour represents the value of $q$, distinguishing each theory shown individually in \fig{bu:tiles}. Inset is the behaviour over a large range of $m$.}
\label{bu:massvscore}
\end{center}
\end{figure}

\begin{figure}[h!]
\begin{center}
\includegraphics[width=0.98\columnwidth]{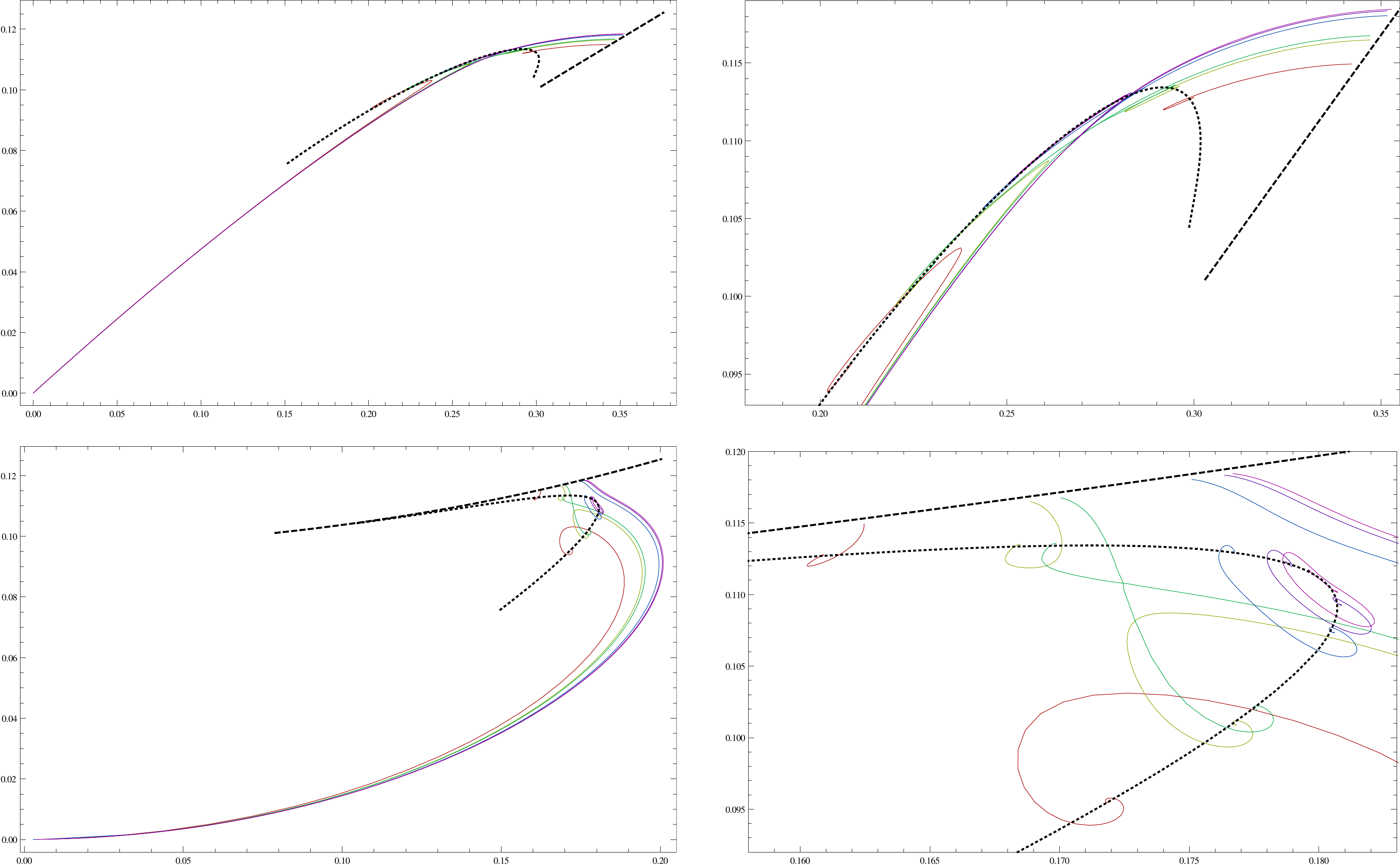}
\setlength{\unitlength}{0.1\columnwidth}
\begin{picture}(1.0,0.45)(0,0)
\put(-4.6,1.7){\makebox(0,0){$\frac{m}{\mu^3}$}}
\put(-4.6,5.0){\makebox(0,0){$\frac{m}{\mu^3}$}}
\put(-2.0,4.0){\makebox(0,0){$\frac{\rho}{\mu^2}$}}
\put(3.5,4.0){\makebox(0,0){$\frac{\rho}{\mu^2}$}}
\put(-2.1,0.3){\makebox(0,0){$\frac{\phi_2}{\mu^2}$}}
\put(3.5,0.3){\makebox(0,0){$\frac{\phi_2}{\mu^2}$}}
\end{picture}
\caption{Scaling invariants illustrating convergence to the planar solutions, for the values of $q$ shown in  \fig{bu:tiles} and  \fig{bu:massvscore}. The black dotted line represents the attractor solutions at large $\phi_c$ plotted for an interval of $q$. The black dashed line represents the low temperature planar hairy black holes plotted for an interval of $q$.
The scaled conserved charges spiral into the attractor lines but smoothly asymptote to the line of planar solitons for large charges. The right panels zoom into the interesting region to illustrate these features. }
\label{bu:globalscalinginvariant}
\end{center}
\end{figure}

These branches of solutions at a fixed value of $q$ can be viewed as contour lines of a function $q(m,\phi_c)$ shown in \fig{bu:massvscore}. 
We  have now constructed unbounded global soliton solutions for all values of $q$. While for  $q<q_c$ the unbounded branch becomes disconnected from global AdS$_4$, we saw that there was a single connected branch when $q > q_c$. Moreover, as in the cases $q>q_c$ where we encountered planar solitons, the unbounded $q<q_c$ solutions match on to the hairy planar black hole solutions at low $T/\mu$, i.e., we have planar solitons for all values of $q$. This is illustrated in \fig{bu:globalscalinginvariant}. From \fig{bu:globalscalinginvariant} it is clear that the conserved charges (normalised by $\mu$ to account for scaling) spiral into the attractor curves; this is of course expected given the presence of damped oscillations about the attractor solutions. 

Another feature clearly visible is the presence of bubbles in the solution space: all solution branches occupy finite volume of $\mu$-normalised parameter space and $m(\phi_c)$ bubbles are closed curves in this space. The latter will be subdominant in the microcanonical ensemble, but their presence is a testament to the non-linearities inherent in the problem under consideration. It is also reassuring that the asymptotic values of the rescaled parameters smoothly merge onto those of the planar hairy black holes irrespective of whether $q<q_c$ or $q>q_c$. 

\begin{figure}[h!]
\begin{center}
\includegraphics[width=0.98\columnwidth]{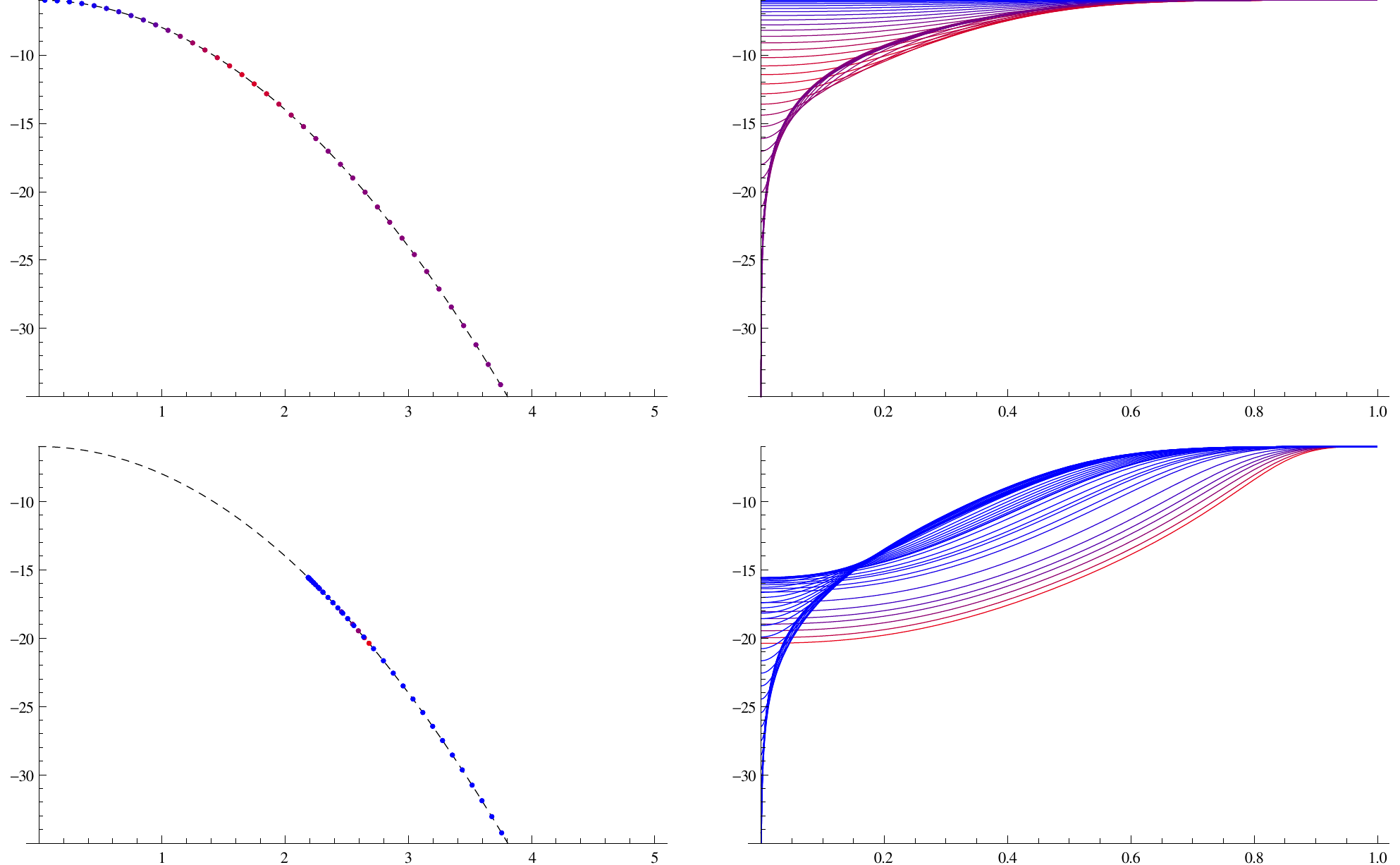}
\setlength{\unitlength}{0.1\columnwidth}
\begin{picture}(1.0,0.45)(0,0)
\put(-4.7,6.2){\makebox(0,0){$V(\phi_c)$}}
\put(0,3.5){\makebox(0,0){$\phi_c$}}
\put(-4.7,3){\makebox(0,0){$V(\phi_c)$}}
\put(0,0.4){\makebox(0,0){$\phi_c$}}
\put(0.5,6.2){\makebox(0,0){$V(r)$}}
\put(5,4){\makebox(0,0){$\frac{r}{r+1}$}}
\put(0.5,3){\makebox(0,0){$V(r)$}}
\put(5,0.4){\makebox(0,0){$\frac{r}{r+1}$}}
\thicklines
\put(2,6.7){\makebox(0,0){\AdS{4}}}
\put(2,2.5){\vector(1,-3){0.3}}
\put(3,1.8){\makebox(0,0){planar limit}}
\put(1.05,2.){\vector(0,-1){1.0}}
\put(1.3,1.5){\makebox(0,0){${\cal X}_\infty^{(2)}$}}
\put(1.1,5.2){\vector(0,-1){1.0}}
\put(1.4,4.8){\makebox(0,0){${\cal X}_\infty^{(1)}$}}
\end{picture}
\caption{\emph{Left column}: $V(\phi)$ with each dot corresponding to the value of $\phi_c$ for solutions shown in the \emph{Right column}: radial plots of the potential $V(\phi(r))$ evaluated on solutions belonging to the two branches at $q^2=1.2<q_c^2$ (as illustrated in \fig{bu:tiles}). \emph{Top row}: Solutions belonging to the branch with a maximum mass. \emph{Bottom row}: Solutions belonging to the branch not connected to the vacuum; solutions are only shown up to a finite mass. Colour indicates the mass.}\label{bu:subcriticalpotential}
\end{center}
\end{figure}

\begin{figure}[h!]
\begin{center}
\includegraphics[width=0.98\columnwidth]{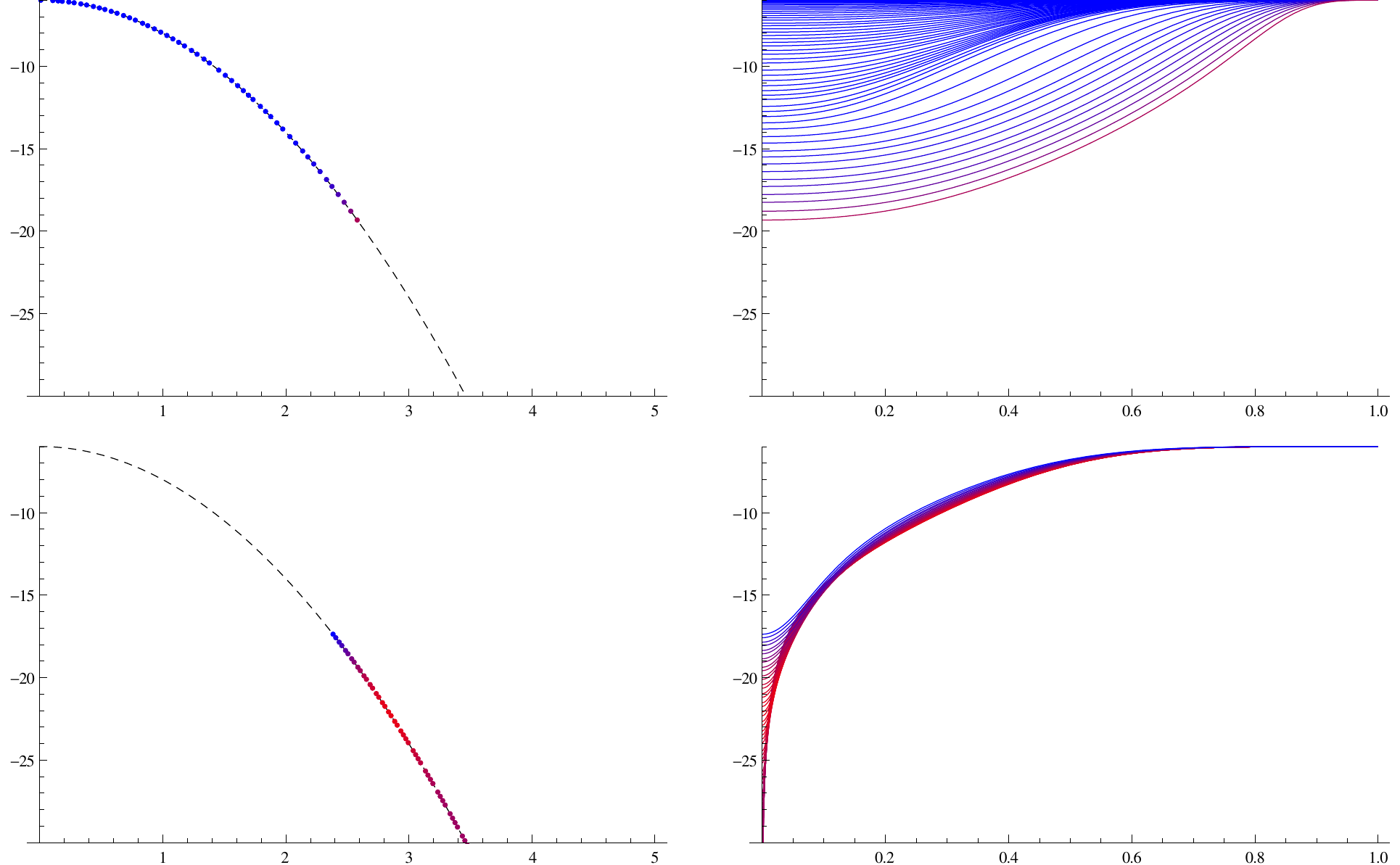}
\setlength{\unitlength}{0.1\columnwidth}
\begin{picture}(1.0,0.45)(0,0)
\put(-4.7,6.2){\makebox(0,0){$V(\phi_c)$}}
\put(0,3.5){\makebox(0,0){$\phi_c$}}
\put(-4.7,3){\makebox(0,0){$V(\phi_c)$}}
\put(0,0.4){\makebox(0,0){$\phi_c$}}
\put(0.5,6.2){\makebox(0,0){$V(r)$}}
\put(5,4){\makebox(0,0){$\frac{r}{r+1}$}}
\put(0.5,3){\makebox(0,0){$V(r)$}}
\put(5,0.4){\makebox(0,0){$\frac{r}{r+1}$}}
\thicklines
\put(2,6.7){\makebox(0,0){\AdS{4}}}
\put(2,5.5){\vector(0,-1){1.0}}
\put(3,4.8){\makebox(0,0){planar limit}}
\put(1.05,2.){\vector(0,-1){1.0}}
\put(1.6,1.5){\makebox(0,0){${\cal X}_\infty^{(1)},{\cal X}_\infty^{(2)}$}}
\end{picture}
\caption{As \fig{bu:subcriticalpotential} but for the two branches at $q^2=1.3>q_c^2$. \emph{Top row}: Solutions belonging to the branch connecting the vacuum to the planar limit; solutions are only shown up to a finite mass. \emph{Bottom row}: Solutions belonging to the branch connecting the two attractor solutions.}\label{bu:supercriticalpotential}
\end{center}
\end{figure}

We will now discuss some of the qualitative features of the solutions themselves, and how the various branches are distinguished. Many of these branches will exhibit behaviour different from the perturbative solitons, with the latter exemplified by the radial profile \fig{bu:orientation}. For this purpose it will be useful to illustrate how the solutions probe the potential $V(\phi)$; this is illustrated in \fig{bu:subcriticalpotential} for the two branches of solution at $q^2=1.2<q_c^2$ and in \fig{bu:supercriticalpotential} for the two branches of solution at $q^2=1.3>q_c^2$.

Each solution shown in \fig{bu:subcriticalpotential} and \fig{bu:supercriticalpotential} begins at $\phi=\phi_c$ in the IR (by construction) and monotonically in $r$ traces out a path to the $\phi=0$ maximum of $V(\phi)$ in the UV (as enforced by our boundary conditions). Those solution branches which are connected to the \AdS{4} vacuum explore the full domain $\phi_c \in [0,\infty)$ at the core. Along this branch solutions either tend to the planar limit as $\phi_c\to \infty$ if $q>q_c$ or to the attractor ${\cal X}^{(1)}_\infty$ if $q<q_c$. On the other hand, the branches of solutions which are not connected to the vacuum are restricted to a bounded domain $[\phi_{min}(q),\infty)$ at the core with the precise nature depending on $q$.
\begin{list1}
\item For $q< q_c$  we have two distinguished end-point solutions: one that becomes planar as $\phi \to \phi_{min}(q)$ and the other that goes to the  attractor ${\cal X}^{(2)}_\infty$ as $\phi_c\to \infty$.
\item $q_\infty > q>q_c$ the solution curve which is double-branched asymptotes to the attractors ${\cal X}^{(1)}_\infty$ and ${\cal X}^{(2)}_\infty$ for large $\phi_c$. The behaviour around $\phi_{min}(q)$ is complex, owing to the bubbles and we restrict attention to the part of the branch connecting the two attractors in \fig{bu:supercriticalpotential}.
\end{list1}

In short, the non-vacuum `ends' of all branches shown are governed by the behaviour of $V(\phi)$ at large $\phi$. This is clearly illustrated by the asymptotic growth of the  charges ${\cal X}$ as a function of $\phi_c$, as illustrated in the left panel of \fig{bu:massvscore}. Since there is a non-trivial behaviour of the region of field space explored for different values of $q$, we may thus anticipate rather different behaviour for the consistent truncations of later sections, at least for large values of $\phi$. In particular as we alter the potential we anticipate qualitatively different growth of the ${\cal X}$ with $\phi_c$. 

Finally, let us consider the planar limit of the solutions for all $q$.  The fact that we attain the planar solitons in the scaling regime indicates that the profiles of the scalar field, gauge potential and the metric should go over to the zero temperature planar superfluid solutions constructed in \cite{Horowitz:2009ij}. It is curious that in all cases this limit is attained with the scalar field pinned at some point down the potential away from the origin -- it would be interesting to understand how this can be related to the non-trivial near-horizon behaviour seen for the zero temperature solutions in \cite{Horowitz:2009ij}.

\subsubsection{Microcanonical phase diagram for states with $\langle {\cal O}_{\phi_1} \rangle \neq 0$}
\label{sec:mcphi2}

So far we have focused on exploring the rich structure of global solitons and their connection with the planar limit. It is also interesting to consider the implications of these solutions for the microcanonical phase diagram, i.e, examine the behaviour of the conserved charges $m(\rho)$. We will sketch the features of this phase digram in brief and discuss the location of  solitonic solutions with respect to the  phase boundary in the microcanonical ensemble.

\paragraph{Behaviour for $q <q_c$:} For values of $q<q_c$ we know from the construction of solitons we only have solutions with mass in the range $[0,m^{(1)}_{max}] \cup [m^{(2)}_{min}, \infty)$, while the charge lies in the range  $[0,\rho^{(1)}_{max}] \cup [\rho^{(2)}_{min}, \infty)$. Recall that our notation here is tied to the attractor solutions which have conserved charges ${\cal X}^{(1)}_\infty$ and ${\cal X}^{(2)}_\infty$: ${\cal X}^{(1)}_{max}$ and ${\cal X}^{(2)}_{min}$ 
are the extremal values along the branch connected to a particular attractor. 

\begin{figure}[h!]
\begin{center}
\includegraphics[width=0.98\columnwidth]{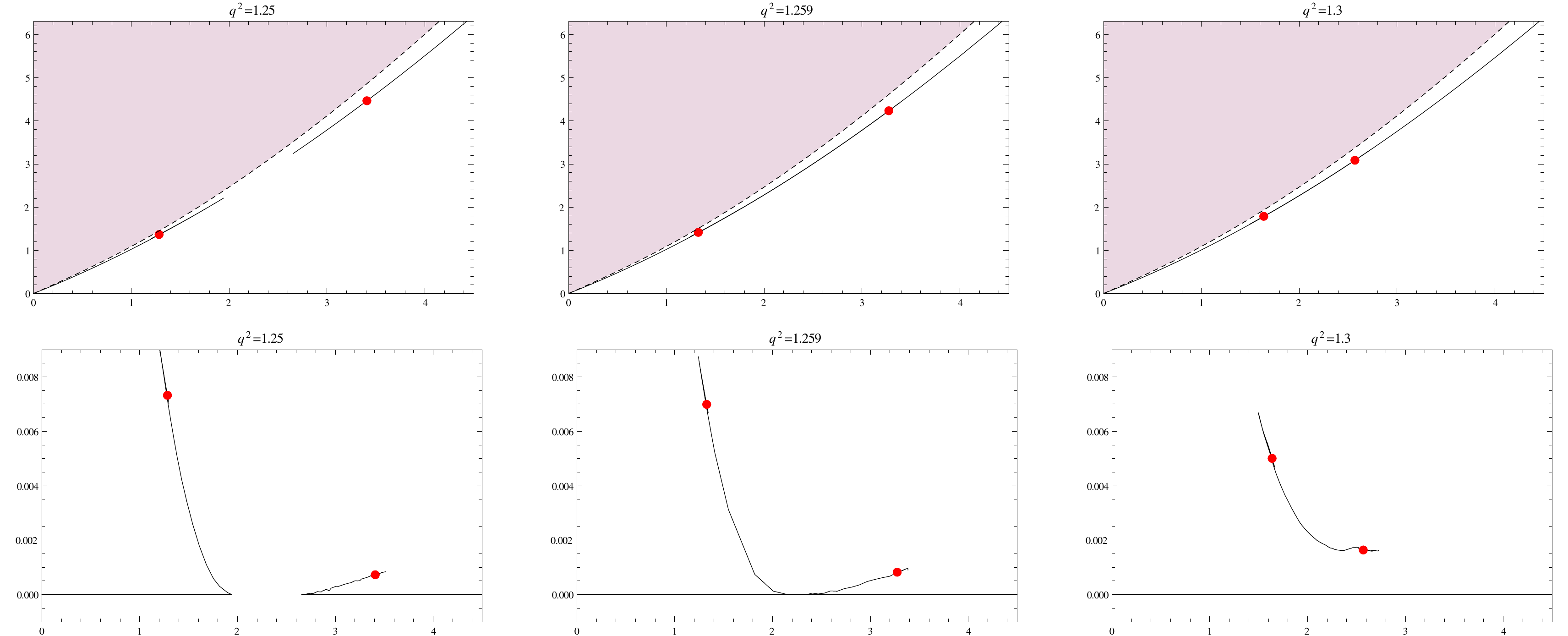}
\setlength{\unitlength}{0.1\columnwidth}
\begin{picture}(1.0,0.45)(0,0)
\put(-4.6,3.7){\makebox(0,0){$m(\rho)$}}
\put(-4.7,1.7){\makebox(0,0){$m(\rho)-\hat{m}(\rho)$}}
\put(0.6,0.4){\makebox(0,0){$\rho$}}
\end{picture}
\caption{Soliton branches in the microcanonical phase diagram for $q<q_c$, $q\simeq q_c$ and $q>q_c$. Top: $m(\rho)$, bottom: $m(\rho)$ with phase boundary function $\hat{m}(\rho)$ \eqref{hatmdef} to illustrate the disfavoured branches. Red dots illustrate the positions of the singular attractor solutions. The black dashed line is the one parameter family of extremal RN-AdS black holes demarcating the region of hair-free black holes. }\label{bu:massvscharge}
\end{center}
\end{figure}
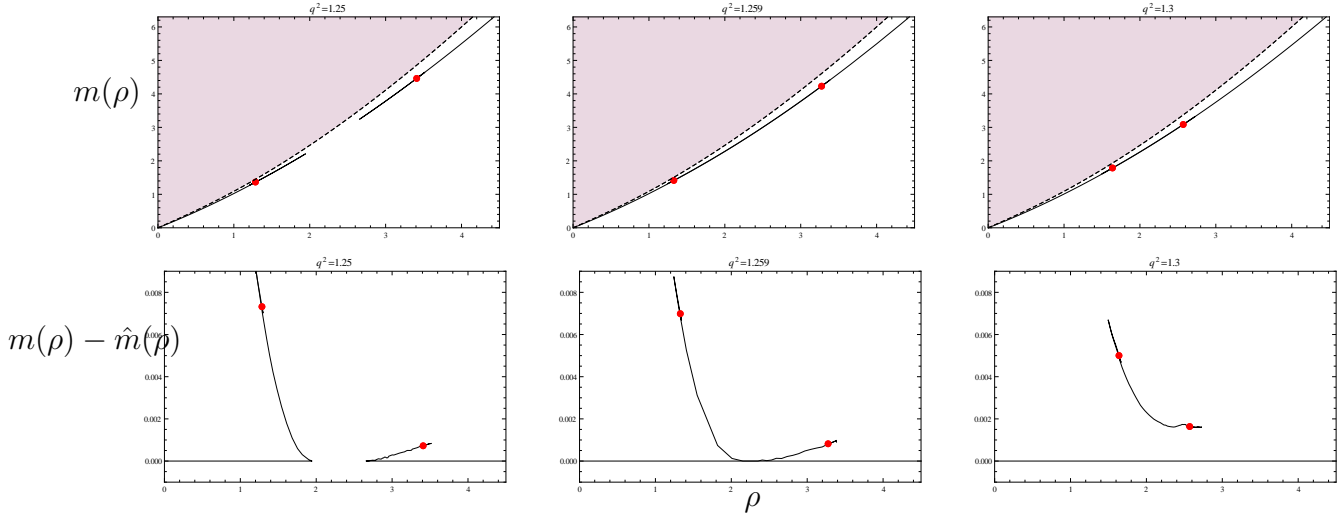

This implies that we have a gap in the microcanonical phase diagram: no smooth solitonic solutions exist for $m \in (m^{(1)}_{max},  m^{(2)}_{min})$ and  $\rho \in (\rho^{(1)}_{max},  \rho^{(2)}_{min})$. The gap in the mass spectrum is clear for example from \fig{bu:massvscore}. We believe that the spectrum of smooth solitonic solutions is exhausted by the solutions we have presented hitherto in \sec{sec:phi2sol}. In particular, we conjecture that the gap in the microcanonical phase diagram is a physical feature. This is  illustrated clearly in the phase curves shown in \fig{bu:massvscharge}.

The end-points of the phase lines viz., $(\rho^{(1)}_{max},m^{(1)}_{max})$ and $ (\rho^{(2)}_{min},m^{(2)}_{min})$ in the $(\rho,m)$ plane are {\em indirectly} controlled by our attractor solutions ${\cal X}^{(1)}_\infty$ and ${\cal X}^{(2)}_\infty$ respectively. This is to be expected; a dynamical systems attractor typically controls the region of the phase plot within its basin of attraction. We know from our numerics that for $q< q_c$ the solutions which stay connected to the \AdS{4} vacuum are within the basin of attraction of  ${\cal X}^{(1)}_\infty$ and those that are connected to the planar zero temperature hairy black hole are within the basin of attraction of   ${\cal X}^{(2)}_\infty$. However, the charges at these end-points points are set by the extremal values of mass and charge obtained along each solution branch. In particular, it is important to note that the solution relevant for these end-points is not the attractor solution; we shall see where the latter lie momentarily.

Let us first define the `phase boundary' curve $\hat{m}(\rho)$ as:
\begin{equation}
\hat{m}(\rho) = \{ m(\rho): (0,0) \to (\rho^{(1)}_{max},m^{(1)}_{max}) \} \cup  \{ m(\rho): (\rho^{(2)}_{min},m^{(2)}_{min}) \to (\infty,\infty )\}
\label{hatmdef}
\end{equation}	
since the presence of oscillations around the attractor solutions implies that the  phase curve $m(\rho)$ as such is a multi-branched zig-zag starting at the extremum mass and charge points  $(\rho^{(1)}_{max},m^{(1)}_{max})$ and $ (\rho^{(2)}_{min},m^{(2)}_{min})$  in the $(\rho,m)$ plane.\footnote{The asymptotic behaviour of this zig-zag which zeroes in onto the attractor solutions can be inferred directly from the approach to the attractor solutions.} In defining $\hat{m}(\rho)$ we have simply removed these zig-zags. The attractor solutions mark the terminus of the zig-zags; we have indicated this explicitly in \fig{bu:massvscharge} and they lie in the region above the phase boundary.

The zig-zag portion of the curve $m(\rho)$ that connects the extremal value of the mass/charge and the attractor points are subdominant in the microcanonical ensemble. We will see in \sec{buholes} that the region above the $\hat{m}(\rho)$ curve is populated with global hairy black holes with non-zero horizon area.  Since black holes carry entropy, it immediately follows that these solutions dominate the microcanonical ensemble (where the entropy or the density of states is to be maximised). The solitonic solutions have no entropy and thus correspond to subdominant saddle point configurations of the corresponding field theory. 

A natural question is what happens in the gap between the extrema: i.e., between the points $(\rho^{(1)}_{max},m^{(1)}_{max})$ and $ (\rho^{(2)}_{min},m^{(2)}_{min})$. Let us first note that global hairy black holes are untouched by this gap in the spectrum; they continue to exist for $m \in (m^{(1)}_{max},  m^{(2)}_{min})$ and  $\rho \in (\rho^{(1)}_{max},  \rho^{(2)}_{min})$. However, as we lower the mass of the black holes for fixed charge in the domain $(\rho^{(1)}_{max},  \rho^{(2)}_{min})$, we should at some point hit a minimal mass solution -- this solution should have vanishing horizon area and thus be horizon-free. Much of this follows from continuity in the space of black hole solutions; there should be no analog  of the non-analytic behaviour seen in the solitonic solutions for black holes.

It is tempting to conjecture that the curve given by $\hat{m}(\rho)$ is the true microcanonical phase boundary of the theory and furthermore argue that the black holes which exist in the soliton gap region terminate on a singular soliton solution. It would be interesting to flesh this out more concretely, but we will refrain from doing so for this phenomenological model. As noted in \sec{sec:Introduction}, a detailed analysis of hairy black holes will appear in \cite{Dias:2011tj} addressing these issues. 
The other models we study which have string theory embeddings (or even known dual field theories) will accord much better control for us to address such issues. We should note that for sufficiently large masses at fixed charge the microcanoical phase digram is dominated by the \RNAdS{4} black holes, which maximise the entropy for given charge.

\paragraph{Behaviour for $q\geq q_c$:} The behaviour of the microcanonical phase diagram for $q = q_c$ is quite interesting. By definition at $q_c$ the two branches of solutions cross. As noted earlier the extremal mass/charge solution of the two branches merge at this value of the scalar charge, implying that we should expect that the gap in the microcanonical phase digram closes off at this point. Indeed this is exactly what we observe. The plot of $m(\rho)$ for the two branches cross and one can smoothly pass between the two branches. 

At $q_c$ we in addition expect to see a kink in the phase curve due to the crossing. This arises because of the zig-zags between the extremal mass/charge solutions on each branch and the corresponding attractor. This behaviour is also clearly visible in the middle panel of the triptych \fig{bu:massvscharge}.

The minimal mass solution for a given charge, i.e., the phase boundary $\hat{m}(\rho)$, now always corresponds to a smooth soliton; one simply jumps from the solution that is connected to the \AdS{4} vacuum to one that is connected to the planar black holes.

For $q_c < q < q_\infty$ we still have the two attractor solutions ${\cal X}^{(1)}_\infty$ and ${\cal X}^{(2)}_\infty$. Now they however control no part of the phase diagram; they have larger mass for a given charge than the solution on the branch of solutions that connects \AdS{4} vacuum to the planar solutions. Moreover, the kink in the phase curve which occurs for $q= q_c$ is now  smoothed out and the attractor solutions control a region above this curve. However, as before everywhere above the $\hat{m}(\rho)$ curve we expect to see global hairy black holes. The latter dominate the ensemble owing to their non-zero entropy; the solitonic solutions which belong to the bubbles or those that stay connected to the attractor solutions are subdominant configurations.

For $q> q_\infty$ we have a simple phase curve; $m(\rho)$ (which now coincides with $\hat{m}(\rho)$) is smooth and characterises the solutions with minimal mass for a given asymptotic charge. We have found no smooth solutions elsewhere in the $(\rho,m)$ plane.

\subsection{Global hairy black holes}\label{buholes}

To round off our discussion about the microcanonical phase diagram we need some information about black hole solutions that exist in the theory. We know a-priori that the Lagrangian \eqref{Sbottomup} has amongst its solutions 
\RNAdS{4} black holes, which attain extremality along a curve $m(\rho)$. Prior to the seminal work of \cite{Gubser:2008px} it was commonly believed that these solutions form the phase boundary of the microcanonical ensemble. We now of course know this to be false: there are scalar hair black holes and charged solitons in the theory. These solutions in fact allow one to attain lower masses.  As explained in \cite{Bhattacharyya:2010yg}, they allow one to close the gap between the extremality and BPS bound for superconformal field theories with holographic duals.  As argued above in \sec{sec:mcphi2}, they facilitate the delineation of the true 
phase boundary.

We will now quickly sketch out the behaviour of the global hairy black holes of \eqref{Sbottomup}. The equations of motion are the same as before, as are the asymptotic boundary conditions. The only change is the IR boundary condition: we require the presence of a non-degenerate Killing horizon. This operationally implies that there is a locus $r = r_+$ where the function $g(r)$ has a simple zero with $g'(r_+)$ being proportional to the surface gravity of the black hole.\footnote{Since $g(r)$ has a simple zero, regularity of the Euclidean solution determines the period of the thermal circle and hence the temperature to be $T= \frac{1}{4\pi}\, g'(r_+) \,e^{-\frac{1}{2}\, \beta(r_+)}$.} 

\begin{figure}[h!]
\begin{center}
\includegraphics[width=0.6\columnwidth]{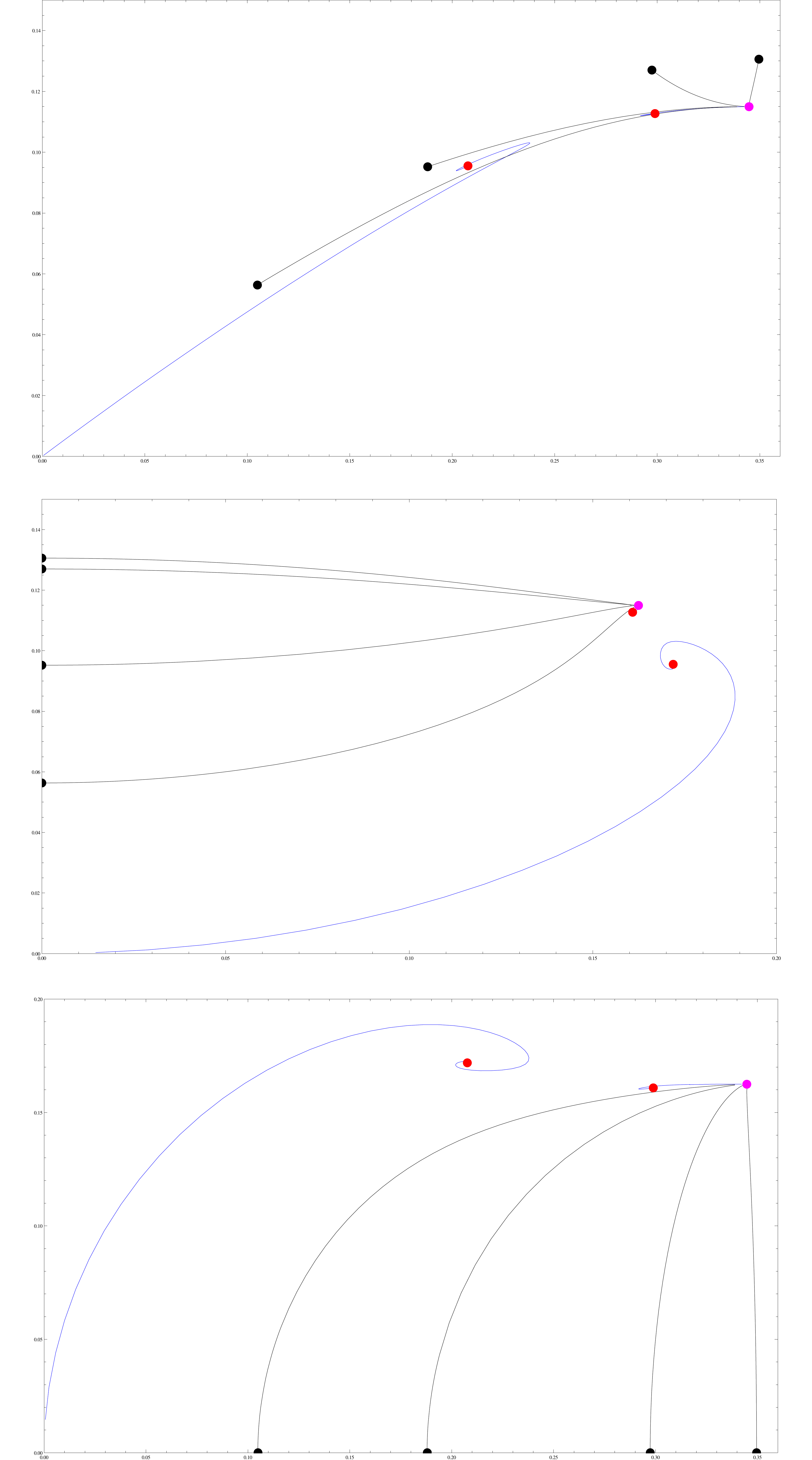}
\setlength{\unitlength}{0.1\columnwidth}
\begin{picture}(0.4,0.45)(0,0)
\put(-6,10){\makebox(0,0){$\frac{m}{\mu^3}$}}
\put(-3,7.25){\makebox(0,0){$\tfrac{\rho}{\mu^2}$}}
\put(-6,6.1){\makebox(0,0){$\frac{m}{\mu^3}$}}
\put(-3,3.6){\makebox(0,0){$\tfrac{\phi_2}{\mu^2}$}}
\put(-6,2){\makebox(0,0){$\frac{\phi_2}{\mu^2}$}}
\put(-3,-0.2){\makebox(0,0){$\tfrac{\rho}{\mu^2}$}}
\put(-4.8,0.2){\makebox(0,0){$r_+=0.2$}}
\put(-3.5,0.2){\makebox(0,0){$r_+=0.4$}}
\put(-1.8,0.2){\makebox(0,0){$r_+=1$}}
\put(0,0.2){\makebox(0,0){$r_+=10$}}
\put(-5.3,4.85){\makebox(0,0){$r_+=0.2$}}
\put(-5.3,5.7){\makebox(0,0){$r_+=0.4$}}
\put(-5.3,6.4){\makebox(0,0){$r_+=1 $ }}
\put(-5.3,6.8){\makebox(0,0){$r_+=10$}}
\put(-4.8,8.7){\makebox(0,0){$r_+=0.2$}}
\put(-3.6,9.5){\makebox(0,0){$r_+=0.4$}}
\put(-1.75,10.3){\makebox(0,0){$r_+=1 $ }}
\put(-0.5,10.5){\makebox(0,0){$r_+=10$}}
\end{picture}
\caption{Convergence to the planar limit of the global hairy black holes for $q^2=1.2$. The black lines are one parameter families of hairy black holes at fixed horizon size $r_+$, for $r_+ = 0.2,0.4,1,10$. The blue lines are the soliton solutions discussed in \sec{sec:phi2sol}. The magenta dot indicates the low temperature planar hairy black hole solution. The red dots illustrate the positions of the putative soliton attractor solutions. The black dots indicate the \RNAdS{4} black hole as given in \eqref{globalschw} at the critical temperature for the onset of the scalar hair instability.}\label{bu:blackholes}
\end{center}
\end{figure}
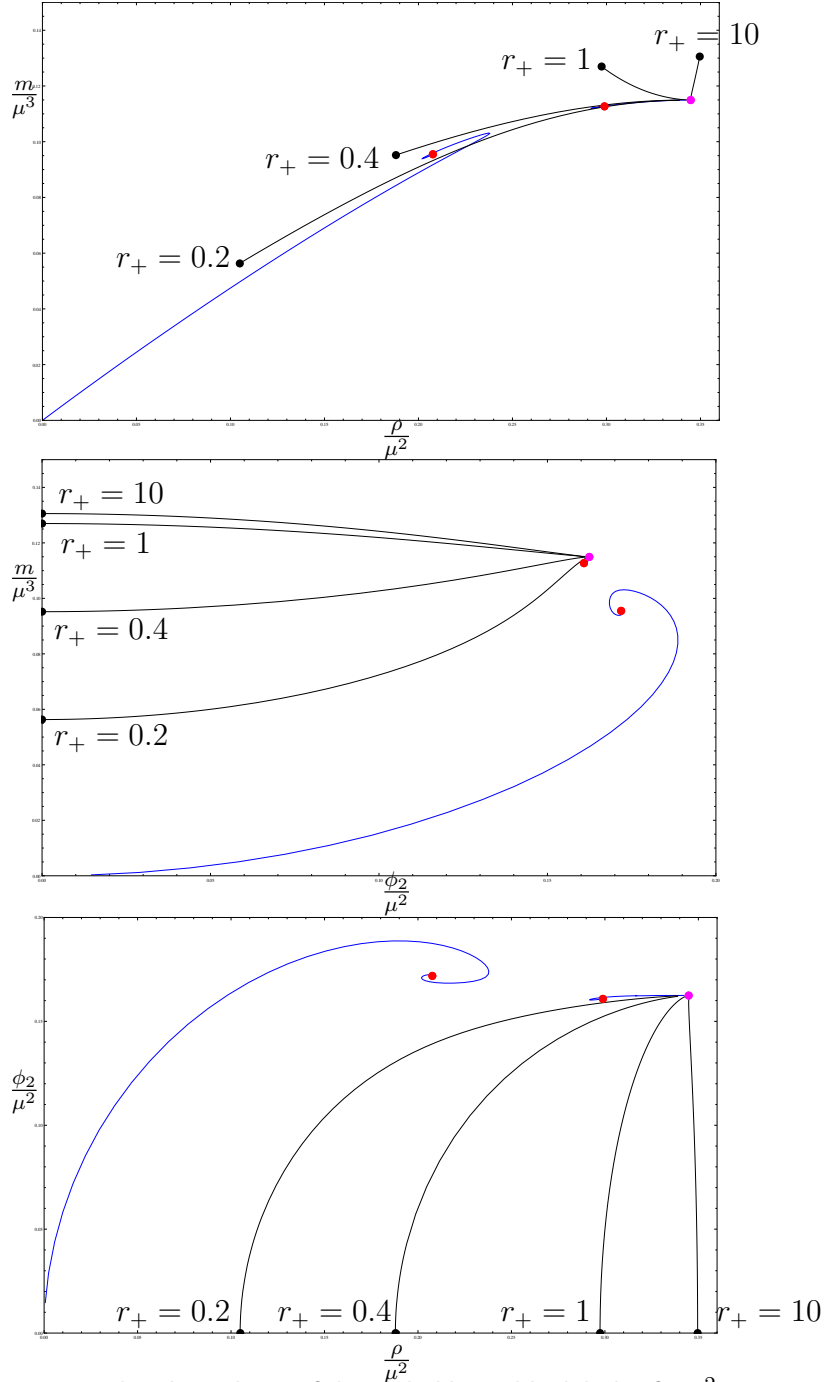

The rest as before involves numerical integration of the field equations. We still have a system of differential order $6$ and our asymptotic data including the scalar boundary condition leaves 4 parameters. We impose $A(r_+) = 0$ to ensure that we have a well-defined one-form on the horizon.  We then have 4 undetermined parameters at the horizon: $A'(r_+)$, $\phi(r_+)$, $\beta(r_+)$ and $r_+$.  We therefore expect to see a two-parameter family of black hole solutions. One of these  parameters without loss of generality can be taken to be the horizon size $r_+$ which makes it easy to talk about small/big black holes by comparing the solution relative to the characteristic \AdS{} scale $\ell$ (here set to $1$). Small black holes with $r_+ \ll 1$ can be constructed perturbatively along the lines discussed in \cite{Basu:2010uz}. We instead construct black hole solutions directly by numerical integration and the result of our numerics is shown in \fig{bu:blackholes}.

The main feature we want to illustrate here is that the large global black holes do indeed pass over nicely to the appropriate planar soliton (i.e., zero temperature limit of the planar hairy black holes). This is achieved by scaling up the parameters of the solution, say $\mu$ whilst holding $r_+$ fixed. Alternately, we could instead consider a scaling where we scale up $r_+$, resulting in a planar black hole with non-zero horizon size.  In both cases the planar solutions  correspond to superfluid configurations with zero superfluid velocity on ${\mathbb R}^{1,2}$. The global hairy black holes in contrast correspond to superfluid configurations for the same field theory on ${\mathbb R} \times {\bf S}^2$.  Furthermore, by explicit construction we have verified that the microcanonical phase diagram is populated by hairy black holes above the soliton phase curve discussed in \sec{sec:mcphi2} at least for the values of $r_+$ plotted here.\footnote{For smaller values of $r_+$ not shown, there appear to be two branches of solution as in the soliton case. One branch connects to the \AdS{4} vacuum, and the other connects with the planar limit. We have been unable to determine numerically whether or not these branches join in the vicinity of the soliton gap.}

\subsection{Exploring scalar boundary conditions}
\label{s:scalarbc}

We have demonstrated that by varying the parameter $q$ we may obtain a rich class of global soliton solutions for the phenomenological Abelian-Higgs models. Later we will concentrate on the theories resulting from consistent truncations where we have no parameters to vary; the Lagrangian has a fixed scalar potential and gauge coupling. 
However, as advertised the theories we consider have the scalar mass lying in the window where both modes are normalizable. We can use this freedom to explore the behaviour of solitons as a function of scalar boundary conditions.

Of main interest to us will be to allow scalar boundary conditions that deform the dual CFT using double-trace deformations as described in \sec{sec:bcscalar}. In order to make comparisons between one of these consistent truncations and the phenomenological theories, we now consider such deformations at fixed $q$. Before turning to this we quickly summarise the situation when we quantise the bulk scalar as a dimension $\Delta =1$ operator on the boundary. This alternate boundary condition is of course the limiting case of double-trace deformations and we should anticipate that the double-trace deformations fall between the cases where $\langle {\cal O}_{\phi_2} \rangle \neq 0$ and $\langle {\cal O}_{\phi_1} \rangle \neq 0$.

\subsubsection{Condensates of $\Delta =1$ operator ${\cal O}_{\phi_2}$}\label{sec:phi1sol}

The behaviour of global solutions where the scalar field is required to behave as $\phi(r) \to \frac{\phi_1}{r}$ as $r \to \infty$, i.e., treating $\phi_2$ as the source and setting it to zero, is qualitatively similar to the $\Delta = 2$ case discussed above. We find that there is a critical charge $q_c$; below $q_c$ the \AdS{4} vacuum is disconnected from the planar hairy solutions, while above $q_c$ the two are connected.

The main feature of interest is the value of $q_c$ itself -- we find it numerically to be 
\begin{equation}
q_c^2 \simeq 0.57\ , \qquad \phi_2 = 0 \ \; (\text{equivalently}\; \Delta =1)
\label{}
\end{equation}	
The fact that $q_c$ is lower for $\Delta =1$ than for $\Delta =2$ boundary condition makes physical sense. In the former cases, the scalar field decays more slowly, implying that the asymptotic field is being held higher. This effectively implies that there is more charge repulsion in this case, and so one encounters the critical behaviour at a smaller value of the scalar charge $q$.\footnote{Small solitons with $\Delta =1$ have $m = \frac{\rho}{2q}$, which implies that $q_{ERN} = 1/2$.  Once again we focus on $q$ larger than this so that our solitons are lighter than the corresponding \RNAdS{} black hole at fixed charge.}

\subsubsection{Double-trace deformations}\label{bucritical}
Having seen the behaviour of the charged scalar solitons for both $\Delta =1$ and $\Delta =2$ boundary conditions, we now turn to the analysis of mixed boundary conditions $\phi_2(\phi_1)$. Specifically, we  consider deforming the field theory by the double-trace operator  ${\cal O}_{\phi_2}^2$ by employing the boundary condition \eqref{dtbc}.
 
In the $\Delta =2$ boundary condition we encounter a critical theory with charge $q_c(\Delta =2) \simeq 1.259$. Above this value of the scalar charge solutions exist for all masses $m \in [0,\infty)$ while for theories with $q<q_c(\Delta =2)$  we found a gap in the soliton spectrum. Based on this observation we shall comment on two interesting cases. 

First, when the theory under consideration lies within the window $q_c(\Delta =1) < q < q_c(\Delta =2)$ there is a gap in the $\Delta = 2$ soliton spectrum but no gap in the $\Delta = 1$ spectrum. In this special window we have the opportunity of studying new critical behaviour as a function of the deformation parameter $\varkappa$ described in \sec{sec:bcscalar}. In particular we can move from an ungapped spectrum ($\varkappa = 0$) via a critical theory at $\varkappa = \varkappa_c(q)$ to a gapped spectrum when $\varkappa > \varkappa_c(q)$. Such phenomena would be interesting to study further.

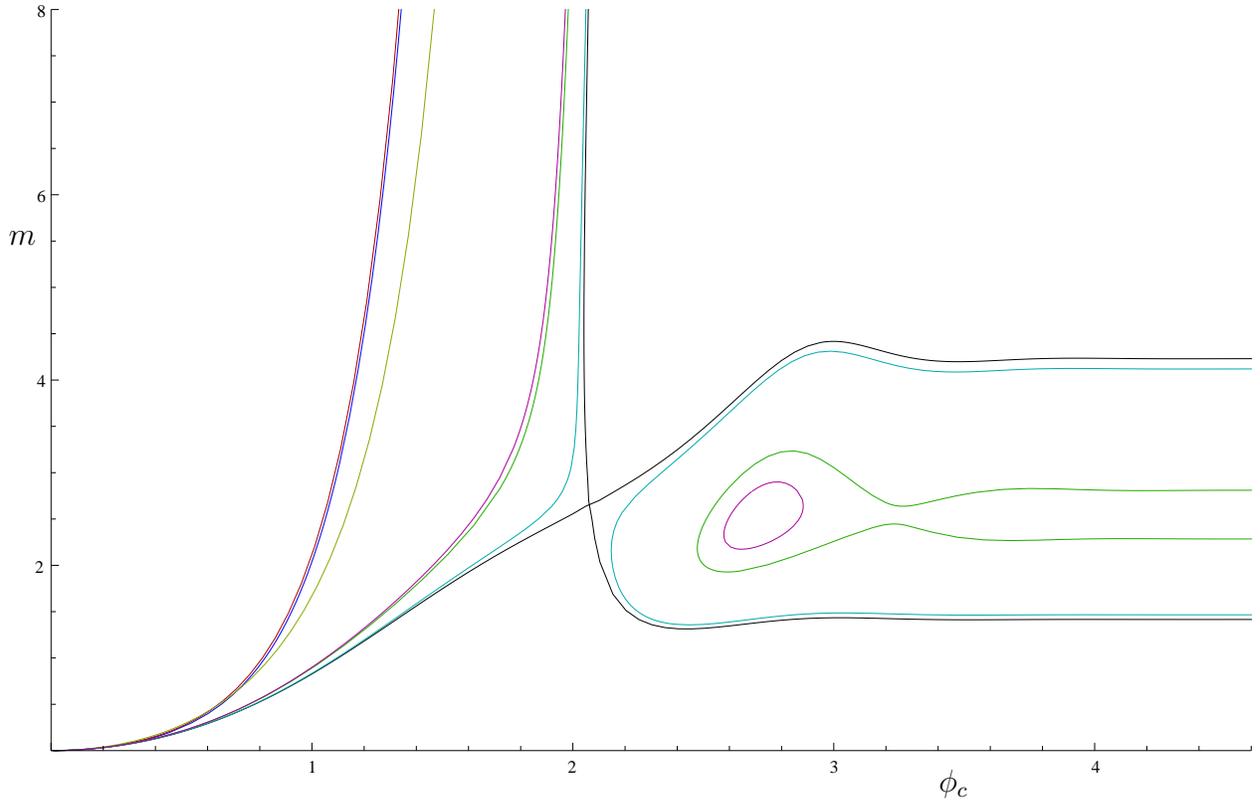
\begin{figure}[h!]
 \begin{center}
\includegraphics[width=0.98\columnwidth]{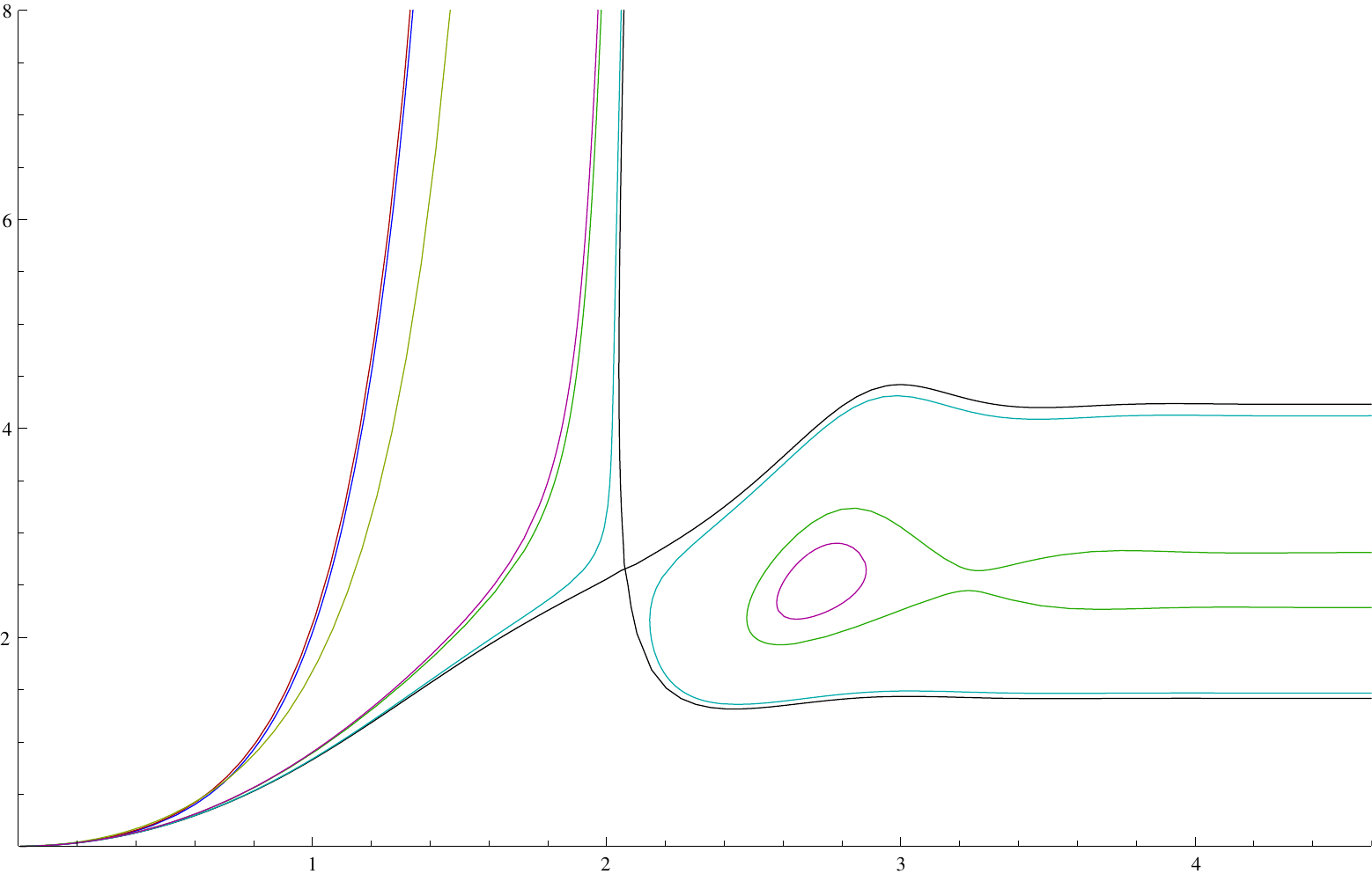}
\setlength{\unitlength}{0.1\columnwidth}
\begin{picture}(1.0,0.45)(0,0)
\put(-4.5,4.8){\makebox(0,0){$m$}}
\put(3,0.4){\makebox(0,0){$\phi_c$}}
\end{picture}
\end{center}
\caption{Global solitons in the double-trace deformed Abelian-Higgs model with $q = q_c(\Delta =2) \simeq 1.259$ with curves shown at fixed $\mu^{-1}\varkappa =0.1,1,10.5,12,100$. For $\varkappa =0$ we have the $\Delta =1$ boundary condition; the solutions and their deformations here smoothly connect the \AdS{4} vacuum to the planar solitons. At large $\mu^{-1}\varkappa$ we encounter the $\Delta =2$ boundary conditions wherein we see the expected merger between the two solution branches. In addition we see closed branches in parameter space for a finite range of $\mu^{-1}\varkappa$. }
\label{bu:doubletrace}
\end{figure}

The second interesting case lies precisely at the critical theory  $q_c(\Delta =2)\simeq 1.259$. This will be of particular interest in the context of the consistent truncation studied in \sec{sec:GSW}, which as we shall show, has qualitatively similar features. At this fixed value of $q$ we can again consider deformations using the parameter $\varkappa$. The results are illustrated in \fig{bu:doubletrace}. In the $\Delta = 1$ case we have only one branch of solutions. Interestingly, we see the emergence of a new branch of solutions not connected with the vacuum at large $\phi_c$ for deviations away from this case. Specifically for deformations satisfying $\varkappa>\varkappa_{bubble} \sim 10.2\mu$ we see the emergence of many familiar features at large $\phi_c$ including the closed bubble solutions previously encountered at supercritical values of $q$. In this sense the relevant deformations appear to be decreasing the effective value of the charge. At larger values of $\varkappa$ both branches converge to the $\Delta = 2$ critical branch configuration, as expected.

There are two important features to note about the double-trace deformed solutions. One is that they seem not amenable to a leading order perturbative construction about the \AdS{4} vacuum (see \App{sec:pertsolitons} for the technical argument). This should be expected; the deformation breaks conformal symmetry and \AdS{4} is not necessarily the true ground state of the system with the double-trace boundary conditions. A second related fact is that the mass density of the bulk solutions also receive contributions from the scalar field; see \eqref{mg1phi}. As noted in various places in the designer gravity literature \cite{Hertog:2004ns,Amsel:2006uf,Amsel:2007im} and more recently in \cite{Faulkner:2010fh} the conserved energy for such deformations can indeed be negative (though still bounded from below hence ensuring a positive energy theorem). This in particular does imply that \AdS{4} vacuum is not the appropriate background for perturbation theory to construct charged solitons; one should rather use the minimum energy designer gravity soliton for this purpose. Since one mostly can only construct the latter numerically, it is simpler therefore to construct charged solitons directly by numerical integration.


\section{Consistent truncation: $SU(3)$}\label{sec:GSW}


Thus far we have focussed on the behaviour of solitons in the phenomenological Abelian-Higgs model. While the system had the luxury of being very simply described by a quadratic scalar potential $V(\phi)$ and Maxwell coupling $Q(\phi)$, it suffers from the drawback of not having a consistent embedding into any known supergravity theory. If we were interested in asking whether the features described above are artefacts of the modeling then we need to generalise our considerations. As described in \sec{sec:Introduction} we can turn to models which arise via consistent truncations of supegravity theories in 10 or 11 dimensions  to ascertain whether the features we see are generically accessible in gravitational systems that might have holographic field theory duals. 
This in addition has the added benefit of allowing us to explore the behaviour  of the solutions for different choices of scalar potentials/Maxwell couplings.

\subsection{Basic facts about the $SU(3)$ truncation}
\label{s:gswpre}

With this motivation in mind, we turn to a very simple model that arises from the consistent truncation of 11D supergravity on a skew-whiffed Sasaki-Einstein manifold \cite{Gauntlett:2009dn}.  We refer to this as the $SU(3)$ truncation in Table \ref{theorytable}. This was the first model that embeds $2+1$ dimensional holographic superconductors into 11D supergravity.  As we shall see it has many interesting features vis-\`a-vis charged solitons.

Since the Lagranagian \eqref{Sgeneral} now has fixed values of scalar potential $V(\phi)$ and Maxwell coupling $Q(\phi)$, given by
\begin{equation}
 V(\phi) = \cosh^2{\frac{\phi}{\sqrt{2}}}\,\left(-7+\cosh{\sqrt{2}\phi}\right) \ , \qquad Q(\phi) = \frac{1}{2} \,\sinh^2{\sqrt{2}\phi}\,
\label{gswdata}
\end{equation}	
and hence $m^2_\phi = -2$ and $q =1$, we have no free parameters to dial. The only freedom we are left with is the choice of boundary condition, which as discussed in \sec{sec:bcscalar} we will generically take to be of the double-trace form \eqref{dtbc}, parameterised by $\varkappa$.\footnote{It can be checked that the theory with the double-trace deformation has a positive energy theorem; the general analysis of \cite{Faulkner:2010fh} applies here and we have explicitly computed the off-shell scalar fake-superpotential to confirm this. We should also note here that the same paper claims to construct neutral designer gravity solitons for the consistent truncation constructed in \cite{Gauntlett:2009dn} and obtains the off-shell fake-superpotential. However, they choose as their scalar potential (in our conventions) $V(\phi) = 5- 12\,\cosh(\phi/\sqrt{2}) + \cosh(\sqrt{2}\phi)$ which is different from what we obtain.  Consequently the cubic term in our fake-superpotential has a slightly different coefficient, i.e., $s_c \simeq 0.7$.}

Before we proceed with the discussion of solitons, let us first recall that the potential $V(\phi)$ in \eqref{gswdata}  has three extrema as noted previously in \sec{sec:exact}.  There is the \AdS{4} vacuum at $\phi =0$ which is the one we are focussed on, but we have in addition local minima at scalar values quoted in \eqref{globalpw}; see  \fig{Vphiplot}. The latter solution is the Pope-Warner (PW) \cite{Pope:1984bd} vacuum with effective \AdS{4} radius $\frac{\sqrt{3}}{2}$. About this vacuum the scalar potential behaves as $V(\phi) \simeq  -8 + 8 \, (\phi-\phi_{PW})^2$, so that the PW vacuum has an irrelevant operator. The presence of a second extremum to the scalar potential implies that we should anticipate domain wall solutions which interpolate between the two vacua. We shall indeed find that such solutions play an interesting role below, along with the other distinguished point $\phi_{zero}$ where $V(\phi)$ vanishes.

\subsection{Global solitons: standard and alternate boundary conditions}
\label{s:gswglobal}

%
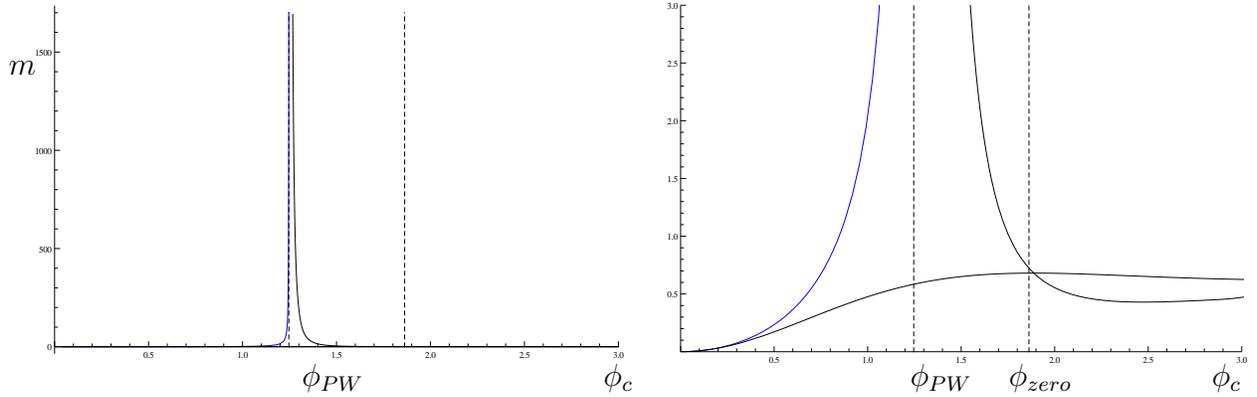
\begin{figure}[h!]
\begin{center}
\includegraphics[width=0.98\columnwidth]{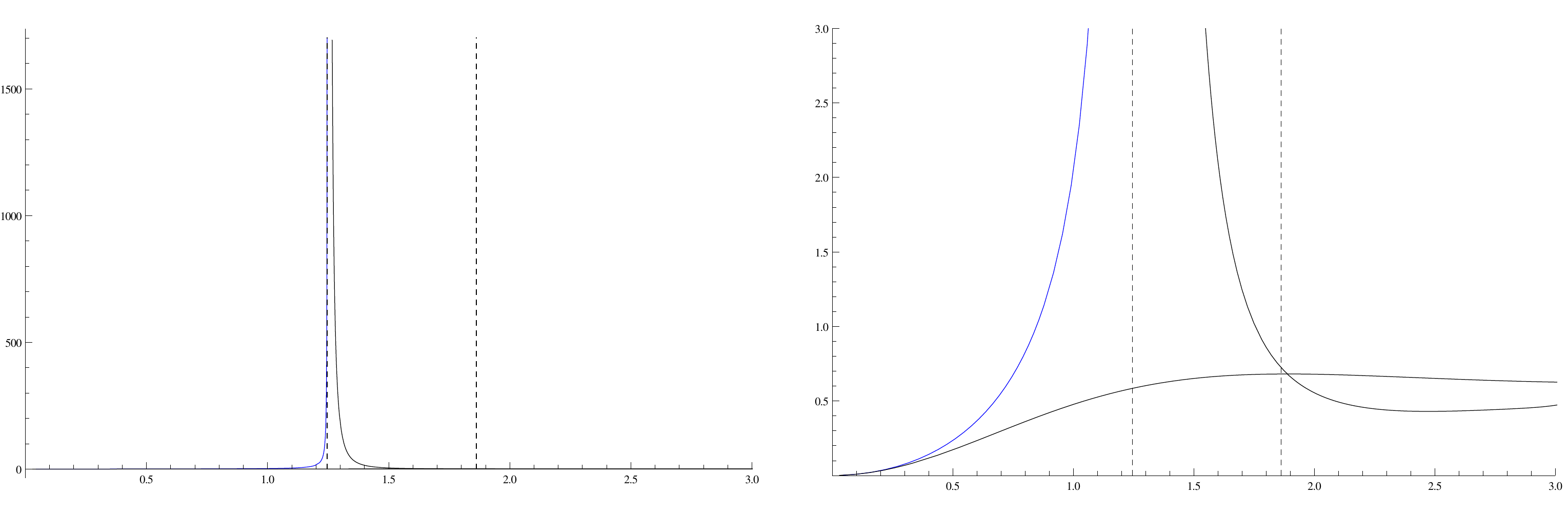}
\setlength{\unitlength}{0.1\columnwidth}
\begin{picture}(1.0,0.45)(0,0)
\put(5.2,0.5){\makebox(0,0){$\phi_c$}}
\put(-2.0,0.5){\makebox(0,0){$\phi_{PW}$}}
\put(2.9,0.5){\makebox(0,0){$\phi_{PW}$}}
\put(3.7,0.5){\makebox(0,0){$\phi_{zero}$}}
\put(0.3,0.5){\makebox(0,0){$\phi_c$}}
\put(-4.5,3.0){\makebox(0,0){$m$}}
\end{picture}
\caption{Charged scalar soliton branches for  dimension $\Delta =2$ ($\varkappa \to \infty$) and dimension $\Delta =1$ ($\varkappa =0$) (blue) for the $SU(3)$ consistent truncation. The vertical dashed line is the value of the scalar $\phi$ in the Pope-Warner vacuum. The right plot is the same data as the left illustrating the low mass behaviour.}\label{gsw:massvscore}
\end{center}
\end{figure}

With these preliminaries out of the way let us now turn to the construction of charged scalar solitons. First we turn to the standard and alternate quantizations of the scalar field, i.e., consider the two special cases $\varkappa =0$ and $\varkappa \to \infty$, or equivalently the $\Delta=1,2$ boundary conditions respectively. Direct numerical integration of the field equations allows one to show the existence of solitons (small solitons as always can be constructed perturbatively). The results of the numerics are reported in \fig{gsw:massvscore} where as in \sec{sec:BU} we have chosen to plot the conserved mass density of the solution as a function of the scalar value $\phi_c$ at the origin. We immediately see that we are on to something interesting:

\begin{list1}
\item For the alternate quantization $\Delta =1$  we note that we have a single branch of solitonic solutions that remains connected to the vacuum global \AdS{4} solution (at $\phi=0$).  The scalar field remains bounded $\phi \in [0,\phi_{PW})$ and we obtain large solitons as $\phi_c \to \phi_{PW}$.
\item For the standard quantization $\Delta =2$ we note that we have two branches of solitonic solutions which cross close to $\phi_c = \phi_{zero}$.  We interpret these solutions in light of the crossing in the following manner:
\begin{list2}
\item[(i)] The branch emanating from vacuum global \AdS{4} solution (at $\phi=0$) is taken to be the one where $\phi \in [0,\phi_{zero})$. 
Along this branch, we find two solutions for a given value of $\phi_c \in (\phi_{PW} , \phi_{zero})$ and for one solution with $\phi_c \to \phi_{PW}$ the conserved charges ${\cal X}$ diverge. This we will call the favoured branch, since we attain all possible values of the conserved charges. We will see in \sec{s:gswmc} that this branch has the lowest mass at any given charge.
\item[(ii)] The second branch is the set of solutions with $\phi_c > \phi_{zero}$ with bounded conserved charges. We note that the region for $\phi>\phi_{zero}$ is governed by an exponentially growing potential, and whilst we see two branches of solution it is unclear that they exhibit damped oscillations as observed in the phenomenological model (where the potential was unbounded from below).\footnotemark 
\end{list2}
\end{list1}
Based on these observations we see that the theory has many features in common with the phenomenological Abelian-Higgs model at the critical charge $q_c$, at least for solutions with $\phi_c \leq \phi_{zero}$.

\footnotetext{In any event, since the two branches cross at finite $\phi_c = \phi_{zero}$, this issue is somewhat irrelevant, at least for the purposes of the microcanonical phase diagram. \label{fattract}}

\begin{figure}[h!]
\begin{center}
\includegraphics[width=0.98\columnwidth]{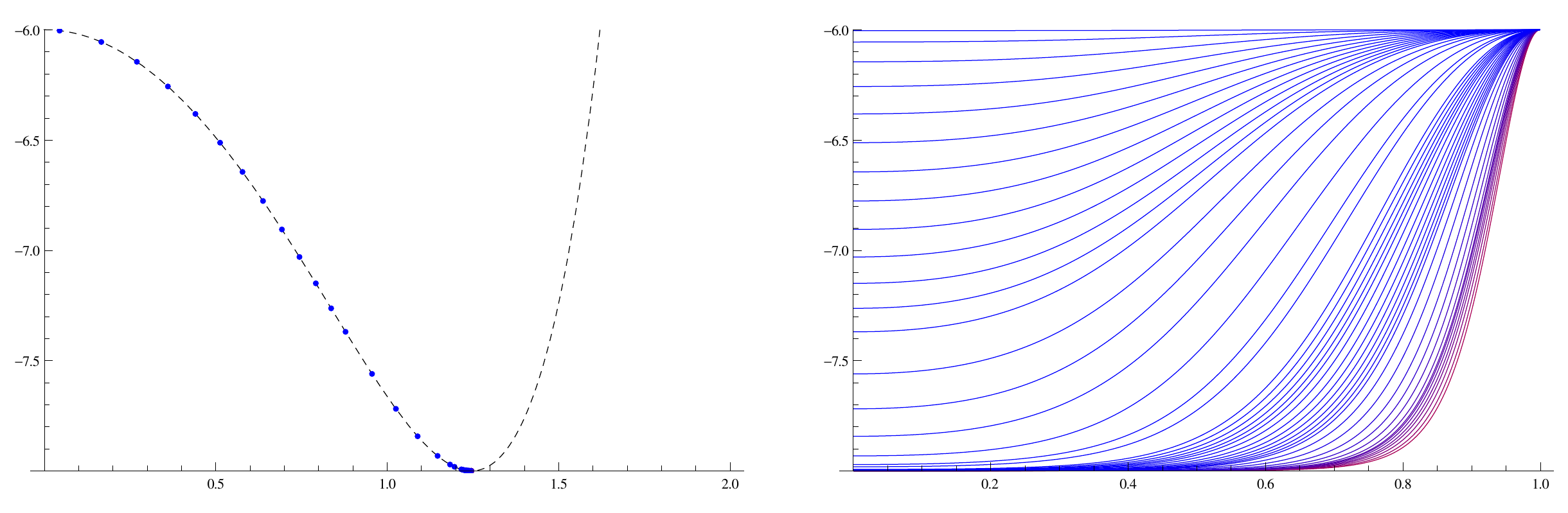}
\includegraphics[width=0.98\columnwidth]{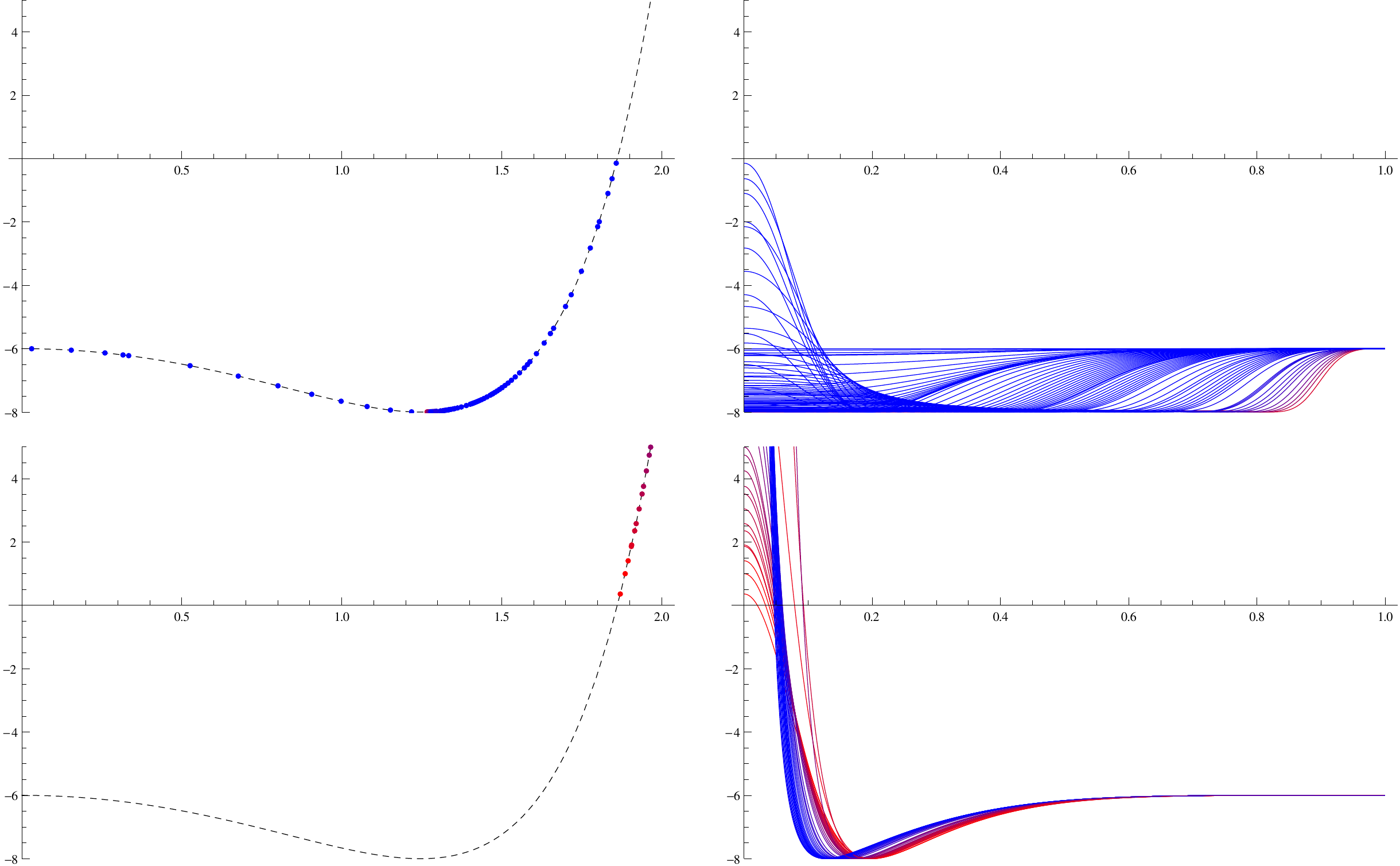}
\setlength{\unitlength}{0.1\columnwidth}
\begin{picture}(1.0,0.45)(0,0)
\put(-4.6,8.5){\makebox(0,0){$V(\phi)$}}
\put(4.8,1.9){\makebox(0,0){$\frac{r}{r+1}$}}
\put(0.2,2.0){\makebox(0,0){$\phi_c$}}
\put(-3.6,7.05){\makebox(0,0){$\Delta = 1$}}
\put(-3.2,5.65){\makebox(0,0){$\Delta = 2,\; \phi_c<\phi_{zero}$}}
\put(-3.2,2.55){\makebox(0,0){$\Delta = 2,\; \phi_c>\phi_{zero}$}}
\thicklines
\put(2,9.7){\makebox(0,0){\AdS{4}}}
\put(2,4.2){\makebox(0,0){\AdS{4}}}
\put(4.3,8.0){\vector(1,0){1.0}}
\put(5.5,7.6){\makebox(0,0){planar shell}}
\put(4.3,3.8){\vector(1,0){1.0}}
\put(5.5,3.5){\makebox(0,0){planar shell}}
\put(1.3,2.4){\vector(0,1){1.0}}
\put(2.4,2.8){\makebox(0,0){singular solutions}}
\end{picture}
\caption{\emph{Left column}: plot of $V(\phi)$ with each dot corresponding to the value of $\phi_c$ for solutions shown in the \emph{Right column}: radial plots of the potential $V(\phi(r))$ evaluated on solutions belonging to each of the three branches shown in \fig{gsw:massvscore}). \emph{Top row}: $\Delta=1$. \emph{Middle row}: $\Delta=2$ for solutions where $\phi_c<\phi_{zero}$. \emph{Bottom row}: $\Delta=2$ for solutions where $\phi_c>\phi_{zero}$. Colour indicates the mass, with red corresponding to higher values.
}\label{gsw:profiles}
\end{center}
\end{figure}
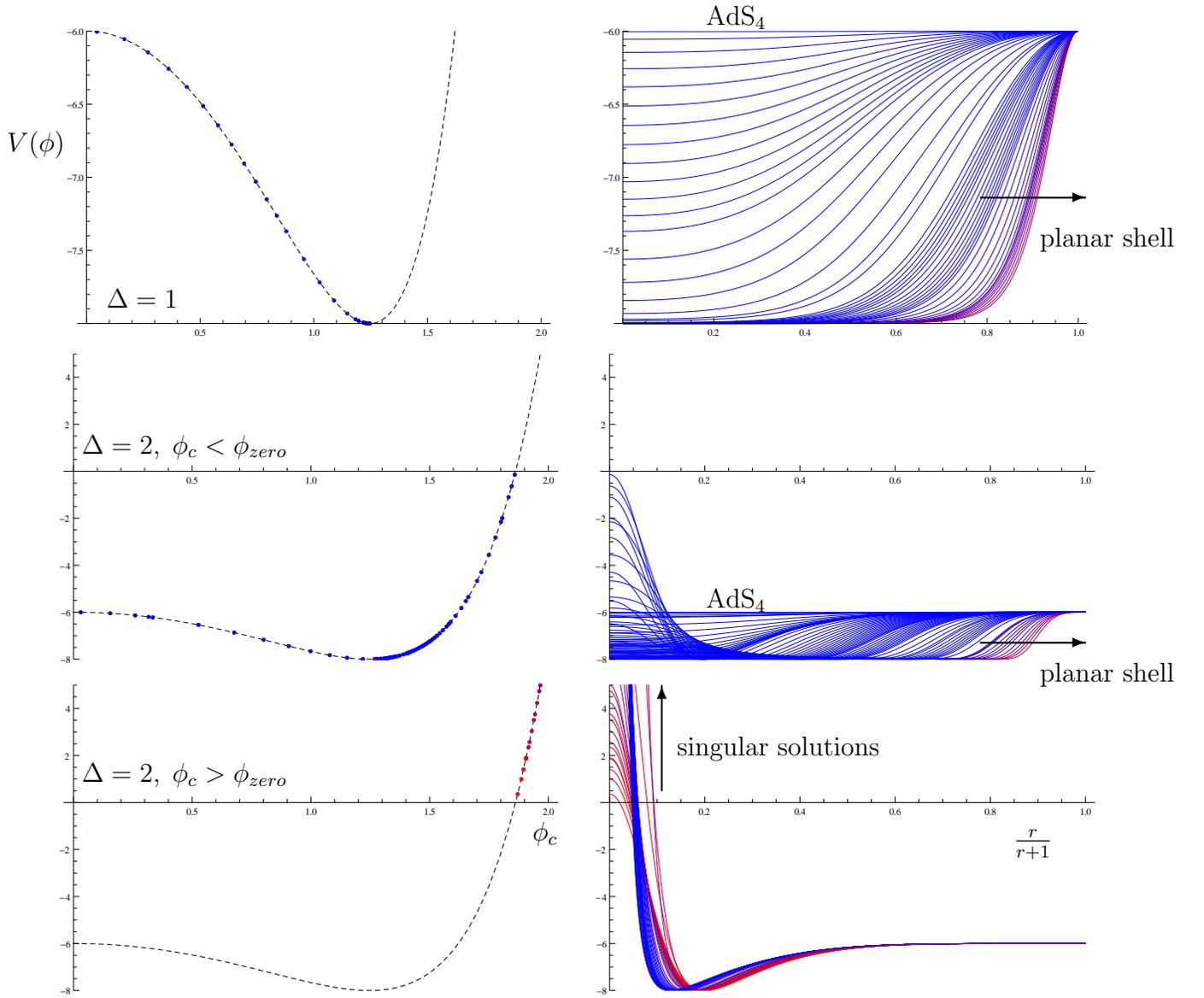

Given the behaviour of $m(\phi_c)$ for the two choices of boundary conditions, let us examine the scalar profiles; these are plotted in \fig{gsw:profiles} using their potential value $V(\phi(r))$ for the three branches of solutions. We note that the branch of solitons which attains a planar limit always has the scalar field approaching the value $\phi_{PW}$ in the limit. As anticipated at the end of \sec{s:gswpre}, in the dimension $\Delta =1$ boundary condition the scalar field explores only the region between the \AdS{4} vacuum at the origin and the Pope-Warner vacuum. While for $\Delta =2$ boundary condition the scalar field explores all of the field space $\phi \in [0,\infty)$, it gets more and more localised to the region $\phi \in [0,\phi_{PW}]$ as we increase the asymptotic charges. 

The singling out of the point $\phi = \phi_{PW}$ is quite natural; once the scalar field enters a local minimum it has to pay a gradient price to get out of the potential well. Hence in the asymptotic limit, the field prefers to stay at the PW vacuum and quickly transit into the \AdS{4} vacuum at the origin. The geometry in the limit ${\cal X}  \to \infty$ starts to resemble very closely a thin shell geometry; we have a bubble of the `true vacuum' which in our consideration is the PW vacuum at $\phi = \phi_{PW}$ inside the `false vacuum' at $\phi=0$. Effectively our large scalar solitons are morphing into the domain wall solutions of planar \AdS{}. 

In the standard holographic lore these planar geometries correspond to RG flows driven by vevs \cite{Freedman:1999gp}. We start in the \AdS{4} vacuum at the origin of field space and turn on vevs for either the $\Delta =1$ operator ${\cal O}_{\phi_2}$ or the $\Delta =2$ operator ${\cal O}_{\phi_1}$. The operators in question being relevant drive us away from the fixed point; in the deep IR we flow to the new conformal fixed point given by the Pope-Warner vacuum. This is precisely the behaviour expected for multi-extrema scalar potentials as discussed in the planar holographic superfluid case by \cite{Gubser:2009cg}, as was indeed confirmed in the original analysis of the model \eqref{gswdata} in \cite{Gauntlett:2009dn}.  The global solitons we have constructed here are the analogs for such RG flows when one considers the field theory on ${\mathbb R} \times {\bf S}^2$; previously such solutions were studied (for neutral scalars) using fake-supergravity techniques in \cite{Elvang:2007ba}.

\begin{figure}[h!]
\begin{center}
\includegraphics[width=0.98\columnwidth]{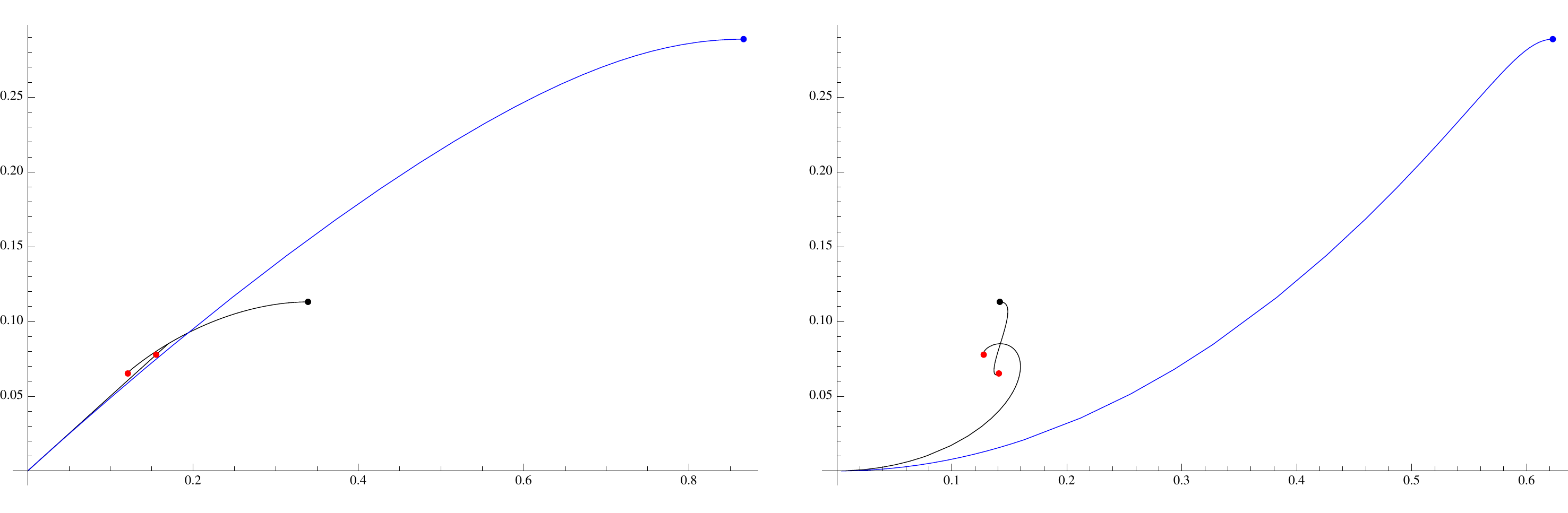}
\setlength{\unitlength}{0.1\columnwidth}
\begin{picture}(1.0,0.45)(0,0)
\put(-4.6,2.5){\makebox(0,0){$\frac{m}{\mu^3}$}}
\put(-1.8,0.4){\makebox(0,0){$\frac{\rho}{\mu^2}$}}
\put(3.5,0.4){\makebox(0,0){$\frac{\phi_2}{\mu^2}$, $\color{blue}{\frac{\phi_1}{\mu}}$}}
\end{picture}
\caption{Scaling invariants illustrating convergence to the planar solutions for the branches shown in \fig{gsw:massvscore}. The black dot is the low temperature planar hairy black hole for $\Delta= 2$ and the blue dot is the corresponding solution for $\Delta =1$ boundary conditions. The red dots represent the largest $\phi_c$ solutions obtained numerically.
}\label{gsw:globalscalinginvariant}
\end{center}
\end{figure}
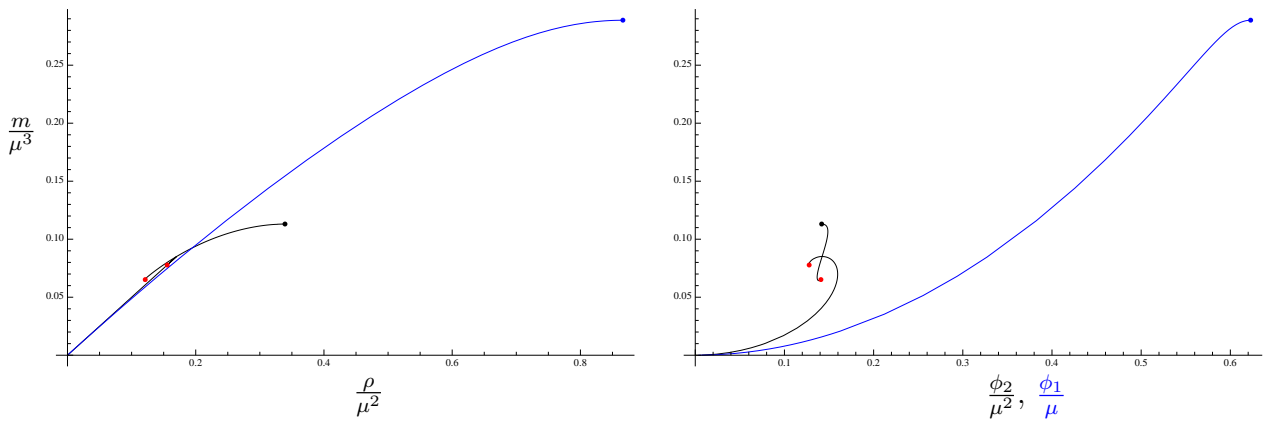

We can also check for consistency that the large solitons we construct here morph smoothly into the planar hairy black holes. In \fig{gsw:globalscalinginvariant} we plot the scaled charges of the theory (in units of $\mu$) to check that as we take $\mu\to \infty$ holding the appropriate dimensionless quantity fixed, the solution becomes a planar soliton. There are again a few distinguished points in this diagram: (i) the global \AdS{4} vacuum, (ii) the large soliton asymptoting to the planar hairy black holes, and (iii) two large $\phi_c$ solutions. This confirms that many features of the picture we had for the case of the Abelian-Higgs model indeed carry through to a model that can be consistently embedded in 11D supergravity.

Aside from the global minimum $\phi_{PW}$, we see that the zero of the scalar potential $\phi_{zero}$ also plays a role in determining the nature of the solutions. This happens only for the $\Delta =2$ case where $\phi_c > \phi_{PW}$ solutions exist.
\begin{list1}
\item For both parts of the favoured branch of solutions, i.e.\ with  $\phi_c \in (\phi_{PW} , \phi_{zero})$,  we find that the scalar field attains $\phi_{zero}$ from below; the field always stays in the region where the potential is negative.
\item It is only for the second  disfavored branch of solutions do we find $\phi_c > \phi_{zero}$.
\end{list1}
Thus the attractor solutions which start out with large values of core scalar correspond to situations where the scalar field begins on the positive arm of the potential in the IR. Solutions which explore this part of the potential are subdominant in the microcanonical ensemble.

\subsection{Exploring criticality: double-trace deformations}
\label{s:gswdt}

One of the interesting features of the $SU(3)$ truncation under consideration is that it appears to sit at the critical value of the charge $q_c$ for the dimension $\Delta =2$ quantization. As noted earlier, this is basically because the two branches of solutions appear to intersect and it would appear that if one were to detune the charge repulsion then the branches would become disconnected. Note that this interpretation is quite natural despite $q =1 < q_c(\Delta = 2)$ for \eqref{gswdata}.  In general we should expect the value of $q_c$ to depend on the details of $V(\phi)$ and $Q(\phi)$, so it is not surprising that there is a smaller value of $q_c$ when we allow non-linearities into the potential.

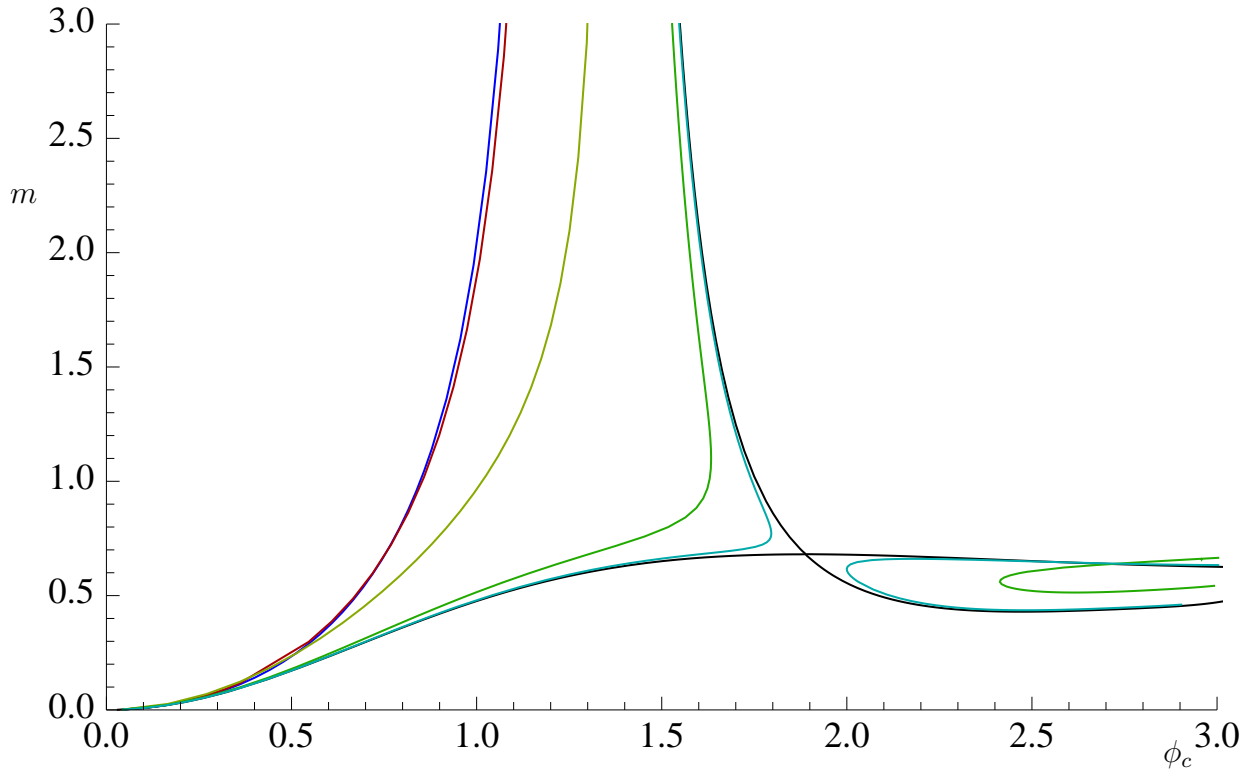
\begin{figure}[h!]
\begin{center}
\includegraphics[width=0.98\columnwidth]{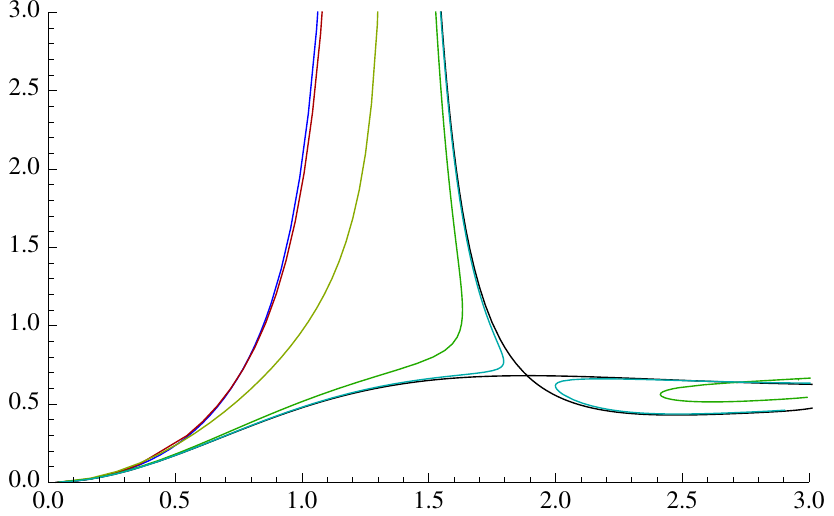}
\setlength{\unitlength}{0.1\columnwidth}
\begin{picture}(1.0,0.45)(0,0)
\put(4.8,0.5){\makebox(0,0){$\phi_c$}}
\put(-4.5,5.0){\makebox(0,0){$m$}}
\end{picture}
\caption{Soliton branches with double-trace deformations at fixed $\mu^{-1}\varkappa=0.1,1,10$ and $100$. For reference $\Delta=1$ is shown in blue and $\Delta=2$ in black, as in \fig{gsw:massvscore}.}\label{gsw:doubletrace}
\end{center}
\end{figure}

One way to verify this of course is to tweak the Maxwell coupling so that we can explore the behaviour of $\Delta =2$ solitons more cleanly. However, this takes us back to the realm of phenomenological models. A more useful strategy then is to explore the behaviour of the system as we tweak the scalar boundary condition, which allows us to stay within the remit of the consistent truncation. This turns out to be straightforward and the result of the numerical exploration is now presented in \fig{gsw:doubletrace}.

As can be inferred directly from this plot we see that the double-trace boundary condition with $\varkappa \in (0,\infty)$ interpolates between the two cases of $\Delta =1$ and $\Delta =2$ quantizations. As we increase $\varkappa$ up from zero, we see that the curves migrate towards the right; the presence of   $\phi_2 \neq 0$ clearly manifests itself. What is even more striking is the emergence of new branches of solutions as we increase $\varkappa$: already for $\varkappa/\mu = 10$ we start to notice a double-valued solution branch for large $\phi_c$ (as previously encountered  in the Abelian-Higgs model).  
We believe that these branches exist all the way down to $\varkappa \to 0$, though at increasing larger values of $\phi_c$, which makes them hard to access.

It is quite reassuring to see that the consistent truncation model incorporates all the crucial physical features  encountered in the phenomenological model. A very natural question here is why is the $\Delta =2$ boundary condition precisely at the critical point of the theory? We believe this has to do with the full details of the non-linear problem; it would be quite interesting to ascertain the physical reason behind this. We leave this issue for future investigation.

\subsection{Microcanonical phase diagram}
\label{s:gswmc}

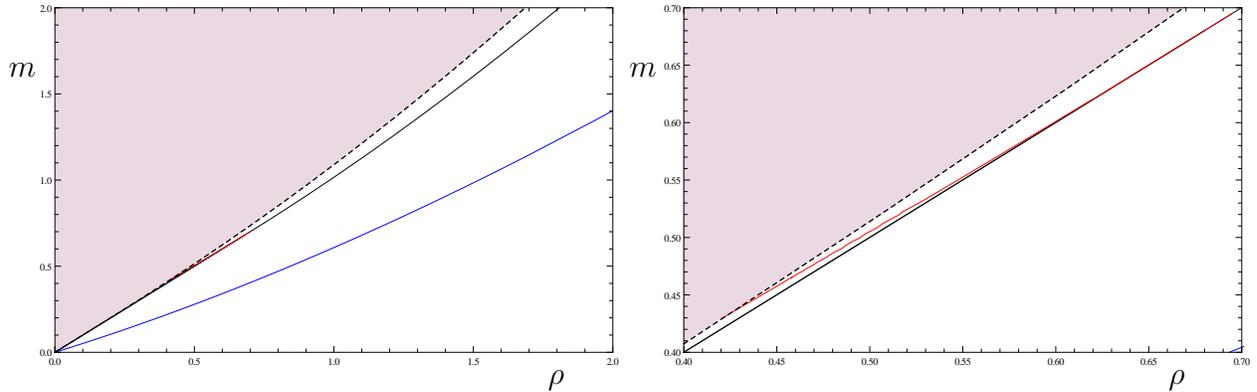
\begin{figure}[h!]
\begin{center}
\includegraphics[width=0.98\columnwidth]{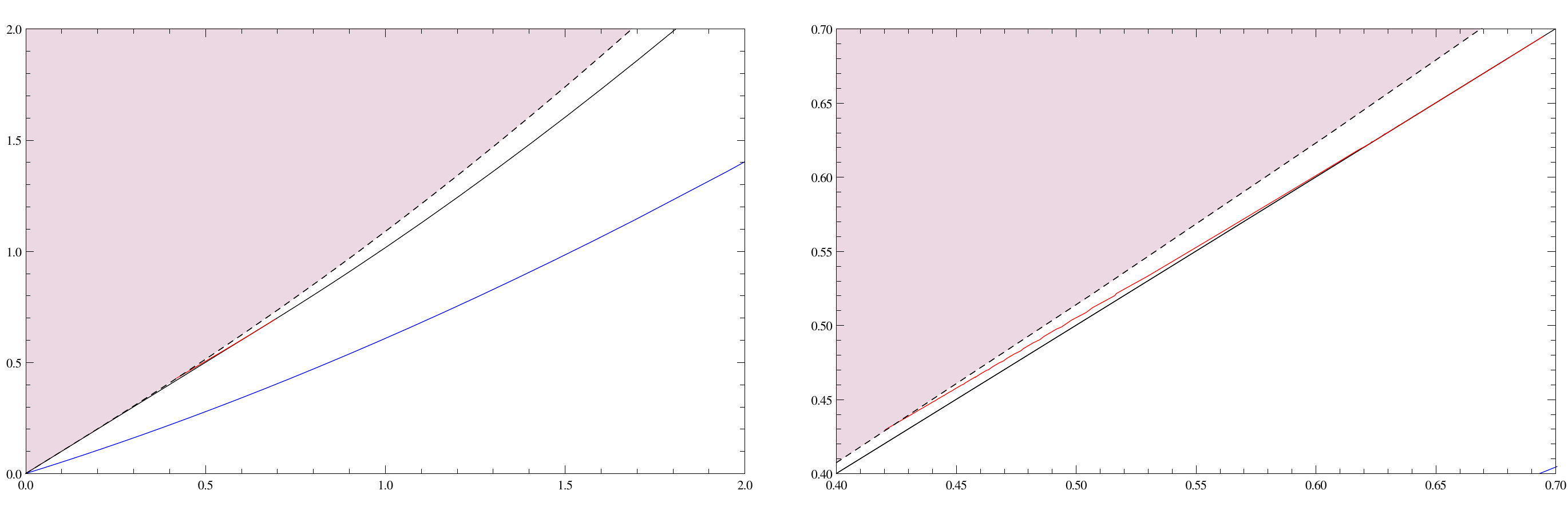}
\setlength{\unitlength}{0.1\columnwidth}
\begin{picture}(1.0,0.45)(0,0)
\put(4.8,0.5){\makebox(0,0){$\rho$}}
\put(-0.2,0.5){\makebox(0,0){$\rho$}}
\put(-4.5,3.0){\makebox(0,0){$m$}}
\put(0.5,3.0){\makebox(0,0){$m$}}
\end{picture}
\caption{Microcanonical phase diagram for the theory \eqref{gswdata}. $\Delta = 1$ global solitons are shown in blue, $\Delta = 2$ for $\phi_c<\phi_{zero}$ in black and $\Delta = 2$ for $\phi_c>\phi_{zero}$ in red. The dashed line indicates extremal \RNAdS{4} black holes. }\label{gsw:micro}
\end{center}
\end{figure}

It is interesting to examine the microcanonical phase diagram of the theory characterised by \eqref{gswdata}, presented in \fig{gsw:micro}. 
We will discuss this for the various choices of boundary conditions in turn:
\begin{list1}
\item For the $\Delta =1$ boundary condition, we have a smooth curve $m(\rho)$ which delineates the phase boundary; these are smooth solitons which we suspect have the minimal mass in a fixed charge sector.\footnote{See \sec{sec:Discussion} for a discussion of this point.} Increasing the mass above this causes one to encounter global scalar hair black holes which dominate the ensemble. For sufficiently high $m$ at fixed $\rho$, hair-free Reissner-Nordstr\"om solutions dominate.

\item The situation is similar for the $\Delta =2$ boundary condition; the main new element is that the phase boundary has a non-analytic point. This corresponds to the value of $m$ and $\rho$ of the soliton which lies at the intersection of the two solution branches. It is hard to tell from a numerical plot that this is the case. It is also possible that we should encounter some zig-zags in the region above the phase boundary controlled by the attractor solutions, which we however have been unable to verify. The rest of the phase diagram is qualitatively similar to the $\Delta =1$ case. A curious feature is that second branch of solutions with $\phi_c > \phi_{zero}$ appears to smoothly merge onto the extremal Reissner-Nordstr\"om phase line. We thus have two solutions with the same mass for low values of charge; of course the extremal black holes dominate by virtue of their non-vanishing horizon area, but it is curious to see non-uniqueness in the microcanonical phase space at zero temperature.

\item Double-trace deformations (not shown) reveal smooth solitonic branches for all $\rho$ between the $\Delta = 1$ soliton branch and the $\Delta = 2$ branch. In addition we have new branches at large $\phi_c$ existing for a finite range of $\rho$. 
\end{list1}


\section{Charged solitons in the M2-brane theory}\label{sec:DG}


Our discussion thus far has focussed on two models, one phenomenological and another that arises from a consistent truncation of 11D supergravity. In neither case though do we actually know any details regarding the candidate dual  field theory. We now turn to an example that which has a known dual CFT, and a simple one at that.

 Consider the maximally supersymmetric theory with 16 supercharges in $2+1$ dimensions, which arises as the low energy limit of the world-volume dynamics of M2-branes. This theory is superconformal and has $SO(8)$ R-symmetry. One can define this theory in the limit $k \to 1$ of a family of $\mathcal{N}=6$ superconformal Chern-Simons theories with a gauge group $U(N)_k\times U(N)_{-k}$ (where the subscripts denote the Chern-Simons levels) and a 't Hooft coupling $\lambda\equiv N/k$ \cite{Aharony:2008ug}.   It is believed that in the case of $k=1,2$, the supersymmetry should get enhanced to $d=3$, $\mathcal{N}=8$ i.e., 16 supercharges. We will refer to this theory as the M2-brane theory for simplicity.
The field theory with $k\to1 $ is dual to M-theory on \AdS{4} $\times \; { \bf S}^7$ with  $G_{4}=3\,(2\,N)^{-3/2}$. 

\subsection{A consistent truncation of the M2-brane theory}

The M2-brane theory has bosonic operators which are charged under the $SO(8)$ R-symmetry. Of interest to us will be operators that carry equal charges under $U(1)^4 \subset SO(8)$; in fact, we will be interested in the lightest chiral primary operator of this type. The operator in question has conformal dimension $\Delta =1$ and since it is a chiral primary has $R$-charge also equal to unity; we will refer to this operator henceforth as ${\cal O}_1$. The conformal dimension immediately implies that we are dealing with a bulk scalar field with mass $m^2_\phi= -2$. 

It turns out that this chiral primary operator, the stress tensor $T_{\mu\nu}$ and the charge current $J_\mu$ (where $J_{\mu}$ is the diagonal Cartan generator of $U(1)^4 \subset SO(8)$) form a closed sub-sector of the theory at large $N$. In particular for ${\cal O}_{x}$ being any other operator in the spectrum of the M2-brane theory we have 
\begin{equation}
\langle T_{\mu_{1} \nu_{1}} \cdots T_{\mu_{j}\nu_{j}}\, J_{\sigma_{1}}\cdots J_{\sigma_{k}}\, {\cal O}_1^n\, {\cal O}_x\rangle = 0
\label{}
\end{equation}	
for  any $j,k,n \in {\mathbb Z}_+$. This statement, which constrains the field theory OPE in the planar limit, arises from the fact that the corresponding supergravity theory admits a consistent truncation to a theory of Einstein-Maxwell-charged scalar dynamics in four dimensions.  We have been calling this the $U(1)^4$ truncation.

The gravitational dynamics of this truncated sector of the M2-brane theory is given by the  Lagranagian \eqref{Sgeneral} with fixed values of scalar potential $V(\phi)$ and Maxwell coupling $Q(\phi)$:
\begin{equation}
 V(\phi) = -2\,\left(2+\cosh{\sqrt{2}\phi}\right) \ , \qquad Q(\phi) = \frac{1}{2} \,\sinh^2\frac{\phi}{\sqrt{2}}
\label{dgdata}
\end{equation}	
We infer from this Lagrangian that $\phi$ corresponds to a field of mass $m^2_\phi = -2$ and $q = \frac{1}{2}$.

The existence of this truncation was first noted in \cite{Chong:2004ce} whose analysis will come to play in a short while and was recently analysed for holographic superconductivity in \cite{Donos:2011ut}. We also note that the corresponding sector  of ${\cal N} =4$ SYM in $d= 3+1$ dimensions was discussed in the context of charged solitons in \cite{Bhattacharyya:2010yg}; we will have occasion to contrast the behaviour of these two systems during the course of our analysis. In the rest of the section we are going to analyse the behaviour of charged solitons and black holes of this truncation.

Note that the connection to the M2-brane theory is contingent on us quantizing the theory \eqref{Sgeneral}, \eqref{dgdata} with $\Delta =1$ boundary conditions for the scalar. We could of course choose to pick $\Delta =2$ or other boundary conditions as discussed before in \sec{sec:bcscalar}. The former corresponds to some non-supersymmetric CFT in $d=2+1$, whose spectrum differs from that of the M2-brane theory by one single-trace operator (assuming all other fields are quantised the similarly in the two cases). Choice of  boundary conditions via $\phi_2(\phi_1)$ can be understood as non-supersymmetric multi-trace deformations of the M2-brane theory. 

\subsection{BPS configurations and supersymmetry equations}
\label{s:dgbps}

One of the advantages of working with a supersymmetric field theory is that we can examine the behaviour of BPS configurations. It turns out that some of the solitonic solutions with the $\Delta =1$ boundary condition on the scalar can be obtained as solutions to BPS equations. The latter have the distinct advantage of being much simpler than the full set of field equations given in \App{sec:generalEoms}. 

The BPS configurations we are interested preserve $\frac{1}{8}$ of the original supersymmetries and were written down originally in \cite{Chong:2004ce}, which we will now exploit. Let us first define a new radial coordinate:\footnote{Comparison with the results of \cite{Chong:2004ce} involves setting their coupling constant $g= \frac{1}{2}$ and rescaling their fields. Specifically, $X_i =1$, $A^i_{there} = \frac{1}{\sqrt{2}} \,A$ and  $\varphi_i = \frac{\phi}{\sqrt{2}}$ for $i = 1,\cdots,4$. The $X_i$ are the familiar scalars encountered in four dimensional $U(1)^4$ gauged supergravity and $\varphi_i$ are the hyperscalars.}
\begin{equation}
r = u\, H(u)
\label{}
\end{equation}	
Consider the following ansatz for the dynamical fields:
\begin{subequations}
\begin{align}
g(r) &=(1+u^2\, H(u)^4) \left(1+u\,\frac{H'(u)}{H(u)}\right)^2\\
A(r) &=  \frac{2}{H(u)} \\
\phi(r) &= \sqrt{2}\, \text{arccosh} \left(H(u) + u\,H'(u)\right)\\
e^{-\beta(r)} &=\frac{1}{\left( H(u)+ u\, H'(u)\right)^2}
\end{align}
\end{subequations}

In terms of these new variables, the equations of motion following from \eqref{Sgeneral} with \eqref{dgdata} turn out to be equivalent to a single non-linear second order ODE for $H(u)$ \cite{Chong:2004ce}:
\begin{equation}
\left(1+ u^2\, H^4\right) \, \frac{d^2}{du^2} \left(u\, H \right) +  u\, H^3\, \left( \left( \frac{d}{du} \left(u\, H \right)\right)^2 -1 \right) = 0
\label{DGsusy}
\end{equation}	
We note that \cite{Chong:2004ce} derive a set of BPS equations which are valid for the M2-brane system with unequal charges in $U(1)^4 \subset SO(8)$; it might be interesting to study the behaviour of solitons in those systems as well, but our current focus will be on the restricted subspace of equal charges.

\paragraph{Asymptopia and conserved charges:}
The equation \eqref{DGsusy} is reasonably easy to analyze. First of all we note that the asymptotic solution takes the form:
\begin{eqnarray}
&&H(u) \to 1 + \frac{h_1}{u} - \frac{h_2}{u^2} +\frac{h_1\,h_2}{u^3} + \cdots \ , \qquad u \to \infty 
\nonumber \\
&& \Longrightarrow\;\; r= u \, H \simeq u + h_1 - \frac{h_2}{u} + \frac{h_1\,h_2}{u^2} + \cdots
\label{}
\end{eqnarray}	
It turns out that the large $u$ behaviour of $H(u)$ is precisely such that the solution is normalizable. To check this let us compute the physical fields in the usual radial variable. We find:
\begin{eqnarray}
\phi(r) &=& \frac{2\,\sqrt{h_2}}{r} - \frac{\sqrt{h_2}\,(3+2\,h_2)}{3\,r^3} +\cdots\nonumber \\
A(r) &=& 2\, \left(1-\frac{h_1}{r} + \cdots \right) \nonumber \\
g(r) &=& r^2 + (1+2\,h_2) - \frac{2\,h_1}{r} + \cdots
\label{}
\end{eqnarray}	
From this information we have a simple identification between the parameters $(h_1,h_2)$ and those used earlier:
\begin{equation}
g_1 = 2\,h_1  \,, \qquad \phi_1 = 2\, \sqrt{h_2} \,,  \qquad\mu = 2 \,,  \qquad \rho = 2\, h_1
\label{}
\end{equation}	
using which we can check that our solution indeed satisfies the correct boundary conditions \eqref{asymfalloffs}. It is also clear from $q = \frac{1}{2}$ that $\mu \,q =1$ as required (see \App{sec:pertsolitons}) and the mass density which for $\Delta =1$ boundary condition is given by $g_1$, scales linearly with the charge density: $m =\, \rho$ along the solution branch.

\paragraph{Regular core analysis:} The other piece of data we need is the behaviour at $u =0$. It is easy to show that there is a smooth solution with series expansion:
\begin{equation}
H(u) = h_c + \frac{1}{6}\, h_c^3\, (1-h_c^2) \, u^2 + \frac{1}{40}\,h_c^5\,(1-6\,h_c^2 +5\,h_c^4)\, u^4 + \cdots
\label{}
\end{equation}	
which then implies that
\begin{equation}
\phi_c = \sqrt{2}\, \text{arccosh}(h_c)
\label{}
\end{equation}	
At this point it is manifestly clear that we have a one-parameter family of solutions. Since there is no constraint coming from the asymptotic boundary conditions, and given that the core behaviour is controlled by a single parameter, we are free to simply pick $h_c$ or equivalently $\phi_c$ and integrate \eqref{DGsusy} out. Every such solution is guaranteed to be a smooth soliton. Before discussing the numerical results let us also pause to note some other interesting features of the equation at hand.

\paragraph{Singular core behaviour:} We could also ask if it is possible to relax the requirement of a regular core to construct other potential solutions. For the moment, lets us call these irregular solitons. The equation \eqref{DGsusy} actually admits  a one parameter family of special solutions given by 
\begin{equation}
H(u) = 1 + \frac{h_s}{u}
\label{splsol1}
\end{equation}	
This solution is nothing but the supersymmetric case of the more general solution presented in \eqref{globalschw} with $m= \rho = 2h_s$. This solution has lower mass than the extremal solution at the same charge and is singular at the core. These solutions are the \AdS{4} analogs of superstars; the fact that static supersymmetric solutions are singular in \AdS{} has been known since the early work of \cite{Romans:1991nq}. We can also study the planar limit of this superstar; the only parameter of the solution is $h_s$, and taking the planar limit by sending $h_s\to\infty$ results in a (zero-parameter) $m=0$, $\rho\neq 0$ singular planar \RNAdS{4} solution.

Another special solution exists if we simplify the ODE in the approximation of large $uH$, that is, by replacing the coefficient of the second derivative term to be $ u^2 \, H^4$ (i.e., dropping the $1$), when
\begin{equation}
H(u) = \sqrt{1+ \frac{c_1}{u} + \frac{c_2}{u^2}}	
\label{}
\end{equation}	
which indicates the possible presence of a two parameter family of solutions with a $\frac{1}{u}$ core singularity. There is indeed a solution with $H(u) \simeq u^{-1}$ at the origin as one can check by a series expansion. It transpires the there are no other singular core solutions for this system; we will return to this issue once we discuss the results of our numerics.

\subsection{Global solitons of the $U(1)^4$ truncation}

We now turn to the results of numerical studies of solitons of the theory \eqref{Sgeneral} and \eqref{dgdata}. We will first discuss the behaviour of solitons for the $\Delta = 1$ boundary condition on the scalar, which as we have just seen can be equivalently studied using the supersymmetry equations.

\begin{figure}[h!]
\includegraphics[width=0.98\columnwidth]{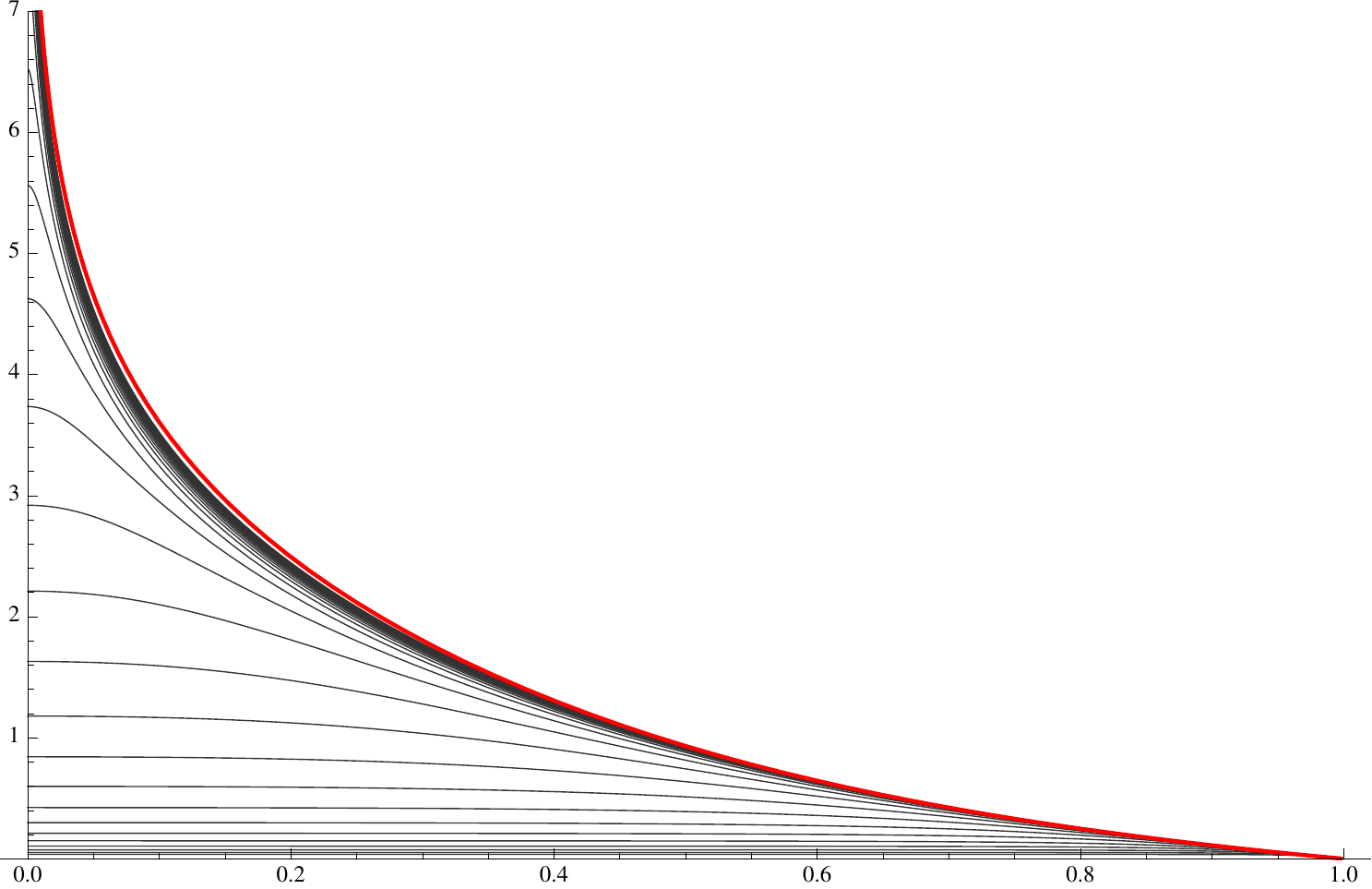}
\setlength{\unitlength}{0.1\columnwidth}
\begin{picture}(1.0,0.45)(0,0)
\put(4.8,0.5){\makebox(0,0){$\frac{r}{r+\phi_1}$}} 
\put(0.0,6.5){\makebox(0,0){$\phi$}}
\end{picture}
\caption{Scalar field $\phi$ profiles for the supersymmetric $\Delta=1$ global soliton for various values of $\phi_c$. The red line is the analytic planar solution \eqref{plneutralsol}, which depends only on $\frac{r}{\phi_1}$ and is the reason for our choice of radial variable (see \sec{sec:PlanarDG}).}\label{dg:susysoliton}
\end{figure}
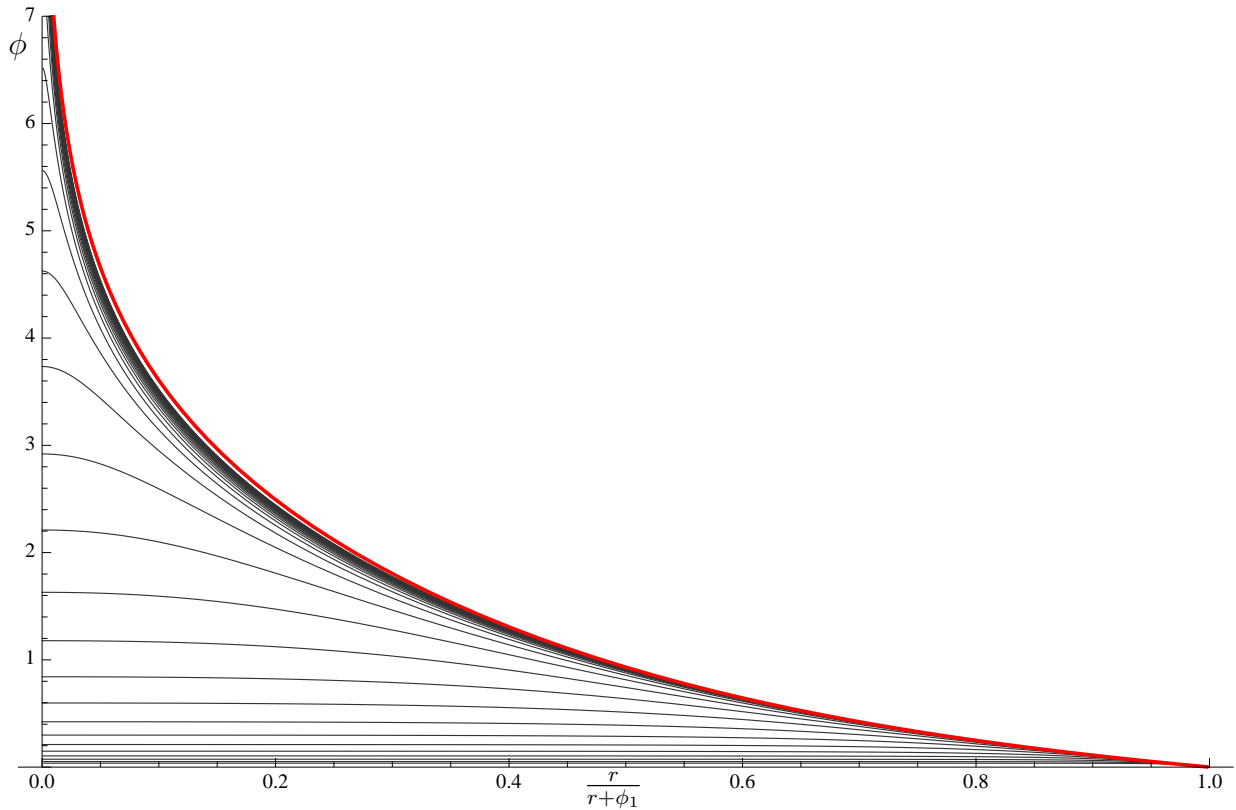

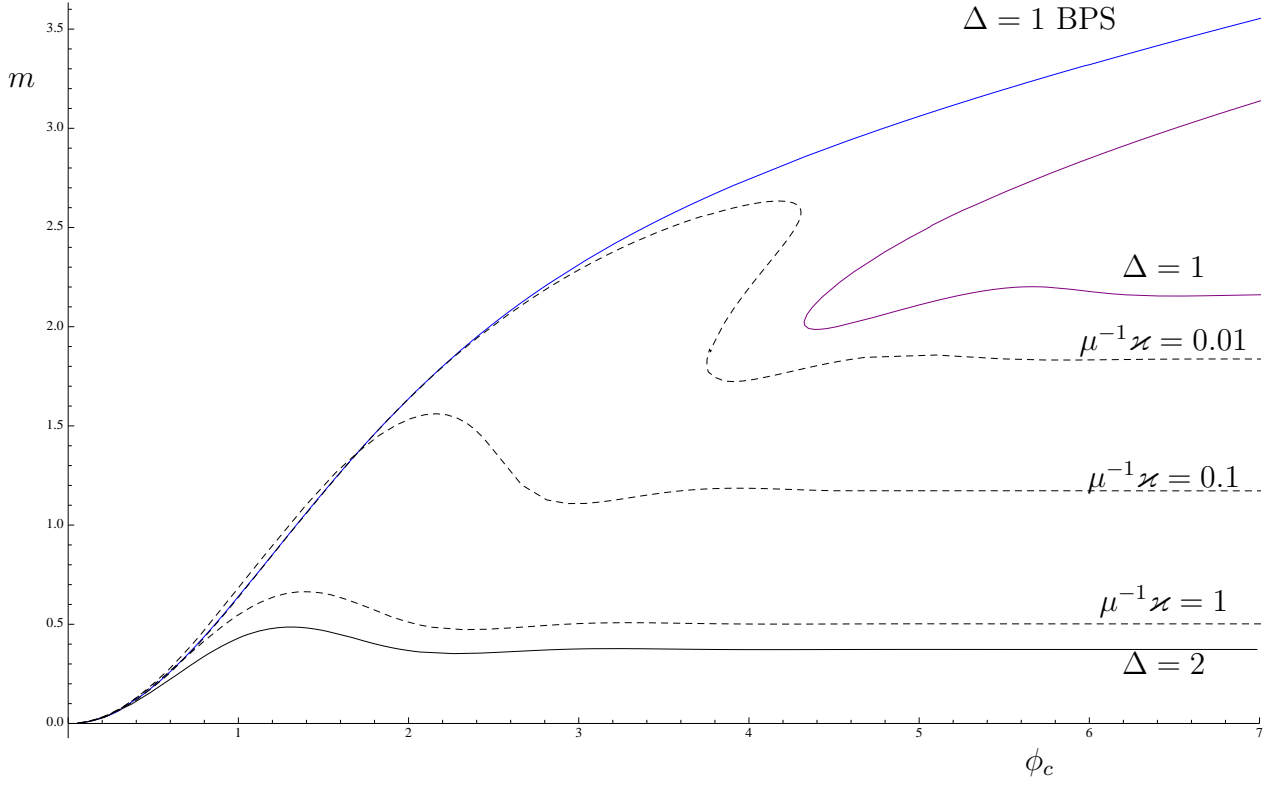
\begin{figure}[h!]
\includegraphics[width=0.98\columnwidth]{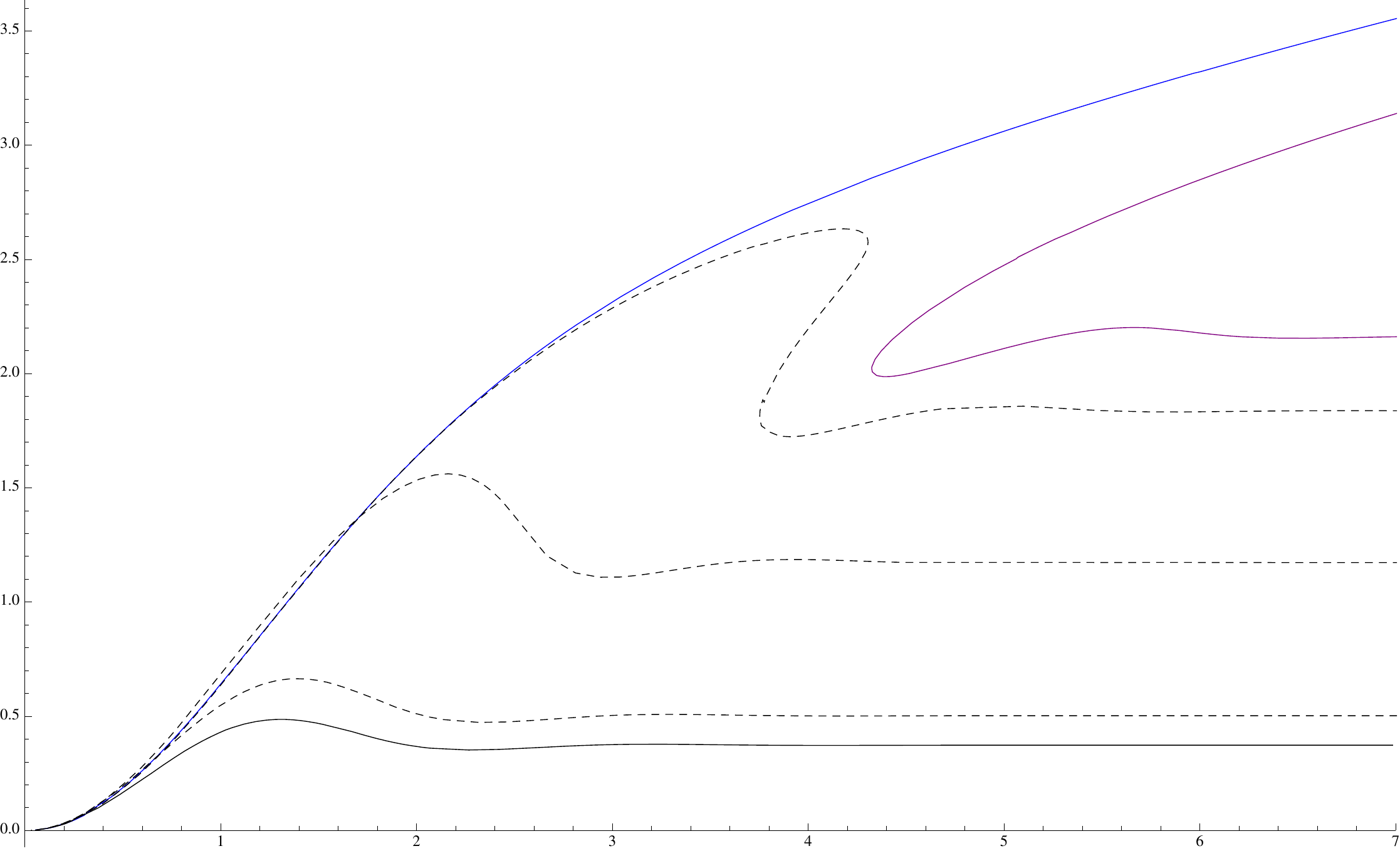}
\setlength{\unitlength}{0.1\columnwidth}
\begin{picture}(1.0,0.45)(0,0)
\put(8,0.3){\makebox(0,0){$\phi_c$}} 
\put(-0.2,5.8){\makebox(0,0){$m$}}
\put(8,6.3){\makebox(0,0){$\Delta = 1$ BPS}}
\put(9,4.3){\makebox(0,0){$\Delta = 1$}}
\put(9,3.7){\makebox(0,0){$\mu^{-1}\varkappa = 0.01$}}
\put(9,2.6){\makebox(0,0){$\mu^{-1}\varkappa = 0.1$}}
\put(9,1.6){\makebox(0,0){$\mu^{-1}\varkappa = 1$}}
\put(9,1.1){\makebox(0,0){$\Delta = 2$}}
\end{picture}
\caption{Soliton branches in the theory \eqref{dgdata}. The blue line is the $\Delta = 1$ supersymmetric soliton and appears together with a non-BPS branch of the full set of equations in purple. The dashed lines are branches in the presence of double trace deformations with $\mu^{-1}\varkappa = 0.01,0.1$ and $1$ showing convergence towards the $\Delta =2$ branch in solid black.}
\label{dg:doubletrace}
\end{figure}

\begin{figure}[h!]
\includegraphics[width=0.99\columnwidth]{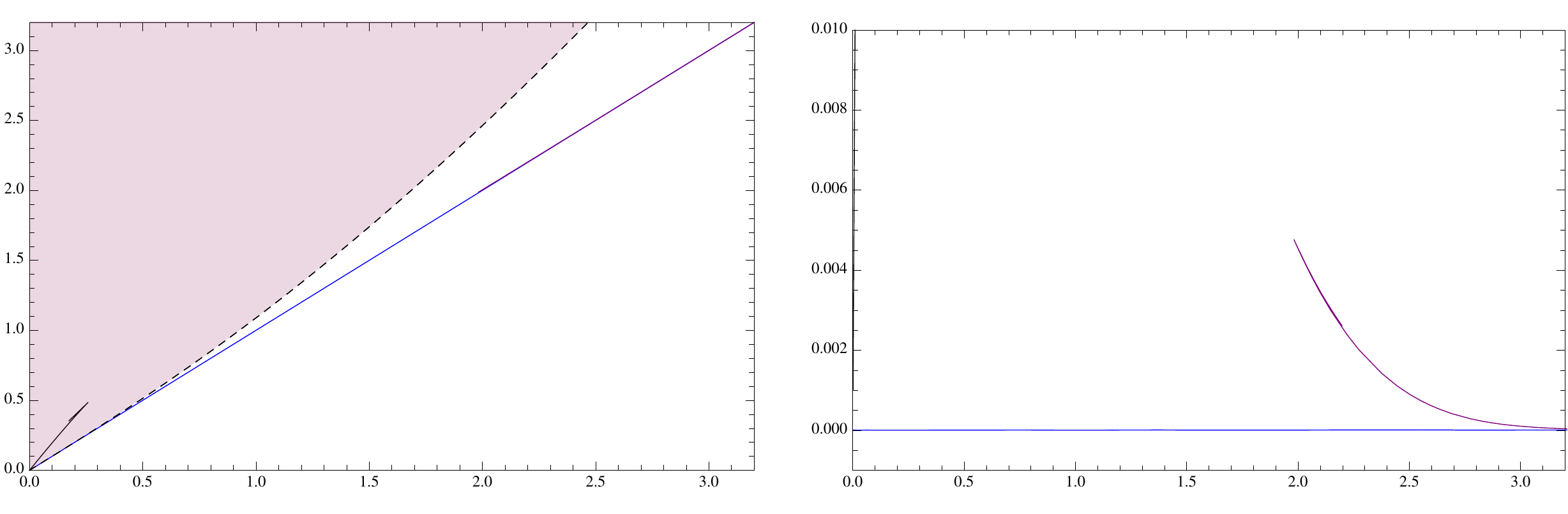}
\setlength{\unitlength}{0.1\columnwidth}
\begin{picture}(1.0,0.45)(0,0)
\put(4.,0.5){\makebox(0,0){$\rho$}} 
\put(0.0,3.3){\makebox(0,0){$m$}}
\put(5.85,3.3){\makebox(0,0){$m-\rho$}}
\put(9,0.5){\makebox(0,0){$\rho$}} 
\end{picture}
\caption{Micro-canonical phase diagram for the theory \eqref{dgdata}, showing both the $\Delta = 1$ supersymmetric soliton saturating the BPS bound (blue) and the $\Delta = 1$ non-BPS branch with the same boundary conditions (purple). The $\Delta = 2$ soliton branch is shown in black and lies above the extremal RN line, as expected based on the perturbative results of Appendix \ref{sec:pertsolitons}.}
\label{dg:micro}
\end{figure}

The result of the numerical integration of the field equations is demonstrated in \fig{dg:susysoliton} -- here we plot the scalar profile for a rescaled radial variable. The solutions are smooth at the origin, though they start to look more and more `spiky' as $\phi_c$ gets large. We have for consistency checked that the results from integrating the full set of field equations produces the same behaviour as the supersymmetric equation \eqref{DGsusy}.

It is interesting to ask what happens to the conserved charges ${\cal X}$ as we tune the core scalar value: given the BPS nature of these solutions, the mass and charge scale together $m=\rho$. What is very curious is that the functions $m(\phi_c)$ and $\phi_1(\phi_c) = \langle {\cal O}_{\phi_2} \rangle (\phi_c)$ are monotone in $\phi_c$, albeit with very slow growth, see \fig{dg:doubletrace}. The curve fits to a power-law profile; we find for $\phi_c \gg 1$ 
\begin{equation}
\phi_1(\phi_c) \simeq 0.8 \sqrt{\phi_c}+0.5 \,, \qquad  m(\phi_c) \simeq 1.1\sqrt{\phi_c}+0.5
\label{}
\end{equation}	
The fact that we have unbounded conserved charges along the soliton branch is not new by now; this behaviour has already been seen, for example in \sec{sec:GSW}, for a similar $\Delta =1$ boundary condition. What is surprising however, is that this slow growth of asymptotic charges is indicative of critical behaviour. We can illustrate this clearly by moving away from the supersymmetric boundary condition and allow an admixture of $\frac{1}{r}$ and $\frac{1}{r^2}$ fall-offs. We can no longer rely on the simple expedient of using \eqref{DGsusy}, but by now we have fairly good handle on constructing solitonic solutions with various boundary conditions. The result of the analysis where we deform via double-trace boundary conditions is shown in \fig{dg:doubletrace}. We note from here that increasing $\varkappa$ causes the $m(\phi_c)$ curve to droop down; the $\varkappa =0$ curve corresponding to the supersymmetric boundary conditions envelopes of all solutions with $\varkappa \neq 0$ for large $\phi_c$.  For small values of $\varkappa$ we encounter a multi-valued curve $m(\phi_c)$; the curve loops back onto itself in a very pronounced manner. At large values of $\phi_c$ is appears to oscillate down towards an attractor solution (though this is hard to ascertain with a great deal of precision). At the other end of the boundary conditions, i.e., as $\varkappa \to \infty$ we see that the loop-back effect has almost disappeared -- we encounter a situation akin to $q<q_c$ for the Abelian-Higgs model studied in \sec{sec:BU}.

It is in this sense that the $\Delta = 1$ boundary condition is critical; for small deformations by $\varkappa$ the branch switches from unbounded to bounded. This criticality, here on the $\Delta = 1$ branch can be compared with the $SU(3)$ truncation \sec{sec:GSW}, where critical behaviour was also observed with the two $\Delta = 2$ solution branches intersecting. In contrast we now seem to have a single solution branch. Furthermore, along this solution branch the field explores the full domain $\phi \in [0,\infty)$; while in earlier examples as we have discussed using the scalar profiles, the field preferentially likes to stay in a bounded region, whereas from \fig{dg:susysoliton} we see that now the scalar field explores the full range of the potential. The main difference of course can simply be attributed to the scalar potential itself; unlike the $SU(3)$ model whose potential given in \eqref{gswdata} had the Pope-Warner extremum, we have a monotonic potential function in \eqref{dgdata}. Note also that in contrast to the bottom up models, the potential in the current $U(1)^4$ truncation is exponential and hence a lot steeper. 

One fascinating aspect of the double-trace deformations is that it reveals a new solution branch even for the $\Delta =1$ boundary condition. This is unprecedented from just the BPS equation \eqref{DGsusy}; as we saw in \sec{s:dgbps} one just picks a core value for the scalar and integrates out. However, nothing prevents non-supersymmetric solitons from existing within the $\Delta =1$ boundary condition. For consistency one anticipates that the mass of solutions along this branch is larger than the BPS value for a given charge $\rho$. Indeed we find that this is true as illustrated in the 
microcanonical phase plot \fig{dg:micro}. The solutions encountered along this new branch are subdominant saddles of the microcanonical ensemble. We should note that it was the ability to tune $\varkappa$ which allowed us to explore this solution branch; one can construct the solitons for $\varkappa\neq 0$ and then relax back towards $\varkappa \to 0$ -- this allows one to explore a larger region of the solution space and reveals new solution branches.

\subsection{Planar theory: analytic neutral soliton and hairy black holes}
\label{sec:PlanarDG}

We have by now a good understanding of the solitonic solutions with various boundary conditions. In the supersymmetric   $\varkappa =0$ case, we learnt that solitons can get arbitrarily large. From our previous experience we would expect the large mass soliton to go into a zero temperature planar black hole with scalar hair -- the new wrinkle in the story is that no such black hole exists!

The theory under consideration with $\Delta =1$ boundary condition was recently studied from a viewpoint of holographic superconductors in \cite{Donos:2011ut}. The authors found something curious -- while a probe scalar analysis reveals the possibility of a critical temperature $T_c$ where planar \RNAdS{4} black holes may become unstable towards condensation of the charged scalar field, it transpires that this linear analysis only reveals the presence of a new branch of hairy black hole solutions. These planar hairy black holes  start their existence at $T=T_c$, but turn out to be sub-dominant saddle points. More to the point, they only exist for $T \ge T_c$  -- there is no solution for lower values of temperature with scalar condensate. The upshot of this analysis (which we have confirmed numerically) is simply that we have a hairy black hole extending off to large temperatures (see Fig.~1 of \cite{Donos:2011ut}).

The absence of candidate planar hairy black holes with $T \to 0$ poses a puzzle for our family of supersymmetric solitons: what happens to them in the limit $\phi_c \to \infty$?  Could it be that they  go over to the planar black hole of arbitrarily large temperature? This bizarre possibility does not come to pass. Rather the solutions go over to a very curious solution: a neutral scalar solution with a singular core.

To understand this let us recall an exact solution of the field equations which we presented earlier in \eqref{globalsingulardg}. This is a neutral  solution since the gauge field is switched off; it also does not carry any energy density, $m = \rho =0$. It is however not supersymmetric since the solution does not fall into the BPS ansatz. All the freedom for this family is encoded in a single parameter which we have chosen to be $\phi_1$ for convenience. The core region has $\phi \simeq -\log r$ indicating that it is a not a regular soliton. Since $\phi_1(\phi_c) \to \infty$ along the branch of supersymmetric solitons, let us ask what happens to \eqref{globalsingulardg} as we take it to large field values. 

Based on the scalings discussed in \sec{sec:planarlimit} and accounting for $\phi_1$ having dimension $1$, 
\begin{equation}
r\to \lambda \,r,\qquad t \to \lambda^{-1}t,\qquad \phi_1\to \lambda\, \phi_1, \qquad  \lambda^2\, d\Omega_2^2 \to d{\bf x}_2^2
\end{equation}
we find that the solution morphs into the following planar \AdS{4} geometry:
\begin{equation}
ds^2 =r^2\, \eta_{\mu\nu}\, dx^\mu\, dx^\nu +  \frac{dr^2}{\frac{\phi_1^2}{2}+r^2} \ , \qquad \phi(r) = \sqrt{2}\text{arcsinh}{\frac{\phi_1}{\sqrt{2}r}}
\label{plneutralsol}
\end{equation}	
The geometry has a singular horizon at $r =0$ (the Ricci scalar diverges there), which goes hand in hand with the diverging scalar field. However, this is the desired end-point of the large supersymmetric solutions. For one, the scalar profile in \eqref{plneutralsol} is the limiting behaviour of the large $\phi_1$ BPS solitons, as we have illustrated in \fig{dg:susysoliton}. Furthermore, by examining the behaviour of $m(\phi_c)$ and $\phi_1(\phi_c)$ along the branch of BPS solutions, we realise that $m \simeq \phi_1$ for $\phi_1 \gg 1$. This implies that in the scaling limit the mass and charge of the supersymmetric solutions are rescaled away and we go over to a non-trivial geometry \eqref{plneutralsol} supported by the scalar field.

The existence of a non-trivial metric with no mass is a very curious phenomenon, but not entirely unprecedented. For one the scaling arguments presented in \sec{sec:planarlimit} as we discussed imply that global supersymmetric solutions must necessarily go over to $m =0$ planar solutions. 
Also as is well-known from the designer gravity constructions, it is possible owing to the presence of non-trivial boundary conditions to attain solutions with zero ADM mass.  As mentioned earlier, once one relaxes boundary conditions we are guaranteed a positive energy theorem (under certain assumptions, cf., \cite{Faulkner:2010fh}), but the global minimum of the energy can indeed be negative. In fact, the solution \eqref{plneutralsol} is also the limiting behaviour of designer gravity solitons originally studied in \cite{Hertog:2004ns}. In the designer gravity language one engineers boundary conditions $\phi_2(\phi_1)$ so as to guarantee existence of solitons. For the theory with potential \eqref{dgdata} the designer soliton curve $\phi_2(\phi_1) \to \text{constant}$ for $\phi_1 \gg 1$. As noted in \cite{Faulkner:2010fh} this is a non-generic case with an exactly determinable fake-superpotential; knowledge of the latter allows us to infer the solution \eqref{plneutralsol}.


\section{Discussion}\label{sec:Discussion}


In this paper we explored the behaviour of charged scalar solitons in asymptotically global \AdS{4} spacetimes.  The main objective for undertaking such an analysis was to obtain a better understanding of when such solutions admit a planar limit and in particular morph smoothly into corresponding solutions in Poincar\'e AdS. As explained in \sec{sec:Introduction}, previous studies of compact objects in AdS geometries (such as stars or solitons) had hitherto only revealed a single branch of solutions, viz., those which were connected to the global \AdS{4} geometry perturbatively. These objects  were confined to an AdS length scale, unless one countered the gravitational attraction by matter repulsion. In contrast, we showed the existence of new branches of solutions even for `small' amounts of matter repulsion.  By examining the scaling-invariant quantities we were able to demonstrate explicitly that the planar limit of global solitons in phenomenological Abelian-Higgs theories matched up with the zero-temperature limit of planar black holes with charged hair. Whilst these results were obtained for a particular value of the scalar mass, we believe them to be generic for other masses and also extend to higher  dimensional \AdS{} geometries. We also showed that the features we see in the phenomenological models are also manifested in models that can be consistently embedded into 11D supergravity with minor modifications to account for the non-trivial matter interactions present in these examples.

An important result of our numerical studies was the unveiling of  critical behaviour in the family of solitonic solutions. This was particularly well-illustrated in the  branch of solitons connected to the global \AdS{4} vacuum. As we increase the charge of the scalar, the solitons go from being bounded in their asymptotic charges to large solitons that connect up to planar configurations. On the other hand one can directly study planar scalar hair solutions, which has been carried out extensively in the past few years, where one finds no peculiar behaviour as a function of the scalar charge. This observation led us to the discovery of a surprising branch of scalar soliton solutions: these have a planar limit but are disconnected from the global \AdS{4} vacuum!  We found a very intricate pattern of branchings in the soliton solution space as a function of the scalar charge; solution branches underwent a series of deformations and even formed closed bubbles in solution space. Such rich structure signals a rather intricate interplay of the non-linearities in the system and it would be intriguing to carry out a direct dynamical systems analysis to shed light on these features.

There is by now a huge literature on asymptotically \AdS{} solutions in both phenomenological models and consistent embeddings of supergravity.  We would like to propose that the method of constructing global solutions then `blowing them up' into the  planar limit is a  useful diagnostic in general. In particular, it serves as an efficient technique to find new connections between various solutions. We leave a study of this richer story to future work.

In the context of models that come from consistent truncations of supergravity/string theory, the ability to choose multiple boundary conditions for the scalar field provided us with a different dial to tune in the consistent truncations we studied. In the context of holographic superconductors, \cite{Faulkner:2010gj} and \cite{Iqbal:2011aj} have already emphasised the role of this dial in providing access to new quantum critical behaviour. For the $SU(3)$ truncation discussed in \sec{sec:GSW} we could show by using the double-trace deformations that the standard quantisation of the scalar field ($\Delta = 2$ boundary condition) was located at a critical point; specifically, it exhibited features seen for the special value of charge, $q = q_c$, in the Abelian-Higgs models.  Similarly, this same method also allowed us to elicit critical behaviour (albeit of a slightly different kind) in the $U(1)^4$ truncation (M2-brane theory discussed in \sec{sec:DG}) for the supersymmetric ($\Delta =1$) boundary condition.  

At the same time it should be emphasised that tuning of  boundary conditions allows one to actually explore the phase space of solutions more completely. For example, in the M2-brane theory we would have not been able to find a new branch of non-supersymmetric solitons for dimensional one operators, without the ability to tweak the boundary conditions. More generally, solutions that are disconnected from the \AdS{4} vacuum are most easily obtained if we can access deformations by tweaking Lagrangian parameters, boundary conditions, etc., and then relaxing back to the undeformed theory. The hysteresis inherent in such relaxations enables one explore non-perturbative segments of the solution space. 

We should note that it is also possible to infer critical behaviour by other means: for example, for the M2-brane theory one can easily show that the supersymmetric solitons lie at a critical point simply by analytically continuing in the `number of $U(1)$ charges'. To understand this consider \eqref{DGsusy} which is the BPS equation in the $\frac{1}{8}$-BPS sector of the M2-brane theory with all four $U(1)$ charges equal. The  general BPS equation for different values of the four $U(1)$ charges was also derived in \cite{Chong:2004ce}. In particular, we can specialise to solutions preserving higher amounts of supersymmetry by setting some of the charges to zero; its effect if we restrict to the locus of equal non-vanishing charges is to modify the exponents in \eqref{DGsusy}: for $p \leq 4$ non-vanishing charges $H^4 \to H^{p}$ and 
$H^3 \to H^{p-1}$. One can however study this theory for any $p$ and show that $p=4$ is a critical point; for $p\leq 4$ we see solutions that exhibit a slow but unbounded growth of the conserved charges as in \sec{sec:DG}, but solutions with $p >4$ have bounded charges in the branch connected with the \AdS{4} vacuum. Moreover,  the toy model clearly shows how attractor solutions at large $\phi_c$ determine the phase space of solutions and also allows one to track the migration of the attractor solution itself as a function of the artificially introduced parameter $p$. While such toy models have merit in enabling analytic studies, it is indeed useful for have more physical dials, such as the scalar boundary conditions, to explore the solution space.

 As an example of the utility of the global solutions, let us briefly mention an example on which we believe our results can shed some light. In a different sector of the $SU(3)$ truncation, studied in \cite{Donos:2011ut}, the authors found charged planar domain walls for one choice of scalar boundary condition but not the other. Based on our results it seems likely that the two different choices should correspond to the planar limits of charged global soliton branches that have unbounded or bounded mass, respectively. It would indeed be interesting to verify this result, for while the specific truncation in question is more complicated than the single field truncations we have studied in this paper, it has the advantage that it allows interpolation between the \AdS{4} vacuum of the M2-brane theory (which preserves $SO(8)$ R-symmetry) to a new supersymmetric vacuum preserving $SU(3) \times U(1)$.  In particular, we should be able to again see emergence of new global domain wall solutions interpolating between the two vacua. 
  
We have also discussed the location of solitons on the microcanonical phase diagram for various theories.  For the supersymmetric quantisation of the M2-brane theory,  it is clear that global solitons form the microcanonical phase boundary because they saturate the BPS bound. We have refrained however from making a similar conjecture for the Abelian-Higgs model because in certain cases one can find global hairy black holes that are lighter than a soliton at a given value of the charge; a more compete discussion of this issue should appear in \cite{Dias:2011tj}.  Whilst we have not explicitly studied global hairy black holes in the $SU(3)$ truncation, we suspect that they will have higher mass in comparison to a solitonic solution at a fixed value of the charge: we in fact conjecture that in this model that microcanonical phase boundary is set by the scalar solitons that we have constructed herein. More generally, it would be interesting to ascertain whether the exponential scalar potentials encountered in consistent truncations models necessitates the solitons being the minimal mass solution in a given charge sector. 

As emphasised in \sec{sec:Introduction} the analysis of solutions in global \AdS{} has the advantage of allowing one to explore finite size effects in a large $N$ field theory. While much of the recent literature of applications of AdS/CFT to condensed matter systems has been restricted to the Poincar\'e patch, it would be useful to generalise the constructions to the global geometry where we have a covariant IR regulator in the spacetime. In particular, it would be interesting to undertake the study of low temperature superfluid behaviour in a strongly-coupled theory on $\mathbb{R} \times {\bf S}^2$. Similarly, it would be of interest to extend consideration of gravitationally interacting fermionic systems to the global setting.

The solitonic solutions discussed here are as elaborated in \sec{sec:Introduction} coherent condensates of bosons, carrying macroscopic charges in the dual field theory. In a strongly coupled system at finite volume, one generically expects that systems should thermalise (in a microcanonical sense at fixed charges). It is rather curious that the non-linear interactions of gravity allow for non-thermal, i.e., non-black hole solutions.\footnote{Note that the solitons are qualitatively different from the CFT states on ${\mathbb R} \times {\bf S}^2$ which exhibit undamped collective oscillations (oscillons) \cite{Freivogel:2011xc}; for one our solutions are globally static.} Generic initial data for matter fields with asymptotically AdS boundary conditions is expected to collapse into a black hole \cite{Dafermos:2006fk}; this has indeed be verified for neutral scalar fields in \cite{Bizon:2011gg}. It would be very interesting to delineate the class of fine tuned initial data which does not thermalise in the strongly coupled field theory dual.

Finally, let us comment on an interesting application of our results to the study of supersymmetric states in the AdS/CFT correspondence. While solutions preserving 4 or more supercharges are by now well-understood, one interesting open issue in the AdS/CFT correspondence relates to the microscopic understanding of supersymmetric black holes, that preserve 2 supercharges. For example in the M2-brane theory discussed in this paper, these solutions would be $\frac{1}{16}$-BPS. It would be interesting to ask whether there are solitonic solutions which preserve such low amounts of supersymmetry and if so what their role is in the dual field theory. For the specific case of \AdS{4} this is an interesting challenge: the solutions by virtue of the BPS condition are required to carry non-vanishing angular momentum, and this in particular makes them co-homogeneity two (the metric and other functions depend on both the radial coordinate and the polar angle of the ${\bf S}^2$).  Preliminary investigations of a simpler set of examples in \AdS{5}, where the presence of two rotation planes allows for co-homogeneity one solutions, was reported in \cite{Bhattacharyya:2010yg}, who studied such within perturbation theory. It would be interesting to extend the analysis to a full numerical investigation to flesh out the intricacies of supersymmetric objects in \AdS{}.

\subsection*{Acknowledgements}
\label{sec:acks}

It is a pleasure to thank Jerome Gauntlett, Sean Hartnoll, Veronika Hubeny, Jorge Santos, Julian Sonner and Toby Wiseman for helpful discussions. MR in addition would like to thank James Lucietti for earlier collaboration and discussions on supersymmetric solutions in \AdS{}. MR would like to thank KITP for hospitality during the ``Holographic Duality and Condensed Matter Physics'' program where part of this work was done. SAG is supported by an STFC studentship. MR is supported in part by an STFC rolling grant  and by the National Science Foundation under Grant No. NSF PHY05-51164. BW is supported by a Royal Commission for the Exhibition of 1851 Research Fellowship.

\appendix
\section{Bulk equations of motion and boundary terms}\label{sec:generalEoms}

In this appendix we present the equations of motion for the general action of \eqref{Sgeneral}, and calculate general boundary counterterms required. The functions $Q(\phi)$ and $V(\phi)$ for the theories of interest are listed in table \ref{theorytable}.  We find
\begin{eqnarray}
R_{ab}-\frac{1}{2} g_{ab}R &=& \frac{1}{2}\left(F_{ac}F_b^{\phantom{b}c} - \frac{1}{4} g_{ab} F^2\right) + \left(\partial_a \phi \partial_b \phi - \frac{1}{2} g_{ab} \left(\partial\phi\right)^2\right)\nonumber\\
&&+\frac{1}{\ell^2} Q(\phi) \left(A_a A_b - \frac{1}{2} g_{ab} A^2\right) - \frac{1}{2 \ell^2} g_{ab} V(\phi)\\
\nabla_a F^{ab} &=& \frac{2}{\ell^2} Q(\phi) A^b\\
\Box \phi &=& \frac{1}{2}\frac{Q'(\phi)}{\ell^2} A^2 + \frac{1}{2}\frac{V'(\phi)}{\ell^2}.
\end{eqnarray}

We must supplement our bulk action \eqref{Sgeneral} with appropriate boundary terms: our total action is
 \begin{equation}
S = S_{bulk} + \frac{1}{8\pi G_4} \int d^3x \,\sqrt{-\gamma} \, K+ S_{ext} + S_{ct}.
\end{equation}
where $S_{ext}$ is the collection of scalar and Maxwell boundary terms necessary for a good variational principle (we have explicitly indicated the Gibbons-Hawking term).  $S_{ct}$ is an expression covariant in intrinsic boundary quantities designed to render the on-shell action finite.

The boundary terms for Einstein-scalar theories with multi-trace boundary conditions for scalars  were originally derived using the covariant phase space formalism in \cite{Amsel:2006uf} and extended to include Maxwell fields in \cite{Faulkner:2010gj}.  By analyzing the asymptotic behaviour of various fields it is easy to convince oneself that the counter terms of interest are independent of $Q(\phi)$ and depend only on the quadratic approximation to $V(\phi)$. Thus, by writing down the counterterms for \eqref{Sbottomup}, we immediately obtain the results of relevance for the more general case. We thus directly borrow the results of these references and quote the final answer without further derivation below.

The main quantity we require is the  ADM energy density $m$ which we use to characterise our solutions. This turns out to  be given by \cite{Hertog:2004ns,Amsel:2006uf}: 
\begin{equation}
m = g_1 + \phi_1 \, \phi_2 + \frac{1}{2}\, \varkappa\, \phi_1^2 \equiv  g_1 + \frac{3}{2}\, \varkappa\, \phi_1^2
\label{mdtrace}
\end{equation}	
for double-trace boundary conditions. Here $g_1$ is the coefficient of the $1/r$ term in the asymptotic behaviour of the metric function $g(r)$ as indicated in \eqref{gasym}, with $\phi_1$ and $\phi_2$ the fall-offs of the scalar field \eqref{phiasym}. Note that in exactly the dimension 1 and dimension 2 cases we have $g_1 = m$.
The total energy of the solutions is $E = {\rm Vol}({\bf S}^2) \, m$; we prefer to use the energy density since it is more naturally suited to analyzing the planar limit. The other conserved charge of interest, the charge density, is obtained simply from the fall-off of the gauge field -- it is given by $\rho$ in \eqref{Aasym}.

\section{Perturbative construction of global solitons}\label{sec:pertsolitons}

In this short appendix we explain the strategy to construct global solitons in a perturbation expansion around the \AdS{4} vacuum. The first construction of such solutions was presented in \cite{Basu:2010uz} and we will effectively be reviewing the same for our models. The general logic of our construction is independent of the details of the scalar field, though both $V(\phi)$ and $Q(\phi)$ will enter into the details of the perturbation theory. 

The starting point for such a construction is the fact that the eigenmodes of the charged scalar field in global \AdS{4} have frequencies $\omega = \Delta +2 \, n$ with $n \in {\mathbb Z}_+$ as long as  we maintain spherical symmetry. Actually, we will be interested in the ground state soliton, so we also have the luxury of setting $n=0$. This ground state is then characterised by the zero-point energy $\Delta$ which is of course determined solely by the scalar mass via the standard formula $\Delta  = \frac{3}{2} \pm \sqrt{\frac{9}{4} + m_\phi^2\, \ell^2}$, where we are restricting attention to $d =3$ i.e., \AdS{4}. We will also have need of the scalar wave-function, which for $m^2_\phi \,\ell^2 = -2$ takes a simple form:\footnote{For generic $\Delta$ the scalar profiles which solve the linearised wave-equation are hypergeometric functions}
\begin{equation}
\phi_{lin}(r) = \frac{1}{(1+r^2)^\frac{\Delta}{2}}
\label{}
\end{equation}	
This wave-function is clearly regular at the origin and satisfies $\phi_{lin}(r) \to r^{-\Delta}$ as $r\to \infty$. Since we noted that the scalar field here has a nontrivial ground state energy, a neutral scalar field will also have a temporal oscillation with frequency $\omega = \Delta$ which is a pure phase. For a charged scalar field this phase can be absorbed into a background gauge field (by the same trick we used in writing the Lagrangian in \sec{sec:tab}). Essentially all we need to do is turn on a constant background gauge field tuned such that we mock up the phase.
The upshot of this discussion is that there is a static perturbation around the global \AdS{4} geometry \eqref{globalschw} (with $m=\rho=0$) where
\begin{equation}
\phi_{lin}(r) = \frac{1}{(1+r^2)^\frac{\Delta}{2}} \ , \qquad \mu = \frac{\Delta}{q}
\label{}
\end{equation}	
assuming that for small $\phi$, $Q(\phi) \simeq q^2\,\phi^2$. This is all the data we need to set up the perturbation expansion. 

Denote by $\varepsilon$ the vev of the dual operator in the field theory; i.e., $\varepsilon = \langle {\cal O}_\Delta \rangle$  -- this will be the small parameter we will use to study perturbation theory. The perturbative solution ansatz can be written as (treating both $\Delta =1$ and $\Delta =2$ together):
\begin{eqnarray}
\phi(r) &=& \frac{\varepsilon}{(1+r^2)^\frac{\Delta}{2}} + \sum_{k=1}^{\infty} \, \varepsilon^{2k+1} \, \phi_{(2k+1)}(r) \nonumber \\
A(r) &=& \frac{\Delta}{q} + \sum_{k=1}^{\infty} \, \varepsilon^{2k} \, A_{(2k)}(r) 
\nonumber \\
g(r) &=& 1+r^2+ \sum_{k=1}^{\infty} \, \varepsilon^{2k} \, g_{(2k)}(r)
\nonumber \\ 
f(r)&=&1+r^2+ \sum_{k=1}^{\infty} \, \varepsilon^{2k} \, f_{(2k)}(r)
\label{}
\end{eqnarray}	
where for brevity we define $f(r) \equiv g(r) \,e^{-\beta(r)} $.

It is a simple matter to iteratively solve for the corrections $\phi_{(2k+1)}(r)$, $A_{(2k)}(r)$, $g_{(2k)}(r)$, $f_{(2k)}(r)$ by recursively solving the equations order by order in powers of $\varepsilon$. Most of the analysis is in fact analytical, because the linearised equations admit closed-form solutions. The integration constants are all fixed by the choice of boundary conditions: regularity at the origin and normalizability  at \AdS{} asymptopia (which we take to also include the scalar boundary condition provided in terms of the fall-off condition).

We now simply quote the results of such a construction for the Abelian-Higgs model discussed in \sec{sec:BU}. The physical parameters characterizing the solution which are of interest are all expressed as functions of $\langle {\cal O}_\Delta \rangle$. For the $\Delta =2$ boundary condition we find
\begin{eqnarray}
\varepsilon &=& \langle {\cal O}_{\phi_1} \rangle
\nonumber \\
\mu &=& \frac{2}{q}+\frac{5 q^2-2}{8q} \varepsilon ^2+ \frac{\left(52 \pi ^2-615\right) q^4+\left(1322-114 \pi ^2\right) q^2+74 \pi ^2-821}{384 q}\varepsilon ^4+ {\cal O}(\varepsilon^6)
\nonumber \\
\rho &=& \frac{\pi}{4}   q \varepsilon ^2-\frac{\pi}{192}   q \left(\left(4 \pi ^2-25\right) q^2+58-8 \pi ^2\right) \varepsilon ^4 +{\cal O}(\varepsilon^6)
\nonumber \\
m &=& \frac{\pi }{4} \varepsilon ^2-\frac{\pi}{384}   \left(\left(8 \pi ^2-65\right) q^2+122-16 \pi ^2\right) \varepsilon ^4+ {\cal O}(\varepsilon^6)
\nonumber \\
\phi_c &=&\varepsilon + \frac{1}{48} \left(\left(16-3 \pi ^2\right) q^2+6 \pi ^2-40\right) \varepsilon ^3+ {\cal O}(\varepsilon^5)
\label{}
\end{eqnarray}	
while the $\Delta =1$ boundary condition leads to
\begin{eqnarray}
\varepsilon &=& \langle {\cal O}_{\phi_2} \rangle
\nonumber \\
\mu &=& \frac{1}{q}+\frac{2 q^2+1}{4q}\varepsilon ^2+\frac{\left(64 \pi ^2-720\right) q^4+\left(660-60 \pi ^2\right) q^2+17 \pi ^2-192}{384 q} \varepsilon^4+ {\cal O}(\varepsilon^6)
\nonumber \\
\rho &=& \frac{\pi}{2}   q \varepsilon ^2-\frac{\pi }{24}   q \left(\left(2 \pi ^2-18\right) q^2+6-\pi ^2\right) \varepsilon ^4+ {\cal O}(\varepsilon^6)
\nonumber \\
m &=& \frac{\pi}{4}  \varepsilon ^2-\frac{\pi}{96}  \left(\left(4 \pi ^2-42\right) q^2+9-2 \pi ^2\right) \varepsilon ^4+ {\cal O}(\varepsilon^6)
\nonumber \\
\phi_c &=& \varepsilon - \frac{1}{16} \left(\pi ^2-8\right) \left(2 q^2-1\right)\varepsilon^3 + {\cal O}(\varepsilon^5)
\label{}
\end{eqnarray}	
Note that we have quoted here the results parameterised by the vev of the CFT operator which is the natural parameter for the perturbation expansion. It is possible to set up the perturbation directly in terms of the core scalar value, $\phi_c$ but that appears to be a bit more cumbersome.

It is interesting to ask whether solitons with multi-trace boundary conditions can be constructed using a similar perturbative expansion. One might naively have thought this to be possible; after all the multi-trace operators correspond to multi-particle states in the bulk via the AdS/CFT dictionary. Consider then the case where we impose a double-trace boundary condition $\phi_2 = \varkappa \, \phi_1$. What we would like to do is find a regular two particle wave-function that in the linear approximation is regular at the origin and satisfies the asymptotic boundary condition mentioned above. 
This turns out to be impossible around the vacuum solution, viz., global \AdS{4}, for reasons we explain below.

Previously, we wrote down the linear eigenmodes for the scalar wave operator, where we had imposed regularity at the origin. It turns out that for $m^2_\phi \,\ell^2 = -2$ the solutions to the linear wave equations for static spherically symmetric configurations about \AdS{4} are 
\begin{eqnarray}
\phi^{(2)}_{lin}(r) &=& \frac{1}{1+r^2} \left(\alpha_2 \, \frac{r^2-1}{r}+ \beta_2 \right) \ , \qquad \mu = \frac{2}{q}\nonumber \\
\phi^{(1)}_{lin}(r) &=& \frac{1}{\sqrt{1+r^2}} \left(\alpha_1+ \frac{\beta_1}{r} \right) \ , \qquad \mu = \frac{1}{q}
\label{linmodes}
\end{eqnarray}	
For the standard boundary condition (i.e., $\Delta =2$) the ground state wave function was given by the first line with $\beta_2 =\phi_2$ and $\alpha_2 =0$ while the alternate boundary condition (i.e., $\Delta =1$) involved a seed solution with $\alpha_1 = \phi_1 $ and $\beta_1 =0$. 

The double-trace deformation of the dual CFT involves ${\cal O}_1^2$ which should correspond to a two-particle state. The naive wave-function for this appears to be simply $\phi_{lin}(r) \simeq \frac{1}{1+r^2}$ which can be viewed for the moment as $(\phi_{lin}^{(1)})^2$. Fortuitously, this is a static mode once we set $\mu = \frac{2}{q}$; this is natural since we are dealing with a two particle state (so the net charge is doubled).  However, this cannot be the full story -- the mode in question does not satisfy the asymptotic boundary condition. Note that double-trace boundary condition requires that we have $\phi_{lin}(r) \to \frac{\phi_1}{r} + \frac{\varkappa\,\phi_1}{r^2}$. Inspecting \eqref{linmodes} it is immediately clear that there is no  linearised  solution that is both regular at the origin and satisfies the asymptotic fall-off (for one we cannot simply set $\alpha_2 = \phi_1$ and $\beta_2=\varkappa\,\phi_1$ as it it irregular at the origin). 

The absence of a seed solution with correct boundary conditions makes it clear that it is not possible to set up a perturbation series analogous to the single trace boundary condition. We believe this is because of the peculiar nature of the multi-trace boundary conditions -- while they naively correspond to multi-particle state in the field theory, such states rather than being described by a new bulk effective field (which should roughly speaking be $\phi^2$) are rather described 
by imposing a local relation between boundary fall-offs. This is somewhat reminiscent of issues arising with double-trace fermonic operators as recently discussed in \cite{Bolognesi:2011un}. We hope to return to a proper resolution of this issue in the future.

Conceptually, the breakdown of perturbation theory around global \AdS{4} is not too hard to understand. While one can easily see this geometry (and the dual CFT vacuum) being the appropriate state to expand around for single trace operators, it is no longer the case when we have an explicit deformation in the CFT Lagrangian. If one has a candidate ground state for the deformed CFT, then we would be able to undertake a perturbation analysis about it.   The trouble is that constructing the appropriate bulk geometry usually involves numerical solving of the field equations. In the bulk of the paper we have side-stepped this question by direct numerical integration of the field equations with the appropriate boundary conditions. It would indeed be interesting to find examples where we can adapt perturbation theory techniques to multi-trace deformed field theories.

%

\providecommand{\href}[2]{#2}\begingroup\raggedright\endgroup

\end{document}